\documentclass[argument]{aastex61}
\usepackage{multirow}
\usepackage{amssymb, amsfonts, amsmath}
\usepackage{fontawesome}

\usepackage{stmaryrd}
\usepackage{longtable}
\usepackage{pifont}

\usepackage{relsize}
\usepackage{csquotes}
\usepackage{units,icomma}
\usepackage{epstopdf}
\usepackage{lineno}

\usepackage{graphicx}
\usepackage{textgreek}
\usepackage{csvsimple,booktabs}
\usepackage{filecontents}
\usepackage{tcolorbox}
\usepackage{xfrac, nicefrac}
\usepackage{hhline}
\usepackage[para,online,flushleft]{threeparttable}
\usepackage{xcolor}
\colorlet{BLUE}{blue}
\colorlet{MAGENTA}{magenta}

\newcommand{\mystar}{\scalebox{1.02}{$\star$}}
\newcommand{\hollowstar}{\text{\ding{73}}}
\newcommand{\myhollowstar}{\scalebox{0.7}{$\hollowstar$}}

\makeatletter
\DeclareRobustCommand{\HI}{%
  \mbox{H\check@mathfonts\fontsize\sf@size\z@\selectfont\,I}%
}

\shorttitle{Search for Extended GeV Sources in the Inner Galactic Plane}
\shortauthors{\textit{Fermi}-LAT Collaboration}

\begin{document}

\title{Search for Extended GeV Sources in the Inner Galactic Plane}

\AuthorCollaborationLimit = 120

\author[0000-0002-6803-3605]{S.~Abdollahi}
\email{abdollahi.soheila@gmail.com}
\affiliation{IRAP, Universit\'e de Toulouse, CNRS, UPS, CNES, F-31028 Toulouse, France}
\author{F.~Acero}
\affiliation{AIM, CEA, CNRS, Universit\'e Paris-Saclay, Universit\'e de Paris, F-91191 Gif-sur-Yvette, France}
\author[0000-0002-2028-9230]{A.~Acharyya}
\affiliation{Center for Cosmology and Particle Physics Phenomenology, University of Southern Denmark, Campusvej 55, DK-5230 Odense M, Denmark}
\author{A.~Adelfio}
\affiliation{Istituto Nazionale di Fisica Nucleare, Sezione di Perugia, I-06123 Perugia, Italy}
\author[0000-0002-6584-1703]{M.~Ajello}
\affiliation{Department of Physics and Astronomy, Clemson University, Kinard Lab of Physics, Clemson, SC 29634-0978, USA}
\author[0000-0002-9785-7726]{L.~Baldini}
\affiliation{Universit\`a di Pisa and Istituto Nazionale di Fisica Nucleare, Sezione di Pisa I-56127 Pisa, Italy}
\author[0000-0002-8784-2977]{J.~Ballet}
\affiliation{Universit\'e Paris-Saclay, Universit\'e Paris Cit\'e, CEA, CNRS, AIM, F-91191 Gif-sur-Yvette Cedex, France}
\author[0000-0001-7233-9546]{C.~Bartolini}
\affiliation{Istituto Nazionale di Fisica Nucleare, Sezione di Bari, I-70126 Bari, Italy}
\affiliation{Universit\`a degli studi di Trento, via Calepina 14, 38122 Trento, Italy}
\author[0000-0002-6729-9022]{J.~Becerra~Gonzalez}
\affiliation{Instituto de Astrof\'isica de Canarias and Universidad de La Laguna, Dpto. Astrof\'isica, 38200 La Laguna, Tenerife, Spain}
\author[0000-0002-2469-7063]{R.~Bellazzini}
\affiliation{Istituto Nazionale di Fisica Nucleare, Sezione di Pisa, I-56127 Pisa, Italy}
% ...

\begin{abstract}
The recent detection of extended $\gamma$-ray emission around middle-aged pulsars is interpreted as inverse-Compton scattering of ambient photons by electron-positron pairs escaping the pulsar wind nebula, which are confined near the system by unclear mechanisms.~This emerging population of $\gamma$-ray sources was first discovered at TeV energies and remains underexplored in the GeV range.~To address this, we conducted a systematic search for extended sources along the Galactic plane using 14 years of \textit{Fermi}-LAT data above 10\,GeV, aiming to identify a number of pulsar halo candidates and extend our view to lower energies.~The search covered the inner Galactic plane ($\lvert l\rvert\leq$\,100$^{\circ}$, $\lvert b\rvert\leq$\,1$^{\circ}$) and the positions of known TeV sources and bright pulsars, yielding broader astrophysical interest.~We found 40 such sources, forming the Second \textit{Fermi} Galactic Extended Sources Catalog (2FGES), most with 68\% containment radii smaller than 1.0$^{\circ}$ and relatively hard spectra with photon indices below 2.5.~We assessed detection robustness using field-specific alternative interstellar emission models and by inspecting significance maps.~Noting 13 sources previously known as extended in the 4FGL-DR3 catalog and five dubious sources from complex regions, we report 22 newly detected extended sources above 10\,GeV.~Of these, 13 coincide with H.E.S.S., HAWC, or LHAASO sources; six coincide with bright pulsars (including four also coincident with TeV sources); six are associated with 4FGL point sources only; and one has no association in the scanned catalogs.~Notably, six to eight sources may be related to pulsars as classical pulsar wind nebulae or pulsar halos.
\end{abstract}
\keywords{Gamma rays: general --- surveys --- catalogs}

\section{Introduction} \label{sec:intro}

The detection of spatially extended $\gamma$-ray emission around several middle-aged pulsars (typically $\gtrsim$\,100 kyr) has been reported in the TeV domain over the past few years~\citep{Abeysekara17, Aharonian21}.~This diffuse emission, which extends over spatial scales of a few tens of parsecs, is interpreted as inverse-Compton (IC) scattering of background radiation fields by high-energy electron-positron pairs accelerated in pulsars and their wind nebulae, and subsequently released into their surroundings.

The observed intensity distribution of these so-called TeV halos or pulsar halos suggests efficient spatial confinement, driven by a physical mechanism yet to be elucidated.~In the case of the halos around pulsars PSR J0633+1746 and PSR B0656+14, as observed by the HAWC collaboration~\citep{Abeysekara17}, and around PSR J0622+3749, as observed by the LHAASO collaboration~\citep{Aharonian21}, the diffusion of particles is significantly inhibited within a few tens of parsecs from the pulsars compared to the surrounding interstellar medium (ISM), being reduced by nearly 2$-$3 orders of magnitude for $\sim$100 TeV electrons.

Pulsar halos offer an opportunity to probe the transport properties of energetic particles in the vicinity of their accelerators.~The origin of the suppressed particle diffusion inferred from TeV observations of some pulsars is still under debate, and several theoretical scenarios have been put forward.~A first option is self-confinement, where the gradient of pairs escaping the nebula triggers kinetic instabilities out of which turbulence develops and scatters particles themselves~\citep{Evoli18, Mukhopadhyay.Linden22}.~A second option is pre-existing external turbulence, in which escaping pairs encounter a region of space where, for some reason like the expansion of the parent supernova remnant (SNR), standard magnetohydrodynamical turbulence has the right properties for efficient confinement of $\sim$100 TeV particles~\citep{LopezCoto18,Fang19}.~A third option, that can be combined with the second scenario, is that the pulsar halo develops in turbulence previously generated by cosmic-ray protons as they escaped the parent remnant~\citep{Schroer21,Schroer22}.~A fourth option, involving perpendicular diffusion as the origin of the confinement, and a favorable magnetic field alignment with respect to the line of sight~\citep{Liu19}, seems increasingly unlikely as more pulsar halos are discovered~\citep{DeLaTorreLuque22}.

This question has consequences on the interpretation of the locally measured cosmic-ray positron flux in relation to nearby pulsars.~In contrast to the assumption in \citet{Abeysekara17} that suppressed diffusion inferred from J0633+1746 and B0656+14 observations extends up to Earth and cuts off their contribution to the positron flux, there seems to be a growing consensus on the idea that inhibited diffusion is restricted to a reasonably small region surrounding the pulsars.~In this scenario, nearby middle-aged pulsars are likely responsible for the observed positron excess~\citep{Profumo18, Fang18, Tang.Piran19, Martin22a}.

As an emerging population of $\gamma$-ray emitters, pulsar halos may constitute a rich source class, owing to the large number of pulsars in the Galaxy and possible long duration of the halo phase (hundreds of thousands of years).~Some models estimated the number of halos detectable by current and future instruments and have shown that a significant fraction of existing TeV sources currently classified as unidentified or as pulsar wind nebulae (PWNe) might actually be halos~\citep{Linden17, Sudoh19, Martin22b}.~Those TeV halos currently below the detection threshold may significantly contribute to the large-scale interstellar diffuse emission in the TeV and sub-PeV domains~\citep{Linden18,Sudoh19,Martin22b}, until they are resolved into individual sources by future surveys with higher sensitivity.

Despite numerous studies on pulsar halos in the TeV domain, their population and properties at lower GeV energies are yet to be explored for the most part.~\citet{DiMauro19} reported the detection of large extended emission around PSR J0633+1746 above 8 GeV with the \textit{Fermi} Large Area Telescope~\citep[LAT;][]{Atwood09}.~For two other pulsars with TeV halos (B0656+14 and J0622+3749), no significant extended emission is detected from LAT data~\citep{DiMauro19, Aharonian21}.~Another search for extended GeV emission around three bright sources, detected with HAWC up to beyond 100\,TeV and coincident with pulsars, resulted in the detection of a $\gamma$-ray halo around two of them, with properties consistent with significant diffusion suppression~\citep{DiMauro21}.~Yet, the corresponding pulsars are relatively young objects with characteristic ages of $\sim$20 kyr, and it is not clear that these observations are probing a late-stage halo phase and not just a more classical early-stage PWN undergoing particle escape \citep{Martin24}.

In this work, we report the results of a comprehensive systematic search for new extended GeV sources in the Galactic plane, using 14 years of \textit{Fermi}-LAT Pass 8 data at energies from 10\,GeV to 1\,TeV.~The analysis consisted in both a search for GeV counterparts of known TeV sources and bright pulsars,~as well as a blind search for extended sources along most of the Galactic plane.~We eventually obtain a catalog of 40 extended sources and present a list of their possible associations with pulsars or higher-energy $\gamma$-ray sources.~This dataset is expected to trigger more in-depth observations and studies to improve our knowledge of particle acceleration and early propagation in a variety of astrophysical setups, for instance pairs released from PWNe, cosmic rays escaping from SNRs, or non-thermal particle transport in star-forming regions.~Compared to the First \textit{Fermi} Galactic Extended Sources catalog~\citep[FGES catalog;][]{Ackermann17},~this study uses a substantially larger dataset and introduces some evolutions in the data analysis strategy, source classification, association, and identification.

The paper is organized as follows.~In Section~\ref{sec:InstObs}, we describe the instrument, data selection and preparation, and analysis framework.~The methodology that was developed to search for extended sources and study some systematic effects relevant to this analysis is outlined in Section~\ref{sec:SysSearchExtSrcs}.~The final catalog of extended sources, labelled the Second \textit{Fermi} Galactic Extended Sources (2FGES) catalog, is presented in Section~\ref{sec:catalog}, where the spatial and spectral information on the 2FGES sources is provided, and their potential associations with known higher-energy $\gamma$-ray sources and strong pulsars are examined.~A discussion of possible pulsar halo candidates is given in Section~\ref{sec:Discussion}.~Finally, we summarize the main findings of this study in Section~\ref{sec:Conclusions}.

\section{\textit{Fermi}-LAT Data And Analysis} \label{sec:InstObs} 

\subsection{Data Selection} \label{subsec:Data}
The LAT on board the \textit{Fermi Gamma-ray Space Telescope} detects photons in the energy range from $\sim$20 MeV to more than 1 TeV using a pair-conversion technique.~It is comprised of a tracker$\slash$converter for direction reconstruction of incident $\gamma$-rays, a CsI(Tl) crystal calorimeter for measurement of the energy deposition, and a surrounding anti-coincidence detector for rejection of the charged particle background.~It predominantly operates in continuous sky-survey mode, during which it scans uniformly the entire sky every $\sim$3 hr (corresponding to two orbits), thanks to its wide field of view ($\sim$2.4 sr at 1\,GeV).~Further description of the LAT instrument and data processing can be found in~\citet{Atwood09} and \citet{Ajello21}, and information regarding the on-orbit calibration is given in~\citet{Abdo09}.

We analyzed LAT data corresponding to the Pass 8 P8R3 event reconstruction scheme~\citep{Atwood13, Bruel18} and collected in the fourteen-year period starting from 2008 August 4 to 2022 July 21 (MJD 54682--59781).~The event selection is based on the low background $\texttt{SOURCEVETO}$ class with the corresponding instrument response functions (IRFs) $\texttt{P8R3\char`_SOURCEVETO\char`_V3}$, which is ideal for diffuse emission analysis.\footnote{\url{https://fermi.gsfc.nasa.gov/ssc/data/analysis/documentation/Cicerone/Cicerone_Data/LAT_DP.html}}~Events with a zenith angle $>$\,105$^{\circ}$ were excluded to reduce contamination from Earth limb $\gamma$-rays.~We selected time intervals when the data quality was good ($\texttt{DATA\char`_QUAL>0}$), and the instrument was in the nominal science data acquisition mode ($\texttt{LAT\char`_CONFIG==1}$).

We performed the data analysis in the 10~GeV--1~TeV energy range.~The 10 GeV lower limit ensures an angular resolution better than $0$\fdg$15$, which allows for better separation of extended sources from the Galactic interstellar diffuse emission.~Additionally, it reduces the relative contribution of pulsars to the emission in some fields of interest, as most pulsars have magnetospheric emission that cuts off beyond several GeV~\citep{Smith23}.

\subsection{Analysis Framework and Tools} \label{subsec:Analysis}

The search for extended halos was run on a list of 544 seed positions (to be introduced in Section~\ref{subsec:Seed_Positions}), each initially located at the center of a $7^{\circ} \times 7^{\circ}$ square region of interest (RoI).~The size of the RoIs will be adapted in subsequent stages of the analysis, depending on the initial results, as explained in~Section~\ref{subsubsec:Results_Init} and Section~\ref{subsec:Ref_Analysis}.~An initial sky model is built for a larger $13^{\circ} \times 13^{\circ}$ region around each position, to account for spillover of signal from sources slightly outside the RoI.~It includes all sources from the 4FGL-DR3 $\textit{Fermi}$-LAT catalog~\citep{Abdollahi22}, along with the standard Galactic diffuse emission model ($\texttt{gll\char`_iem\char`_v07.fits}$) and isotropic template ($\texttt{iso\char`_P8R3\char`_SOURCEVETO\char`_V3\char`_v1.txt}$).~We accounted for the effect of energy dispersion on all model components except the isotropic diffuse template (this operates on spectra with one extra bin below and above the analysis threshold when using the parameter $\texttt{edisp\_bins = -1}$).

We developed a semi-automated extended source detection pipeline based on the $\texttt{Fermipy}$ python package\footnote{\url{https://fermipy.readthedocs.io/en/latest/}}~\citep{Wood17} version 1.0.1, which uses the underlying $\texttt{Fermitools}$\footnote{\url{https://fermi.gsfc.nasa.gov/ssc/data/analysis/software/}} version 2.0.8.~With these tools, we implemented a maximum-likelihood estimation of the source model parameters that best describe the data in each RoI.~In practice, we performed a three-dimensional likelihood analysis for binned data and Poisson statistics, using $0\fdg05 \times 0\fdg05$ spatial pixels and 10 logarithmic bins per decade in energy.

The statistical significance of source detection is quantified via the test statistic $\textrm{TS}\mathrel{=}2\,\log\,(\nicefrac{\mathcal{L}_{\rm test}}{\mathcal{L}_{0}}$)~\citep{Chernoff54}, where $\mathcal{L}_{\rm test}$ and $\mathcal{L}_{0}$ represent the maximum likelihood values for an emission model with and without the test source, respectively.~Here and elsewhere, the term \enquote{$\log$} refers to the natural logarithm.~The statistical significance of source extension is assessed via the likelihood ratio $\textrm{TS}_{\rm ext}\mathrel{=}2\,\log\,(\nicefrac{\mathcal{L}_{\rm ext}}{\mathcal{L}_{\rm ps}}$), where $\mathcal L_{\rm{ps}}$ and $\mathcal L_{\rm{ext}}$ represent the likelihoods for point-like and extended source models, respectively.~A source is considered significantly extended if the extension detection criterion (TS$_{\rm{ext}}$\,$>$\,16, corresponding to a significance level greater than 4$\sigma$) is met.

The analysis begins with an initial optimization of the model for each RoI by fitting all sources iteratively (\textit{optimize} method in $\texttt{Fermipy}$).~This is accomplished in three sequential steps: i.~simultaneously fitting the normalization of up to five sources, selected based on the largest expected number of photons, which together account for more than 95\% of the total predicted counts in the model, ii.~fitting individually the normalization of the remaining sources in the model in descending order of their predicted counts $N_{\rm pred}$ down to 1.0, iii.~fitting individually the normalization and spectral shape parameters of all sources with TS\,$\geq$\,25, sorted in descending order of their TS.~The first two steps of the RoI optimization are used to calculate the TS of sources.~Then, following any major step in the analysis process, a likelihood fit is performed to readjust all free parameters of the model (\textit{fit} method in $\texttt{Fermipy}$).~The set of free parameters for each RoI is described in the next section.

After the initial optimization, subsequent steps in the analysis pipeline involve searching for new sources in the field (\textit{find\_sources} method in $\texttt{Fermipy}$) and testing of their extension (\textit{extension} method in $\texttt{Fermipy}$).~Source finding is achieved using an iterative TS map peak finding method, and the localization of the source is then performed in two steps.~First, a preliminary localization is performed by computing a TS map centered on the initial position and then performing a fit to locate the maximum TS peak.~The source position is then further refined by scanning the likelihood surface in the vicinity of the best-fit position found in the previous step, using a grid of $\textit{nstep} \times \textit{nstep}$ positions with $\textit{nstep}\mathrel{=}7$ and a size set by 99\% positional uncertainty determined from the TS map peak fit.~The testing of source extension is done under the assumption of a two-dimensional (2D) azimuthally-symmetric Gaussian intensity distribution and involves scanning a grid of possible extension radii and performing a likelihood profile scan over the source width to determine the optimal extension.~In order to find the best model for the RoI, the search for source extension includes alternating optimization of the spatial extension and the source position, along with all other free parameters in the RoI.~New statistically significant point sources and 2D Gaussian extended sources (TS and TS$_{\rm ext}$ above 16) are modeled with a power-law ($\texttt{PL}$) spectrum $\nicefrac{dN}{dE}\mathrel{=}N_{0}(\nicefrac{E}{E_{0}})^{-\Gamma}$.

As explained below, different scenarios are considered in the search for new extended emission components.~Alternative source models for a given RoI are then compared via their likelihoods when the models are nested.~Otherwise, we used the Akaike's Information Criterion~\citep[AIC;][]{Akaike74}, which is given by $\textrm{AIC}\mathrel{=}2\,k - 2\,\rm{\log}\,\mathcal L$, where $k$ is the number of degrees of freedom in the fitted model and $\mathcal L$ is the maximum likelihood.

\section{Systematic Search for Extended Sources} \label{sec:SysSearchExtSrcs}

In the following, we describe the two-step methodology that we have developed for creating the catalog of Galactic extended sources emitting above 10\,GeV.~We start by introducing the selection of seed positions used at the beginning of the process.

\subsection{Seed Positions} \label{subsec:Seed_Positions}

The search was conducted at the positions of 99 bright pulsars selected from the Australia Telescope National Facility (ATNF) pulsar catalog~\citep{Manchester05} with $\dot{E}/d^{2}\,\geq\,10^{34}$ erg s$^{-1}$ kpc$^{-2}$, irrespective of their identification as $\gamma$-ray pulsars.~These pulsars are obvious targets to search for detectable pulsar-powered emission, although maybe not necessarily in the form of a pulsar halo since some of these objects are relatively young.~In this work, millisecond pulsars are excluded due to their lower energy outputs and advanced ages, which render their contribution to formation of pulsar halos or PWNe likely minimal~\citep{Abdalla18}.

We also considered a set of TeV sources, because pulsar halos were initially discovered in this domain, and most of these sources are currently unidentified.~We used as seed the positions of 143 TeV sources from the H.E.S.S. Galactic Plane Survey (HGPS) catalog~\citep{Abdalla18} and the third catalog of HAWC sources (3HWC;~\citealt{Albert20}).\footnote{We note that one of the unidentified HGPS sources (HESS J1943+213;~\citealt{Archer18}) is likely an extragalactic object, and two of the 3HWC sources (Markarian 421 and Markarian 501) are extragalactic bright blazars.}~A number of these sources are firmly identified as objects that are not pulsar-related (e.g., SNRs), but we included them anyway since the detection of $\gamma$-ray halos around them is of similar scientific interest: they could allow us to study the transport properties of energetic particles in the vicinity of their accelerators.

In complement to these 242 seed positions, an evenly spaced grid of 302 positions was employed at Galactic longitudes $\lvert l\rvert\leq$\,100$^{\circ}$ and latitudes $\lvert b\rvert\leq$\,1$^{\circ}$ in order to uniformly cover the low-latitude regions of the inner Galactic plane, where the majority of young and middle-aged pulsars are expected.~Seed positions are spaced every 2$^{\circ}$ in longitude and 1$^{\circ}$ in latitude.~At Galactic latitudes $b=\pm1^{\circ}$, seeds are shifted by $1^{\circ}$ in longitude relative to the $b=0^{\circ}$ grid, creating a staggered pattern.

\subsection{Initial Analysis} \label{subsec:Init_Analysis}

When searching for extended emission around a seed position located at the RoI center, a source may already exist there and potentially be connected to the seed or to a previously detected extended emission.~In the initial analysis, a 4FGL source is therefore paired with a seed position if their angular separation is less than 0.1$^{\circ}$.~As explained below, we considered two different hypotheses to model extended emission around the seed position: a single-component model that links it to the central pre-existing source, if any, and a two-component model allowing it to be independent.

Based on an initial investigation of a subset of seed positions early in the pipeline development, we decided to set the initial source detection threshold TS and the extension detection threshold TS$_{\rm ext}$ to 16, to have the normalization of isotropic diffuse emission fixed to 1, and to keep faint, sub-threshold sources within 3$^{\circ}$ of the RoI's center fixed at their best-fit values obtained from the initial optimization of each RoI model, as described in Section~\ref{subsubsec:Method_Init}.~These reasonable criteria aim to stabilize the fit process and limit the cross-talk between components.

Then, the other free parameters of the source model in each RoI were specified in relation to two concentric circular regions with radii $r_{\rm inner}$ and $r_{\rm outer}$ centered at each RoI:~i.~normalization of sources with TS\,$>$\,16 within the outer circular region, ii.~normalization and spectral shape parameters of sources with TS\,$>$\,50 within the inner circular region, iii.~normalization and spectral shape parameters of all sources with TS\,$>$\,500 in the RoI, iv.~normalization of the Galactic diffuse background component.~Based on this, the spectral parameters of sources in the annulus region with $r_{\textrm{inner}}<r<r_{\textrm{outer}}$ are fitted only if they are exceptionally bright (TS\,$>$\,500); otherwise, just the normalizations of sources with TS\,$>$\,16 are fitted.~The fitted spectral parameters for sources with curved spectra modeled with $\texttt{LogParabola}$ (LP) or $\texttt{PLSuperExpCutoff4}$ (PLEC4)\footnote{\url{https://fermi.gsfc.nasa.gov/ssc/data/analysis/scitools/source_models.html}} depend on their peak energy, given by $\texttt{LP\char`_EPeak}$ and $\texttt{PLEC\char`_EPeak}$ in the 4FGL-DR3 catalog: only parameter $\alpha$ for the LP model and parameter $\gamma_{0}$ for the PLEC4 model of a source is fitted if its peak energy is less than 10 GeV; otherwise, all spectral parameters are fitted.~Initially, we have set $r_{\rm{inner}}\mathrel{=}2^{\circ}$ and $r_{\rm{outer}}\mathrel{=}3^{\circ}$, but these regions will be adapted in a second stage of the analysis.

This baseline set of free parameters hereafter called $P_{\rm free}$ was readjusted in any global fit throughout the various stages of the analysis pipeline.~However, when optimizing the spatial parameters of an extended component simultaneously with the free background parameters, we reduced this set by fixing all sources with TS\,$<$\,50 within the outer circular region $r < r_{\textrm{outer}}$ except the source initially linked to the seed position.~We will refer to this modified set of free parameters as $P_{\rm{free,\,ext}}$ in the following.

\subsubsection{Initial Analysis: Analysis Method and Procedure} \label{subsubsec:Method_Init}

The initial optimization of the sky model in each RoI, following the procedure detailed in Section~\ref{subsec:Analysis}, yields an AIC value AIC$_{1}$.~On top of this first model, we searched for extended emission originating from or close to the center of the RoI.~If no known 4FGL source is already coincident with the center of the field and can be paired with the seed position, we first test for the existence of a significant centrally located source.~In the positive case, a new point source is added at the center with a $\texttt{PL}$ spectrum of an initial index of 2.3, which is the typical photon index for firmly identified PWNe in the 1--10 TeV range, as discussed in~\citet{Abdalla18}.~A global fit is then performed to find the best-fit spectral parameters of the $P_{\rm free}$ set.~We then test the existence of significant extended emission, assuming a 2D Gaussian morphology with an initial angular size defined by a 68\% containment radius of $r_{68}=0.5^{\circ}$ and a $\texttt{PL}$ spectrum with an initial index of 2.3, from or close to the center of the RoI using the two independent hypotheses described below:
\begin{enumerate}
   \item \textit{Single-component model}: The extended $\gamma$-ray emission at the RoI center is attributed to the central source, either a 4FGL source spatially coincident with the seed position (and possibly already extended) or a newly added point source at the seed position.~The spatial parameters of the central source were optimized by jointly fitting its angular extension and position, along with all free parameters in the $P_{\rm{free,\,ext}}$ set.~If significant extension is found (i.e., if the TS$_{\rm ext}$ of the central source is above 16), a global fit is performed to readjust the $P_{\rm free}$ set, resulting in an AIC value AIC$_{2,1}$.
   \item \textit{Two-component model}: The emission at the RoI center is represented by the central source, either a 4FGL source spatially coincident with the seed position (and possibly already extended) or a newly added point source at the seed position, and an additional 2D Gaussian component.~The spectral parameters of the 2D Gaussian component are first optimized together with those of the $P_{\rm free}$ set, and then its spatial parameters are fitted along with those in the $P_{\rm{free,\,ext}}$ set.~If a significant extension is found (i.e., if the TS$_{\rm ext}$ of the additional extended component is above 16), a global fit is performed to readjust the $P_{\rm free}$ set, resulting in an AIC value AIC$_{2,2}$.
\end{enumerate} 

If a significant extended component is detected, i.e., if AIC$_{2,1}$ or AIC$_{2,2}$ is smaller than AIC$_{1}$, the source model with the minimum AIC value is adopted as the optimal model for the subsequent stage, and the newly detected extended source in each RoI is then referred to as the \enquote{central extended component} in the following.~In the last stage, aiming for the most complete description of each field, we search for new point sources and other (2D Gaussian) extended sources elsewhere in the RoI, starting with an assumed $\texttt{PL}$ spectrum with an index of 2.0 and an initial angular size of $r_{68}\mathrel{=}0.5^{\circ}$ for testing extended sources.~If one or more sources with TS\,$>$\,16, or even TS$_{\rm{ext}}>$16, are found, a global fit is performed to readjust the $P_{\rm free}$ set.~We then revisit each RoI's model and further refine the spatial and spectral features of the central extended component, along with other free parameters considered in the second stage.

\subsubsection{Initial Analysis: Results} \label{subsubsec:Results_Init}

The initial search resulted in 253 extended components, of which 203 are located within their 7$^{\circ}\times7^{\circ}$ RoIs and have an angular size of $r_{68} \leq 3.5^{\circ}$.~The remaining 50 extended components, which have larger sizes, may not be solidly constrained as they are not well-contained within their respective RoIs.~Discrepancies in detecting the GeV halo around the Geminga pulsar~\citep{DiMauro19,Xi19} underscore the challenges of accurately characterizing very large extended sources and highlight the need for more tailored treatments.~Consequently, very large extended sources, like the halo around the nearby Geminga pulsar, are excluded from the subsequent stages of this study.~Assuming a typical radius of 30\,pc for extended objects, as observed for the halo around Geminga~\citep{Abeysekara17}, our analysis setup restricts the probing of sources to distances beyond about 500\,pc.

Due to the high seed density in our search region and the resulting multiple overlapping regions of interest, a given extended component may have been detected multiple times.~In order to extract a list of unique extended components, we identified duplicates in our list of 203 components from the following criteria: 
\begin{equation}
   \textrm{separation} \leq 0.5 \cdot \textrm{max}(r_{\rm i}, r_{\rm j})~\textrm{\&}~\textrm{surface overlap } (S_{\rm ec_{\rm i}\cap \rm ec_{\rm j}})/(S_{\rm ec_{\rm i}\cup \rm ec_{\rm j}}) \geq 0.25\label{eq1}
\end{equation}
where $r_{\rm i}$ and $r_{\rm j}$ are the 68\% containment radii of each pair of extended components ${\rm ec_{\rm i}}$ and ${\rm ec_{\rm j}}$.~Using these criteria, we eliminated all duplicates but the one with the highest TS in each group.~The end result is an initial catalog of 71 extended components, with a small number of overlapping sources that do not meet the criteria given above.~We found that the two-component model was preferred for 62 of them, representing $\sim$87\% of the total.

\subsection{Refined Analysis} \label{subsec:Ref_Analysis}

In the initial analysis, detected extended components were investigated individually, one at a time, in each given RoI.~However, their spectral and spatial properties may be affected or biased by the presence of other, nearby extended components in the same field.~We therefore revisited each extended component by incorporating other newly detected extended components into the corresponding source model, if they lie entirely or partially inside the RoI.~For each extended component, modeled with a $\texttt{PL}$ spectrum and a 2D Gaussian spatial model, a five-step refined analysis (T$_{0}$$-$T$_{4}$, detailed in Section~\ref{subsubsec:Method_Ref}) was performed in a new RoI centered on $p_{\rm init}$ with a modified side size of $6^{\circ} + 2 \cdot r_{\rm init}$, where $p_{\rm init}$ and $r_{\rm init}$ represent the best-fit position and 68\% containment radius of the extended component, respectively, as extracted from the initial catalog of 71 extended components.

The baseline set of free parameters is defined as before, with the sizes of the inner and outer circular regions being adapted to each RoI as $r_{\rm inner}=2^{\circ} + r_{\rm init}$ and $r_{\rm outer}=3^{\circ} + r_{\rm init}$, respectively.~In all steps except the final one (T$_{0}$$-$T$_{3}$), the normalization and spectral shape parameters of the central extended component and the associated 4FGL source, if there is any, are fitted only if their TS values are greater than 50; otherwise, only their normalizations are allowed to vary.~In the final step of the refined analysis (T$_{4}$), however, both the normalization and spectral shape parameters of the central extended component are fitted irrespective of its TS value, while the handling of the associated 4FGL source remains consistent with the previous steps.

Sources are kept in the model even if they fall below the detection threshold.~To verify the robustness of this approach, we conducted an additional evaluation in the final step of the refined analysis (T$_{4}$) by removing low-significance sources coincident with extended components from the RoIs.~This check assessed whether these sources had a notable impact on the detection significance of the extended components.~The results of this evaluation are discussed in Section~\ref{subsec:Final_list}.

If a surrounding extended component overlaps with the central extended component and has a TS greater than 50, its normalization and spectral shape parameters are fitted.~Otherwise, if any of these criteria are not met, only the normalization is allowed to vary in the likelihood maximization.~When adding an extended component that is identified as a single component in the initial analysis, the associated 4FGL source is eliminated from the source model.~During the refined analysis, we refrained from searching for new sources within the RoIs for two reasons:~first, only very few new sources were detected in certain RoIs during the initial analysis, likely due to the use of the up-to-date 4FGL-DR3 catalog; and second, we wanted to prevent any potential interference with nearby extended sources.

Our refined process is intended to provide a more precise assessment of the extended components' properties.~In the following section, the details of the five-step evaluation process for each extended component, taken from the initial catalog, are given.

\subsubsection{Refined Analysis: Analysis Method and Procedure} \label{subsubsec:Method_Ref}

The five steps of the refined analysis are summarized as follows (where the acronym ``EC'' stands for ``extended component''):
\begin{itemize}
\item T$_{0}$ - Baseline model (no EC)
\item T$_{1}$ - Central EC (initial morphology)
\item T$_{2}$ - Central EC (initial morphology) + Nearby ECs (initial morphologies)
\item T$_{3}$ - Central EC (revised morphology) + Nearby ECs (initial morphologies)
\item T$_{4}$ - Central EC (revised morphology) + Nearby ECs (revised morphologies) 
\end{itemize}
The related data analysis details and results are as follows:

\textit{T$_{0}$}: The process starts with a baseline model that includes only 4FGL-DR3 sources as well as the standard isotropic and Galactic diffuse emission templates, without any new extended components.~The model is fitted to the data in the new RoI and the results serve as a reference for subsequent tests.

\textit{T$_{1}$}: The central extended component is added to the source model, with initial spectral and spatial parameters taken from the initial catalog, and a fit is performed.~If the likelihood of this fit is degraded with respect to that of test T$_{0}$, the extended component is discarded.~This occurred for 3 out of 71 extended components, with degradation ranging from 4 to 11 in $\log\mathcal{L}$.~The observed discrepancies compared to the initial analysis are likely due to the complex optimization trajectories, not starting from exactly the same conditions.~This was particularly the case for an extended component detected at the position of point source 4FGL J1427.8$-$6051, which is surrounded by three confused sources.\footnote{Confused sources flagged with the \enquote{c} suffix in the 4FGL-DR3 catalog are potentially affected by the Galactic diffuse emission, making their spatial and spectral properties systematically uncertain.}~In the other two cases, our initial analysis did recover previously known extended sources 4FGL J0534.5+2201i and 4FGL J1443.0$-$6227e, with morphologies almost identical to those in the 4FGL-DR3 catalog, ensuring that no information was lost.~As a result, these three components are excluded, leaving 68 extended components for the subsequent analyses.

\textit{T$_{2}$}: Other extended components from the initial catalog are incorporated into the source model if their centroid is inside the region, whether they are entirely or partially within the RoI.~Then, a fit is performed to adjust all free parameters in the RoI, resulting in updated spectral properties, fluxes, and significance values for the extended components.

\textit{T$_{3}$}: The spatial parameters of the central extended component, including both the position and extension of the 2D Gaussian model, are revisited.~A fit is then conducted to further adjust all other free parameters of the sky model, including those of the central associated 4FGL source if present.~During revising the properties of central extended components, four components ended up with unreliable angular sizes of $r_{68} > 3.5^{\circ}$ due to difficulties in the fitting procedure.~Three of these had large initial angular sizes of $r_{68}$ between 3.0$^{\circ}$ and 3.5$^{\circ}$, and the last one ended up near the edge of the search region, with a large offset of 3.96$^{\circ}$ from its initial centroid, thus not fully contained within the RoI.~Eventually, this stage yields revised morphologies for 50 out of the 68 extended components.~In addition to the four very large components mentioned above, 14 extended components no longer meet the minimum TS$_{\rm ext}$ criterion, and 10 of these fall below the initial TS detection threshold of 16.~For the subsequent steps, the morphological and spectral properties of these 18 components are therefore reverted to those from step T$_{2}$.\footnote{We retained these 14+4 components with their initial properties to be maximally inclusive, as it was impossible to anticipate what would happen to these components in the subsequent T$_{4}$ step.~Moreover, even with their initial morphologies, these components appeared to significantly improve the overall fit likelihood when included as neighboring components in other RoIs (see Fig.~\ref{fig:LRT}).~Nevertheless, the majority of these 18 components either fell below the detection threshold in step T$_{4}$ or resulted in fragile detections (low TS and/or failure to persist under an alternative interstellar emission model).~Consequently, none of these components appear in the final catalog.}

\textit{T$_{4}$}: The extended components with spectral and spatial properties individually refined in step T$_{3}$ are added to each respective RoI.~At this stage, the central extended component may be offset from the RoI center.~We then perform a final fit to further refine the parameters of the extended components in each RoI.~Out of the 68 extended components analyzed, 56 are detected and characterized within the 10 GeV$-$1\,TeV energy range, with both TS$_{\rm{ext}}$ and TS values exceeding the initial threshold of 16, and will be the focus of the subsequent section.~The remaining 12 components, however, no longer meet the TS$_{\rm{ext}}$\,$>$\,16 criterion and are excluded from further consideration as extended sources.

Further refinement will be achieved through the systematic analysis of the initial list of 68 extended components, as detailed in Section~\ref{subsec:Systematics} and discussed in Section~\ref{subsec:Final_list}.

\subsubsection{Refined Analysis: Results} \label{subsubsec:Results_Ref}

\begin{figure}[t]
\centering
\includegraphics[width=0.55\textwidth]{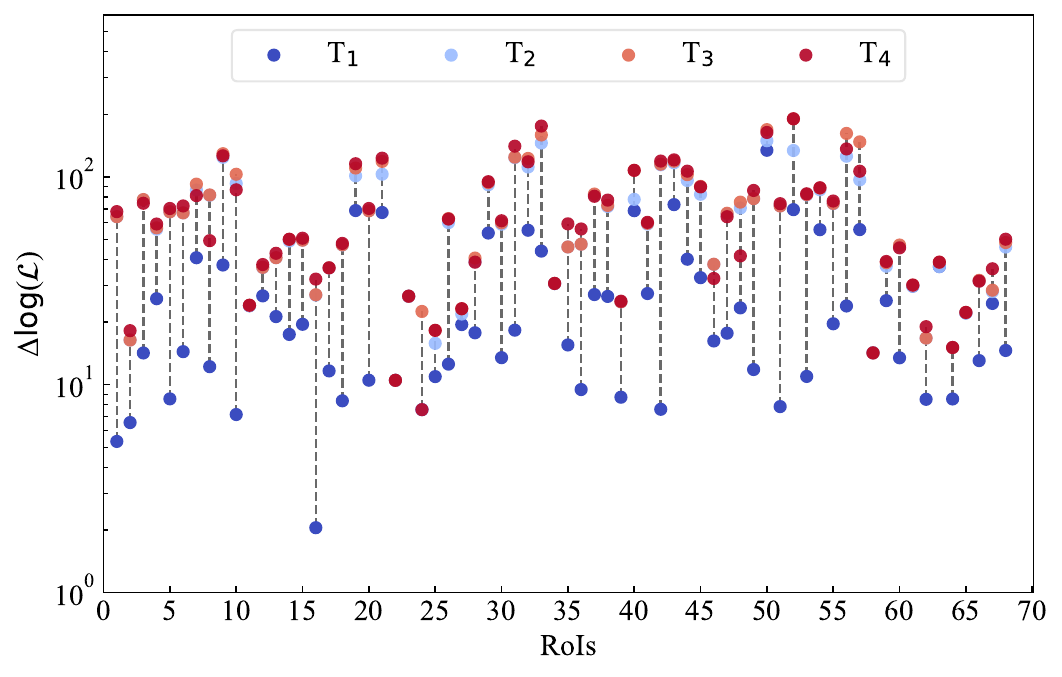}
\caption{\label{fig:LRT} Variations in $\log$-likelihood for the 68 extended components analysis over steps T$_{1}$ to T$_{4}$ of the refined analysis with respect to step T$_{0}$.}
\end{figure}

The variations in fit likelihood values for the 68 extended components during the five-step refined analysis are presented in Figure~\ref{fig:LRT}.~Throughout this process, the inclusion of other nearby components within each RoI and their sequential optimization has led to a noticeable increase in the model likelihood for the vast majority of RoIs.~This improvement is particularly evident in the transition from step T$_{1}$ to T$_{2}$, with the likelihood values clearly converging in the final steps from T$_{3}$ to T$_{4}$.~The likelihood remains nearly unchanged in eight RoIs throughout the refined analysis, as each of these regions encompasses only a single extended component, while a slight degradation in likelihood is observed in seven RoIs between steps T$_{3}$ and T$_{4}$.

Figure~\ref{fig:TS_Flux_init_ref} (left) illustrates the TS and TS$_{\rm{ext}}$ values of the 56 extended components in the initial analysis compared to the refined analysis.~A general reduction in both TS and TS$_{\rm ext}$ values after the five-step refined analysis results from the inclusion of other newly detected extended components within the RoIs, allowing for a higher degree of flux transfer between extended components and ensuring a better fit of the interstellar emission.~Eventually, the TS values of the detected components range from the initial threshold of 16 to over 1000, with a median value of 61.~The highest TS values correspond to previously known extended sources, flagged with the \enquote{e} suffix in the 4FGL-DR3 catalog, which are re-detected by our pipeline with slightly modified morphologies compared to those prescribed in the catalog.~Similarly, the TS$_{\rm ext}$ values range from the threshold of 16 to over 400, with a median value of 53.~Figure~\ref{fig:TS_Flux_init_ref} (right) presents a comparison of the photon flux of extended components integrated over the analysis energy band (10 GeV--1 TeV), $F_{>\,\rm{10\,GeV}}$, between the initial and refined analyses.~Overall, the mean flux value of the extended components is reduced by $\sim$15\% in the refined analysis, with the photon flux ranging from 6.2$\times10^{-11}$ ph cm$^{-2}$ s$^{-1}$ to 5.9$\times10^{-9}$ ph cm$^{-2}$ s$^{-1}$.

\begin{figure}[t]
\centering
\includegraphics[width=0.85\textwidth]{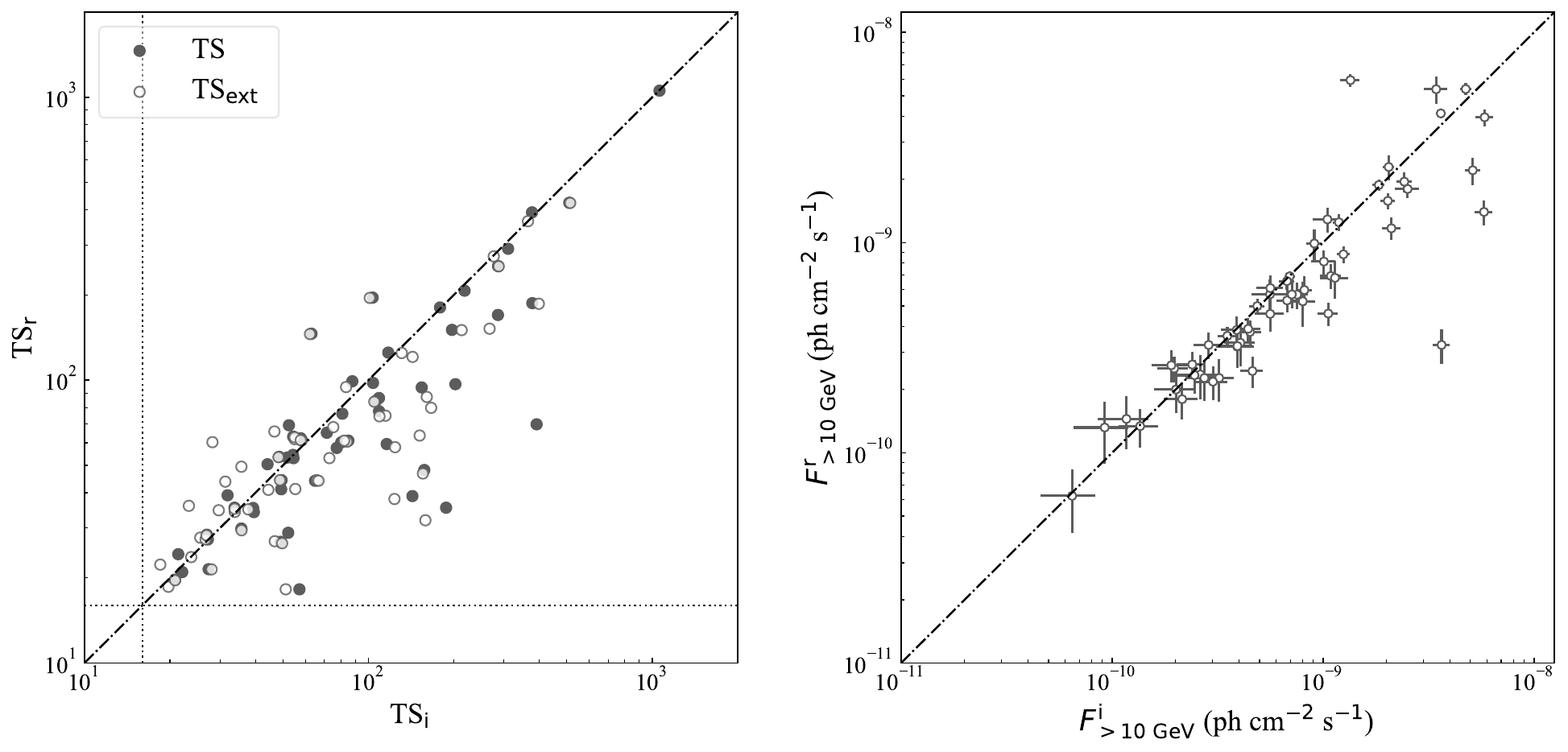}
\caption{\label{fig:TS_Flux_init_ref}Left:~Comparison of the TS (filled circles) and TS$_{\rm ext}$ (open circles) values for the 56 extended components in the initial analysis (TS$_{\rm{i}}$) with those in the refined analysis (TS$_{\rm{r}}$).~The vertical and horizontal dotted lines indicate the initial threshold values for TS and TS$_{\rm ext}$, and the diagonal line represents a slope of 1.~Right:~Same as the left panel, but for the integrated photon flux of the 56 extended components in the initial ($F^{\rm{i}}_{\rm{>\,10~GeV}}$) and refined ($F^{\rm{r}}_{\rm{>\,10~GeV}}$) analyses.~The diagonal line also represents a slope of 1.~\label{fig:TS_Flux_init_ref}}
\end{figure}

\subsection{Systematic Uncertainties} \label{subsec:Systematics}

Two important sources of systematic uncertainties in the detection of Galactic extended emission and the estimation of their spectral properties are the imperfect modeling of the strong Galactic interstellar emission and the incomplete knowledge of the IRFs.~In the following, we describe the approach used to estimate the impact of the former and discuss it in relation to the latter.

To assess the systematic uncertainties from the Galactic interstellar emission model (IEM), we developed an alternative model for each RoI.~We recall that the standard IEM used so far is one single model covering the whole sky, developed in a data-driven way with the goal of globally reducing residuals as much as possible in the context of the catalog creation.\footnote{\url{https://fermi.gsfc.nasa.gov/ssc/data/analysis/software/aux/4fgl/Galactic_Diffuse_Emission_Model_for_the_4FGL_Catalog_Analysis.pdf}}~When analyzing a specific, restricted region of the sky, this model can be fitted to the data (together with other components of a sky model) in terms of an overall and possibly energy-dependent normalization, but the internal composition of the model, i.e., the relative weight of its different components, is not modified.~Here, we developed alternative IEMs tailored to each RoI by allowing the different components of the standard IEM to be adjusted specifically to each sky region.

The standard Galactic diffuse emission is modeled as a linear combination of several interstellar components decomposed into 10 concentric annuli spanning from 0 to 35 kpc, with boundaries set at 0, 0.15, 0.6, 2, 4, 6, 7, 9, 12, 15, and 35 kpc.~This applies to atomic and molecular gas templates (coming from $\HI$ and CO line observations) and to the IC emission template (coming from a cosmic-ray transport model in the Galaxy).~Other components include a so-called dark neutral medium (DNM) gas represented by two components, a positive and a negative correction to the $\HI$ and CO gas templates, and an ad-hoc component called~\enquote{patch} to include the non-template components such as the Galactic center excess, Loop I, or the \textit{Fermi} bubbles.

The innermost part of the Galaxy ($R \lesssim 200~\rm{pc}$) contains a substantial amount of molecular gas due to the presence of the so-called Central Molecular Zone, with no significant signal from atomic hydrogen.~As a result, gas in this region is entirely represented by the two innermost CO annuli 0 and 1, whereas no emission is assigned to the two $\HI$ annuli 0 and 1.~We combined the three CO annuli 1 through 3 as a single component CO$_{\rm{in}}$, owing to the relatively weak and geometrically thin (in latitude) CO emission in this region ($0.15~\rm{kpc} < R < 4~\rm{kpc}$), compared to that in annuli 4--6.~In the outer Galaxy, we merged the two CO annuli 7 and 8, which lack measurable CO emission, with the three outermost $\HI$ annuli 7 through 9 and the insignificant negative DNM component, which does not have a specific distribution over annuli, into a single component named \enquote{Others}.~In contrast to the clumpy structure of the atomic and molecular gases along the Milky Way, the IC emission is remarkably smoother, so we merged all 10 IC annuli together.~This reduces the number of IEM components from 33 to the following 14, while separating the inner and outer Galaxy components: a combined IC component, five $\HI$ annuli (2 through 6), a combined CO$_{\rm{in}}$ component for the inner Galaxy, four CO annuli (0, and 4 through 6), the positive component of DNM gas, the \enquote{Others} combined component for gas in the outer Galaxy, and the patch component.~Each of these components is actually a model cube, including a spectral distribution retained from the standard IEM.

When analyzing data in a given sky region, the alternative model is built as follows.~In the first step, the relative flux of the 14 IEM components is computed in each RoI, for an initial normalization of one.~Those components with a relative flux of less than three percent of the total are merged into a single component to further reduce the number of components (and avoid keeping empty components, for instance inner Galaxy gas templates when looking at an outer Galaxy field).~Consequently, the number of IEM components in each RoI is reduced to $N + 1$, where $N$ is the number of components with a relative flux greater than three percent.~In the second step, we fitted each RoI model following the procedure described in $T_{0}$ of the refined analysis, replacing the standard IEM model by the set of $N + 1$ templates appropriate for the field.~During this process, each template can be rescaled via a power-law in energy, thereby allowing to adjust both the overall normalization and the spectral shape of the component.~This fit yields a likelihood $\log\mathcal{L}^{\rm{i}}_{\rm{alt}}$.~In the third step, all IEM components are added together, taking into account the possible modification of their normalization and spectral shape, and this produces the alternative IEM for the given RoI.~From that point on, the alternative IEM is treated as the standard one was, and individual components are not altered anymore.~In the fourth and last stage, we repeated step $T_{4}$ of the refined analysis, replacing the standard IEM by the alternative IEM, and this yields likelihood $\log\mathcal{L}^{\rm{f}}_{\rm{alt}}$.~We emphasize that source extent was not revisited with the alternative IEMs due to computational reasons.

\begin{figure}[tbp]
\centering
\includegraphics[width=0.49\textwidth]{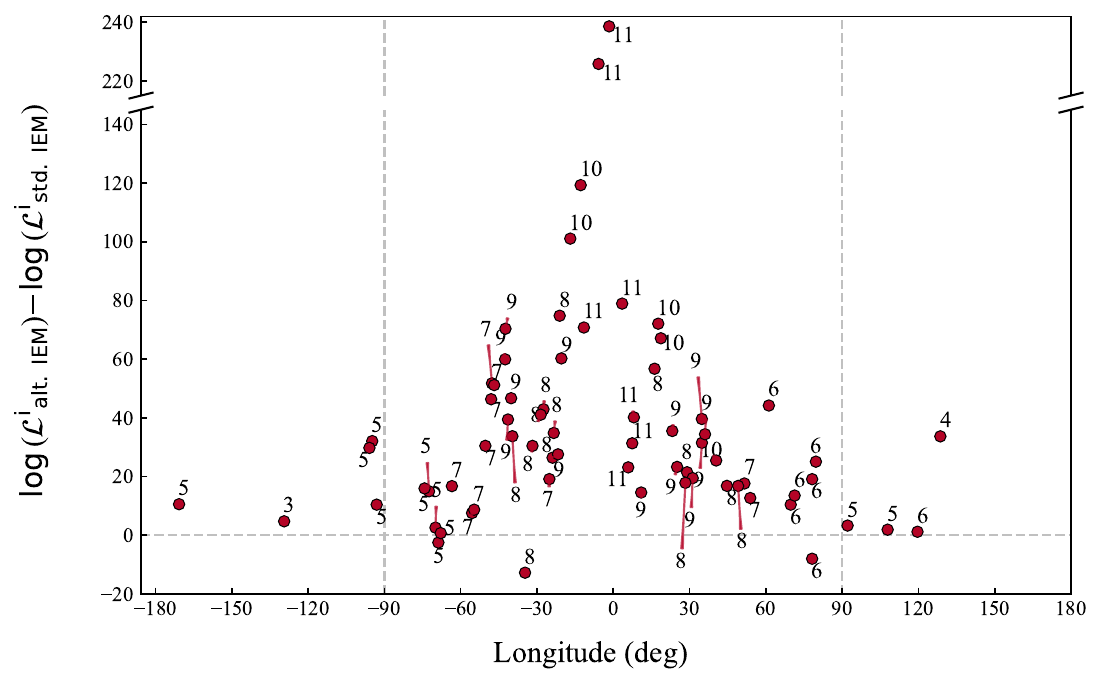}
\hfill
\includegraphics[width=0.49\textwidth]{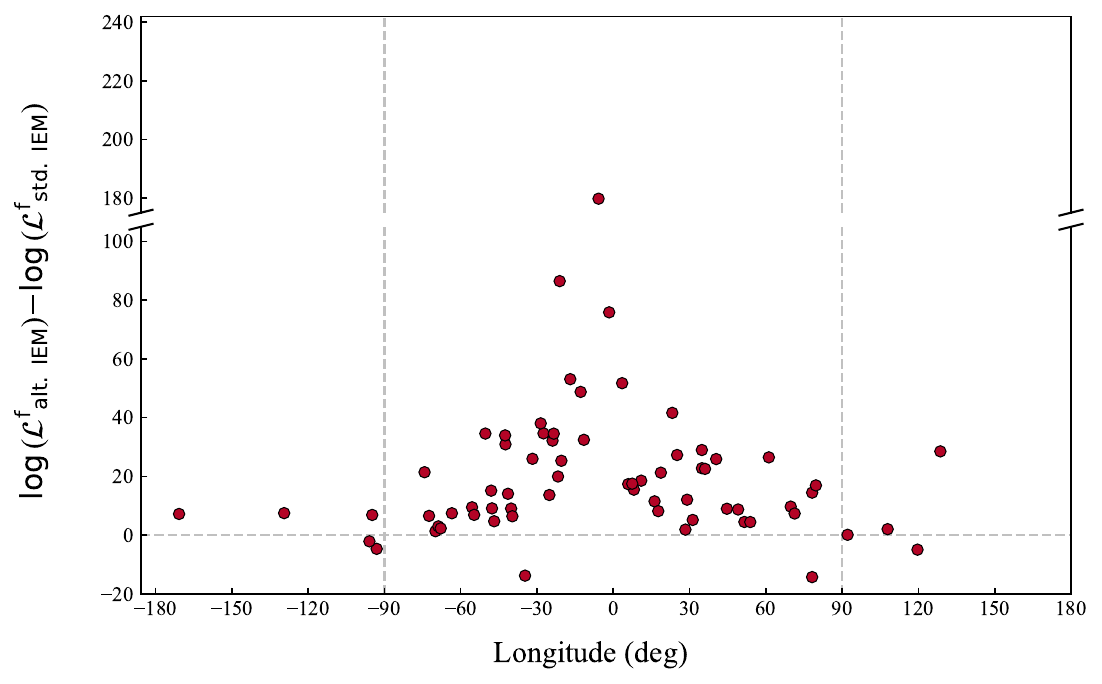}
\caption{\label{fig:LL_IEM_opt_std}Variation in $\log\mathcal{L}$ from using an RoI-specific alternative IEM in step $T_{0}$ (left) and step $T_{4}$ (right) for the 68 extended components.~The annotated values in the left panel represent the number of IEM components with a relative flux above the threshold of three percent ($N$ out of 14 in step $T_{0}$ of the analysis, as outlined in the text).~The vertical lines indicate the inner Galaxy region ($\lvert l \rvert\,\leq 90^{\circ}$).}
\end{figure}

The variations in $\log\mathcal{L}$ values from using an alternative IEM for each of the 68 RoIs in steps $T_{0}$ and $T_{4}$ are represented in Figure~\ref{fig:LL_IEM_opt_std}.~The results for step $T_{0}$, when only 4FGL sources are included, show that the alternative IEM provides a better description of the data than the standard IEM overall, with a modest decrease in likelihood $\lesssim13$ observed in only three RoIs (which is an optimization issue since the procedure should normally allow to recover exactly the standard IEM if fitting its components individually does not provide any benefit).~The results for step $T_{4}$, now including extended components in the RoIs, also show an improvement in $\log\mathcal{L}$, except in five RoIs where decreases in $\log\mathcal{L}$ $\lesssim14$ were obtained.~For the majority of RoIs, the $\log\mathcal{L}$ increase is in the range of 5$-$40 for about 5$-$10 additional degrees of freedom, indicating a significant, although not always substantial, improvement.~In a handful of cases, mostly RoIs in the inner Galaxy, the improvement is very significant.

This set of RoI-specific IEMs can therefore be considered a reasonable alternative to the standard IEM, warranting an examination of its impact on the detection and properties of the 68 extended components identified after step $T_{1}$ in the refined analysis.~Of these 68 extended components, 50 remained above the initial detection threshold of 16 after applying the alternative IEM, while 18 fell below it.~Among these 18 extended components, 10 were already sub-threshold detections at step T$_{4}$ with the standard IEM.~The implications of these findings, including the robustness of the sources and their TS maps, will be discussed in detail in the following section, where the final list of extended components is established.

Another source of systematic uncertainty in such analyses is the imperfect knowledge of the IRFs, especially the LAT effective area, for which we employed the bracketing IRFs method as described in~\citet{Ackermann12}.~For binned analysis with energy dispersion correction enabled, the effective area systematic uncertainties are estimated to be 3\% in the energy range 10 GeV--100 GeV, and then to increase as a function of energy as $ 3\% + 7\%\,\times\left[\log_{10}(E/1\,\rm{MeV})-5\right]$.\footnote{\url{https://fermi.gsfc.nasa.gov/ssc/data/analysis/scitools/Aeff_Systematics.html}}~The above assessment of systematic uncertainty from the IEM resulted in much larger effects, $>$\,20\% on average, so we henceforth neglect systematic uncertainties from the IRFs.

\begin{figure}[h!]
\centering
\includegraphics[width=0.95\textwidth]{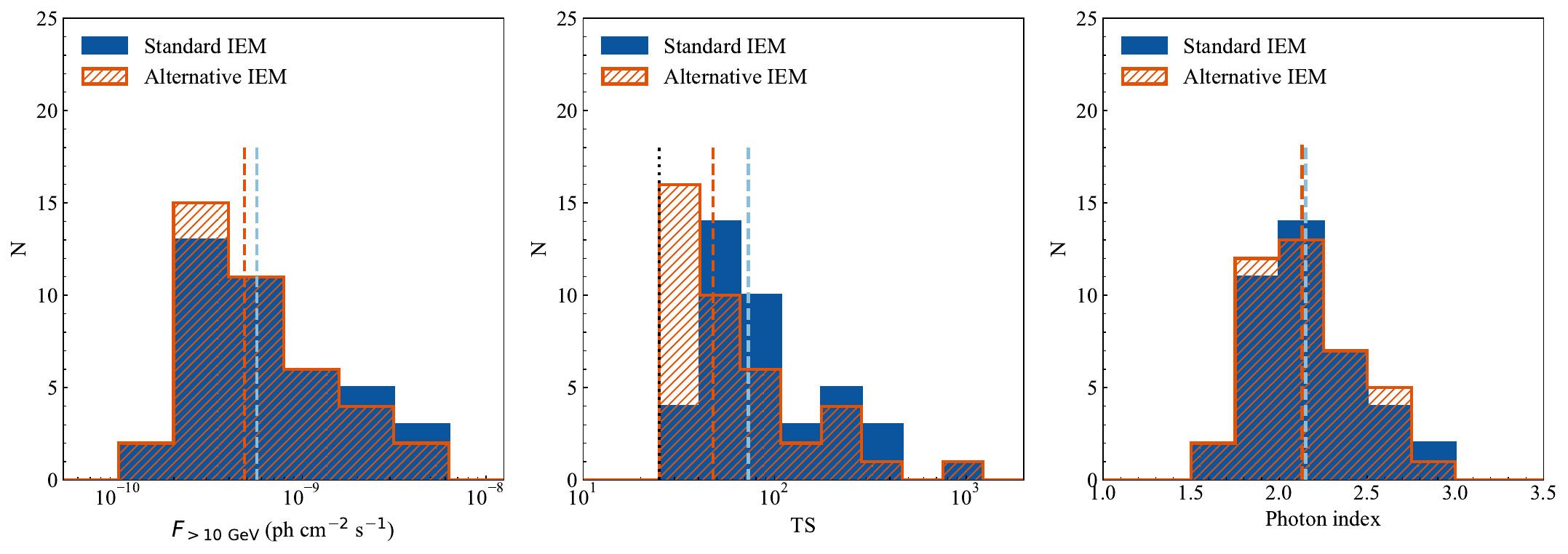}
\caption{\label{fig:IEM_Sys}Distributions of the integrated photon flux (left), TS (middle), and spectral index (right) for the 40 extended components that meet the TS $\geq 25$ threshold using the standard IEM (blue) and the RoI-specific alternative IEMs (orange and hatched).~The dashed lines represent the median values, and the dotted black line in the middle panel indicates the TS threshold of 25.}
\end{figure}

\subsection{Final Selection and Characterization of Extended Sources} \label{subsec:Final_list}
Following a detailed analysis of the 68 extended components using the standard IEM in the refined analysis and RoI-specific alternative IEMs in our systematic study, we thoroughly inspected the TS maps to identify the most robust detections.~This process revealed that the most reliable extended components are those with TS values exceeding 25 and TS$_{\rm{ext}}$ values exceeding 16, allowing us to refine the list accordingly.~Of the 68 extended components, 40 were consistently detected with both IEMs, while 5 were detected with neither.~Additionally, 11 components were detected exclusively with the standard IEM, and 12 no longer met the TS$_{\rm{ext}}$ criterion.~This comprehensive process resulted in a final list of 40 reliable extended sources, meeting the criteria of TS $\geq 25$ and TS$_{\rm ext}$ $\geq 16$, forming the 2FGES catalog.~The 16 sources with TS$_{\rm ext}$ $\geq 16$ but intermediate TS values (between 16 and 25) using the standard IEM are now listed in Appendix~\ref{sec:appendix}, which may be of interest for future studies. 

Figure~\ref{fig:IEM_Sys} illustrates the flux, TS, and spectral index for these 40 extended sources using both the standard IEM and the RoI-specific alternative IEMs.~The median flux and TS for these sources are $4.8\times10^{-10}~\rm{ph}~\rm{cm}^{-2}~\rm{s}^{-1}$ and 48, respectively, when using the RoI-specific alternative IEMs, representing reductions of 15\% and 34\% compared to the values obtained with the standard IEM (see Table~\ref{tab:ecs_properties}).

\begin{figure}[t]
\centering
\includegraphics[width=0.9\textwidth]{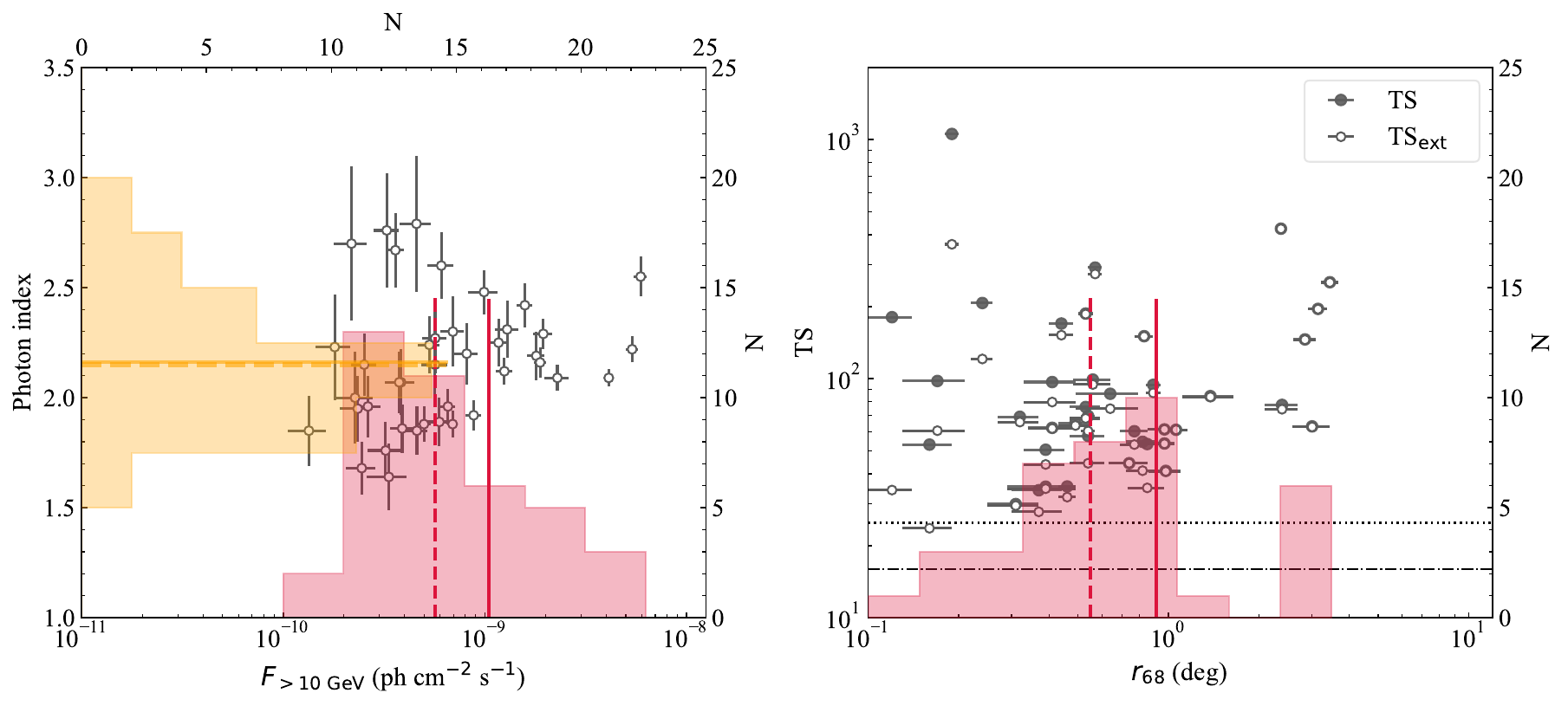}
\caption{\label{fig:Flux_index_corr}Left:~Relation between integrated photon flux and photon index for the 40 extended sources.~The solid and dashed lines represent the mean and median values of the respective histograms for each parameter.~Right:~Relation of the 68\% containment radius with TS and TS$_{\rm ext}$.~The vertical solid and dashed lines indicate the mean and median angular extension values, respectively, while the horizontal dotted line represents the detection TS threshold of 25, and the horizontal dash-dotted line represents the TS$_{\rm ext}$ threshold of 16.} 
\end{figure}

We investigated potential correlations among the measured properties of the 40 extended sources in the 2FGES catalog, including flux, spectral index, angular extension, and TS values.~Our analysis revealed no significant correlations, except for a clear relationship between flux and TS values.~Figure~\ref{fig:Flux_index_corr} (left) illustrates the relationship between the photon index and flux.~The spectral index spans a broad range from 1.6 to 2.8, with a median value of $\Gamma\mathrel{=}2.2$.~Notably, the majority of these extended sources exhibit spectra that are considerably harder than the typical photon index of approximately 2.7 observed for large-scale interstellar emission.~This finding supports the robustness of the detections, suggesting they are unlikely to be artifacts resulting from known imperfections in the modeling of interstellar emission.~Figure~\ref{fig:Flux_index_corr} (right) shows the relationship between angular extension and both TS and TS$_{\rm ext}$ values, indicating that most sources have angular sizes smaller than the mean value of $r_{68}\mathrel{=}0.9^{\circ}$, with only six sources exhibiting very large angular sizes ($>2^{\circ}$).~As we will discuss in Section~\ref{subsec:dubious_srcs}, most of these larger sources are classified as dubious.~Table~\ref{tab:ecs_properties} provides a summary of the mean, median, minimum, and maximum values for the physical properties of the 2FGES sources, including photon flux, spectral index, angular extension, TS, and TS$_{\rm ext}$.

As part of our additional evaluation, as described in Section~\ref{subsec:Ref_Analysis}, we also examined the effects of removing low-significance sources that were overlapped with the extended sources.~This alternative analysis confirmed that, in most cases, the significance of the extended sources remained stable.~Notably, for six extended sources in the final catalog, however, the TS values increased when these overlapping 4FGL sources were removed, which is consistent with expectations in complex regions where such sources may obscure or reduce the apparent significance of the extended source (these six sources are flagged in Table~\ref{ext_comps_properties}).

\begin{deluxetable}{lccccc}[t]
\tablecaption{Summary of the Descriptive Statistics of the 2FGES Source Properties\label{tab:ecs_properties}}
\tablenum{1}
\tablehead{
\colhead{Physical Parameters} & \colhead{$F_{>\,\rm{10\,GeV}}$ (ph cm$^{-2}$ s$^{-1}$)} & \colhead{$\Gamma$} & \colhead{$r_{68}$ ($^{\circ}$)} & \colhead{TS} & \colhead{TS$_{\rm{ext}}$}
}
\startdata
Mean & 1.05$\times10^{-9}$ & 2.16 & 0.91 & 135.48 & 97.80 \\
Median & 5.67$\times10^{-10}$ & 2.15 & 0.55 & 72.95 & 62.36 \\
Minimum & 1.34$\times10^{-10}$ & 1.64 & 0.08 & 29.89 & 23.68 \\
Maximum & 5.93$\times10^{-9}$ & 2.79 & 3.45 & 1054.86 & 423.09 \\
\enddata
\tablecomments{$F_{>\,\rm{10\,GeV}}$ represents the photon flux integrated over the 10 GeV$-$1\,TeV energy band and $\Gamma$ is spectral photon index.~The catalog sources are modeled with PL spectra.~The spatial size ($r_{68}$), TS, and TS$_{\rm{ext}}$ of the sources are given in columns 4$-$6.}
\end{deluxetable}

\section{\textit{Fermi}-LAT Galactic Extended Sources Catalog} \label{sec:catalog}

The final 2FGES catalog comprises 40 extended sources that meet the criteria of TS $\geq 25$ and TS$_{\rm ext}$ $\geq 16$, as established through our detailed selection process.~The spatial and spectral properties of these sources are summarized in Table~\ref{ext_comps_properties}.~To verify the reliability of our catalog-making process and gain insights into the nature of these sources, we conducted a search for possible associations between the 2FGES sources and known $\gamma$-ray sources, as well as a subset of bright pulsars from the ATNF catalog, using a spatial coincidence approach as described in Section~\ref{subsec:MWL_Associations}.~The results of this association process are detailed in Table~\ref{ext_comps_associations}.

Based on these associations and data analysis properties, the 2FGES sources are categorized into seven main groups.~A summary of these categories is provided in Table~\ref{tab:det_categories}.~In the following sections, we first describe the association methodology and then provide a detailed discussion of the 2FGES sources within each category.~We also present in Figures~\ref{fig:4fgle} to \ref{fig:orphan} the TS maps for all sources, organized according to these categories.~These maps were produced by computing the TS over the full energy range and across the entire RoI, with the 2FGES source excluded from the final best-fit model, using a point-source kernel to reveal the small-scale structure of the excess emission.~During this process, the background model was held fixed at its final parameters to reveal the full signal associated with each 2FGES source within each RoI.~This approach prevents coincident 4FGL point-like sources from compensating for the exclusion of the 2FGES source.~When the background model is allowed to vary freely, 4FGL point sources very close to the centroid of 2FGES sources can partially or fully compensate for the removal of the extended sources.~This makes a test point source close to the centroid a partially degenerate component and resulting in a low or zero TS value.~Consequently, as shown in Table \ref{ext_comps_properties}, the TS$_{\rm{ext}}$ values are often very close to the TS values when the background model already includes a point source close to the centroid of the 2FGES source.~For very extended sources without a coincident point source, there is minimal signal to be collected by a point source close to the centroid, also contributing to this effect.

Additionally, we checked the TS maps produced with extended-source kernels and confirmed that, in the majority of cases, no other significant excesses of similar extent and magnitude were present elsewhere in the RoIs.

\subsection{Search for Associations with Bright Pulsars and Gamma-Ray Sources} \label{subsec:MWL_Associations}

We searched for possible associations of the detected extended sources with previously known $\gamma$-ray sources in the 4FGL, HGPS and 3HWC catalogs, and also in the first LHAASO catalog of $\gamma$-ray sources, which was released during the writing of this paper~\citep{Cao24}.~Given our primary interest in pulsar halos, we also considered possible associations with bright pulsars that were initially used as the seed positions (see Section~\ref{subsec:Seed_Positions}).~A simple spatial coincidence approach was applied to establish these associations, in which each 2FGES source was paired with the nearest object in each list of sources if their angular separation was less than the source's 68\% containment radius.~This method, however, may result in incorrect associations, particularly for extended sources that are spatially large or located in crowded regions.~Conversely, some relevant associations might be missed due to energy-dependent morphologies (e.g., in PWNe) or significant present-day offsets from the power sources that provided the bulk of radiating particles in the past (e.g., in pulsar halos).~In the end, 39 out of 40 2FGES sources were associated with at least one $\gamma$-ray source or bright pulsar.~The remaining source, 2FGES~J1626.1$-$4710, could not be associated with any objects in the considered catalogs and is illustrated in Fig.~\ref{fig:orphan}.

In the following subsections, the TS maps include information on associations with TeV sources, with their contours or spatial extensions overlaid where available, as well as associations with bright pulsars and 4FGL sources when applicable.\footnote{The HGPS contours were obtained from the significance maps available at \url{https://www.mpi-hd.mpg.de/hfm/HESS/hgps/}, using the map corresponding to a 0.1$^{\circ}$ correlation radius.~The 3HWC contours were derived from the significance maps accessible at \url{https://data.hawc-observatory.org/datasets/3hwc-survey/fitsmaps.php}.~In regions observed by both HAWC and H.E.S.S., only H.E.S.S. contours were included to avoid overloading figures.~The spatial information of 1LHAASO sources associated with 2FGES sources is taken from~\citet{Cao24}. The 3HWC contours were derived from the significance maps using the map for a point-source search for 2FGES J1945.0+2506, and the map for a 0.5$^\circ$ extended-source search for 2FGES J0618.3+2227.}

\clearpage
\startlongtable
\tabletypesize{\scriptsize}
\begin{deluxetable*}{lccccccc}
\tablecaption{Summary of the Properties of the 2FGES Sources\label{ext_comps_properties}}
\tablecolumns{9}
\tablenum{2}
\tablehead{[-1.ex]
\colhead{2FGES Name} & \colhead{$l$} & \colhead{$b$} & \colhead{$r_{68}$} & \colhead{$\Gamma$ (\%)} & \colhead{$F_{>\,\rm{10\,GeV}}$ (\%)} & \colhead{TS$_{\rm{ext}}$} & \colhead{TS$_{\rm{std.}}$ (TS$_{\rm{alt.}}$)} \\ 
[-1ex]
\colhead{} & \colhead{($^{\circ}$)} & \colhead{($^{\circ}$)} & \colhead{($^{\circ}$)} & \colhead{} & \colhead{($\times\,10^{-10}\,\rm{cm}^{-2}\,\rm{s}^{-1}$)} & \colhead{} & \colhead{}}
\startdata
J0006.0+7319 & 119.64$\pm$0.11 & 10.74$\pm$0.16 & 0.85$\pm$0.12 & 1.85$\pm$0.16 (0) & 1.34$\pm$0.28 ($-$16) & 34.93 & 53.27 (33.10) \\
J0618.3+2227 & 189.28$\pm$0.12 & 3.21$\pm$0.10 & 0.97$\pm$0.08 & 2.07$\pm$0.15 (0) & 3.84$\pm$0.63 ($-$10) & 53.42 & 53.56 (42.86) \\
J0837.7$-$4534 & 264.13$\pm$0.20 & $-$2.69$\pm$0.17 & 2.85$\pm$0.24 & 2.19$\pm$0.11 (0) & 18.00$\pm$1.76 ($-$20) & 145.51 & 145.91 (88.29) \\
J1034.0$-$5832$^{\mystar}$ & 285.81$\pm$0.07 & $-$0.31$\pm$0.06 & 2.37$\pm$0.07 & 2.09$\pm$0.04 (1) & 41.20$\pm$2.18 ($-$17) & 423.09 & 423.67 (196.70) \\
J1045.1$-$5937 & 287.58$\pm$0.05 & $-$0.58$\pm$0.04 & 0.41$\pm$0.07 & 2.76$\pm$0.26 (0) & 3.26$\pm$0.46 ($-$13) & 61.88 & 62.29 (46.08) \\
J1101.7$-$6110 & 290.10$\pm$0.19 & $-$1.07$\pm$0.12 & 0.98$\pm$0.12 & 2.79$\pm$0.31 ($-$4) & 4.58$\pm$0.80 (3) & 40.95 & 41.15 (44.45) \\
J1112.6$-$6059 & 291.24$\pm$0.06 & $-$0.39$\pm$0.06 & 0.39$\pm$0.06 & 1.95$\pm$0.15 ($-$2) & 2.34$\pm$0.44 ($-$3) & 43.77 & 50.43 (51.19) \\
J1119.3$-$6126 & 292.16$\pm$0.04 & $-$0.51$\pm$0.03 & 0.32$\pm$0.05 & 2.15$\pm$0.14 (0) & 2.52$\pm$0.36 (1) & 65.84 & 69.27 (70.39) \\
J1156.5$-$6205 & 296.56$\pm$0.11 & 0.10$\pm$0.25 & 3.01$\pm$0.43 & 2.31$\pm$0.13 (0) & 12.90$\pm$1.72 ($-$21) & 62.84 & 63.15 (37.83) \\
J1312.3$-$6253 & 305.31$\pm$0.07 & $-$0.11$\pm$0.06 & 0.39$\pm$0.10 & 2.70$\pm$0.35 (0) & 2.18$\pm$0.40 ($-$5) & 34.67 & 35.34 (31.77) \\
J1404.3$-$5835 & 312.33$\pm$0.13 & 2.95$\pm$0.20 & 1.38$\pm$0.27 & 2.24$\pm$0.13 (0) & 5.31$\pm$0.64 ($-$41) & 83.88 & 84.52 (26.77) \\
J1408.4$-$6126 & 312.00$\pm$0.06 & 0.07$\pm$0.04 & 0.89$\pm$0.05 & 2.29$\pm$0.07 ($-$2) & 19.50$\pm$2.13 ($-$18) & 87.22 & 94.22 (62.06) \\
J1417.5$-$6057 & 313.19$\pm$0.03 & 0.18$\pm$0.02 & 0.17$\pm$0.04 & 1.68$\pm$0.12 ($-$2) & 2.45$\pm$0.41 ($-$10) & 60.59 & 97.77 (70.07) \\
J1452.1$-$5942 & 317.62$\pm$0.13 & $-$0.32$\pm$0.10 & 1.06$\pm$0.10 & 2.30$\pm$0.16 (0) & 6.94$\pm$0.94 ($-$31) & 61.10 & 61.11 (27.70) \\
J1505.6$-$5814 & 319.85$\pm$0.08 & 0.15$\pm$0.06 & 0.53$\pm$0.06 & 2.07$\pm$0.14 (0) & 3.76$\pm$0.52 ($-$29) & 68.21 & 76.11 (36.97) \\
J1514.2$-$5909 & 320.34$\pm$0.01 & $-$1.20$\pm$0.01 & 0.19$\pm$0.01 & 1.88$\pm$0.06 (0) & 6.92$\pm$0.38 ($-$1) & 364.55 & 1054.86 (1014.18) \\
J1554.6$-$5322$^{\mystar}$ & 328.25$\pm$0.05 & 0.24$\pm$0.03 & 0.46$\pm$0.03 & 2.48$\pm$0.10 (0) & 9.91$\pm$1.69 ($-$9) & 31.96 & 35.41 (29.02) \\
J1614.5$-$5154 & 331.48$\pm$0.03 & $-$0.67$\pm$0.03 & 0.44$\pm$0.04 & 1.92$\pm$0.07 (1) & 8.80$\pm$0.79 ($-$5) & 151.99 & 169.97 (148.09) \\
J1617.8$-$5052 & 332.58$\pm$0.07 & $-$0.28$\pm$0.06 & 0.49$\pm$0.06 & 2.20$\pm$0.14 ($-$13) & 8.13$\pm$1.08 ($-$31) & 63.70 & 65.16 (33.28) \\
J1626.1$-$4710 & 336.14$\pm$0.07 & 1.37$\pm$0.07 & 0.37$\pm$0.07 & 2.23$\pm$0.24 (0) & 1.80$\pm$0.35 (5) & 27.74 & 34.15 (38.13) \\
J1633.5$-$4743$^{\mystar}$ & 336.61$\pm$0.05 & 0.08$\pm$0.03 & 0.57$\pm$0.03 & 2.16$\pm$0.07 (0) & 18.80$\pm$1.23 ($-$6) & 273.88 & 291.28 (249.36) \\
J1640.6$-$4633 & 338.29$\pm$0.01 & $-$0.03$\pm$0.01 & 0.08$\pm$0.01 & 1.88$\pm$0.08 ($-$1) & 4.99$\pm$0.40 ($-$2) & 49.43 & 391.79 (378.51) \\
J1647.6$-$4551 & 339.61$\pm$0.07 & $-$0.49$\pm$0.08 & 0.54$\pm$0.07 & 2.15$\pm$0.15 ($-$2) & 5.66$\pm$0.82 (13) & 44.27 & 57.59 (76.07) \\
J1712.8$-$3950 & 347.19$\pm$0.26 & $-$0.41$\pm$0.14 & 3.45$\pm$0.22 & 2.22$\pm$0.06 (0) & 53.80$\pm$3.57 ($-$33) & 252.55 & 253.04 (105.02) \\
J1714.3$-$3829 & 348.46$\pm$0.01 & 0.14$\pm$0.01 & 0.12$\pm$0.02 & 2.67$\pm$0.17 ($-$2) & 3.60$\pm$0.37 (3) & 34.21 & 180.68 (192.20) \\
J1741.4$-$3015 & 358.40$\pm$0.18 & 0.09$\pm$0.11 & 3.15$\pm$0.23 & 2.55$\pm$0.09 (0) & 59.30$\pm$4.33 ($-$60) & 195.30 & 195.75 (25.82) \\
J1754.4$-$2600 & 3.51$\pm$0.06 & $-$0.17$\pm$0.07 & 0.31$\pm$0.06 & 2.00$\pm$0.21 (0) & 2.27$\pm$0.50 (7) & 29.46 & 29.89 (34.97) \\
J1759.4$-$2355 & 5.89$\pm$0.07 & $-$0.11$\pm$0.05 & 0.83$\pm$0.06 & 2.42$\pm$0.10 ($-$1) & 15.80$\pm$1.39 ($-$15) & 150.24 & 150.52 (104.03) \\
J1829.3$-$1556 & 16.25$\pm$0.12 & $-$2.45$\pm$0.08 & 0.56$\pm$0.08 & 1.86$\pm$0.11 (1) & 3.89$\pm$0.50 ($-$39) & 94.53 & 99.10 (35.15) \\
J1829.9$-$1423 & 17.70$\pm$0.11 & $-$1.87$\pm$0.11 & 0.64$\pm$0.15 & 1.85$\pm$0.11 (1) & 4.60$\pm$0.59 ($-$30) & 74.88 & 86.45 (40.30) \\
J1834.7$-$0846 & 23.23$\pm$0.02 & $-$0.31$\pm$0.02 & 0.24$\pm$0.02 & 1.96$\pm$0.08 ($-$1) & 6.56$\pm$0.59 (5) & 120.84 & 207.14 (234.54) \\
J1836.4$-$0652$^{\mystar}$ & 25.11$\pm$0.02 & 0.18$\pm$0.03 & 0.54$\pm$0.03 & 2.09$\pm$0.06 ($-$1) & 22.90$\pm$3.07 ($-$4) & 60.37 & 69.80 (63.35) \\
J1857.2+0248$^{\mystar}$ & 36.09$\pm$0.04 & $-$0.02$\pm$0.04 & 0.41$\pm$0.08 & 1.89$\pm$0.11 (0) & 5.95$\pm$0.97 ($-$5) & 79.83 & 96.76 (54.44) \\
J1907.8+0623 & 40.48$\pm$0.12 & $-$0.71$\pm$0.12 & 0.74$\pm$0.11 & 1.64$\pm$0.15 (0) & 3.34$\pm$0.76 ($-$9) & 44.30 & 44.30 (37.65) \\
J1913.5+1024 & 44.70$\pm$0.12 & $-$0.12$\pm$0.08 & 0.77$\pm$0.08 & 2.27$\pm$0.16 (0) & 5.67$\pm$0.80 ($-$19) & 53.01 & 60.27 (38.25) \\
J1923.0+1410 & 49.11$\pm$0.01 & $-$0.40$\pm$0.01 & 0.16$\pm$0.03 & 2.60$\pm$0.15 (0) & 6.08$\pm$0.88 ($-$4) & 23.68 & 52.98 (49.54) \\
J1930.3+1838 & 53.89$\pm$0.13 & 0.18$\pm$0.11 & 0.82$\pm$0.10 & 1.76$\pm$0.13 (0) & 3.22$\pm$0.59 ($-$10) & 41.22 & 54.43 (41.57) \\
J1945.0+2506 & 61.19$\pm$0.21 & 0.38$\pm$0.18 & 2.39$\pm$0.30 & 2.25$\pm$0.11 (0) & 11.70$\pm$1.44 ($-$35) & 74.49 & 77.66 (30.61) \\
J2020.9+4030$^{\mystar}$ & 78.21$\pm$0.02 & 2.21$\pm$0.02 & 0.53$\pm$0.03 & 2.12$\pm$0.06 (2) & 12.50$\pm$1.19 (18) & 186.21 & 187.14 (155.32) \\
J2252.4+5828 & 107.90$\pm$0.12 & $-$0.88$\pm$0.12 & 0.97$\pm$0.12 & 1.96$\pm$0.14 (1) & 2.63$\pm$0.40 (1) & 61.33 & 61.33 (60.43) \\
\enddata
\tablecomments{Columns 2--4 provide the Galactic longitude, Galactic latitude, and the spatial extension ($r_{68}$) of the 2FGES sources, respectively, derived from a 2D Gaussian fit used as the spatial model.~The photon index $\Gamma$ and the 10\,GeV--1\,TeV photon flux, along with their statistical uncertainties, are presented in columns 5 and 6, respectively.~The values in parentheses indicate the relative impact of using the alternative IEM, expressed as a percentage and derived as $\nicefrac{(X_{\rm{alt.}} - X_{\rm{std.}})}{X_{\rm{std.}}}$, where $X_{\rm{std.}}$ and $X_{\rm{alt.}}$ represent the measured quantities using the standard and alternative IEMs, respectively.~Columns 7 and 8 provide the TS$_{\rm{ext}}$ and TS values obtained using the standard IEM in the analysis, with the TS values obtained using the alternative IEM shown in parentheses.~Sources flagged with a star indicate those for which the TS values increased upon removal of coincident low-significance sources during the additional evaluation.}
\end{deluxetable*}

\clearpage
\startlongtable
\tabletypesize{\scriptsize}
\begin{deluxetable*}{lcccccc}
\tablecaption{Summary of the Potential Associations of the 2FGES Sources\label{ext_comps_associations}}
\tablecolumns{7}
\tablenum{3}
\tablehead{[-1.ex]
\colhead{2FGES Name} & \colhead{4FGL Name ($^{\circ}$)} & \colhead{Class} & \colhead{PSR Name ($^{\circ}$)} & \colhead{HGPS Name ($^{\circ}$)} & \colhead{3HWC Name ($^{\circ}$)} & \colhead{1LHAASO Name ($^{\circ}$)} 
}
\startdata
J0006.0+7319 & J0007.0+7303 (0.28) & PSR & J0007+7303 (0.28) & ... & ... & J0007+7303u (0.28, 0.17) \\
J0618.3+2227 & J0618.9+2240c (0.25) & ... & ... & ... & J0617+224 (0.19) & J0617+2234 (..., 0.25) \\
J0837.7$-$4534 & J0835.3$-$4510 (0.59) & PSR & B0833$-$45 (0.59) & J0835$-$455 (0.40) & ... & ... \\
J1034.0$-$5832 & J1036.3$-$5833e (0.30) & ... & J1028$-$5819 (0.76) & J1026$-$582 (0.98) & ... & ... \\
J1045.1$-$5937 & J1045.1$-$5940 (0.04) & BIN & ... & $\eta$ Carinae (0.08) & ... & ... \\
J1101.7$-$6110 & J1102.0$-$6054 (0.27) & bcu & J1101$-$6101 (0.15) & ... & ... & ... \\
J1112.6$-$6059 & J1112.2$-$6055 (0.09) & ... & J1112$-$6103 (0.07) & ... & ... & ... \\
J1119.3$-$6126 & J1119.1$-$6127 (0.04) & PSR & ... & J1119$-$614 (0.04) & ... & ... \\
J1156.5$-$6205 & J1152.6$-$6207 (0.46) & spp & J1151$-$6108 (1.10) & ... & ... & ... \\
J1312.3$-$6253 & J1312.3$-$6257 (0.08) & ... & ... & ... & ... & ... \\
J1404.3$-$5835$^{\dag}$ & J1403.2$-$5758 (0.62) & ... & ... & ... & ... & ... \\
J1408.4$-$6126 & J1409.1$-$6121e (0.13) & ... & J1413$-$6205 (0.89) & ... & ... & ... \\
J1417.5$-$6057 & J1417.7$-$6057 (0.03) & ... & J1418$-$6058 (0.15) & J1418$-$609 (0.06) & ... & ... \\
J1452.1$-$5942 & J1450.2$-$5937c (0.25) & spp & ... & J1457$-$593 (0.74) & ... & ... \\
J1505.6$-$5814 & J1503.7$-$5801 (0.34) & bcu & ... & J1503$-$582 (0.31) & ... & ... \\
J1514.2$-$5909 & J1514.2$-$5909e (0.01) & PWN & ... & J1514$-$591 (0.02) & ... & ... \\
J1554.6$-$5322 & J1553.8$-$5325e (0.13) & ... & ... & ... & ... & ... \\
J1614.5$-$5154 & J1615.3$-$5146e (0.18) & SNR & ... & J1614$-$518 (0.07) & ... & ... \\
J1617.8$-$5052$^{\dag}$ & J1616.2$-$5054e (0.26) & PWN & ... & J1616$-$508 (0.15) & ... & ... \\
J1626.1$-$4710 & ... & ... & ... & ... & ... & ... \\
J1633.5$-$4743$^{\dag}$ & J1633.0$-$4746e (0.10) & spp & ... & J1632$-$478 (0.28) & ... & ... \\
J1640.6$-$4633 & J1640.7$-$4631e (0.03) & spp & ... & J1640$-$465 (0.01) & ... & ... \\
J1647.6$-$4551 & J1648.4$-$4611 (0.36) & PSR & J1648$-$4611 (0.34) & J1646$-$458 (0.40) & ... & ... \\
J1712.8$-$3950 & J1713.5$-$3945e (0.16) & SNR & J1718$-$3825 (1.76) & J1713$-$397 (0.13) & ... & ... \\
J1714.3$-$3829 & J1714.4$-$3830 (0.02) & PSR & ... & J1714$-$385 (0.04) & ... & ... \\
J1741.4$-$3015 & J1742.0$-$3020 (0.16) & ... & B1737$-$30 (0.19) & J1741$-$302 (0.13) & ... & ... \\
J1754.4$-$2600 & J1755.4$-$2552 (0.26) & snr & ... & ... & ... & ... \\
J1759.4$-$2355 & J1759.7$-$2354 (0.08) & ... & ... & J1800$-$240 (0.32) & J1757$-$240 (0.54) & ... \\
J1829.3$-$1556 & J1829.3$-$1614 (0.30) & spp & ... & ... & ... & ... \\
J1829.9$-$1423 & J1830.1$-$1440 (0.29) & ... & ... & ... & ... & ... \\
J1834.7$-$0846 & J1834.5$-$0846e (0.04) & spp & ... & J1834$-$087 (0.04) & ... & ... \\
J1836.4$-$0652 & J1836.5$-$0651e (0.03) & pwn & J1838$-$0655 (0.40) & J1837$-$069 (0.27) & J1837$-$066 (0.39) & J1837$-$0654u (0.19, 0.28) \\
J1857.2+0248 & J1857.7+0246e (0.14) & PWN & J1856+0245 (0.11) & J1857+026 (0.05) & J1857+027 (0.01) & J1857+0245 (..., 0.07) \\
J1907.8+0623 & J1907.9+0602 (0.36) & PSR & J1907+0631 (0.23) & J1908+063 (0.15) & J1908+063 (0.10) & J1908+0615u (0.16, 0.16) \\
J1913.5+1024 & J1913.3+1019 (0.09) & PSR & J1913+1011 (0.22) & J1912+101 (0.24) & J1912+103 (0.34) & J1912+1014u (0.10, 0.23) \\
J1923.0+1410 & J1923.2+1408e (0.06) & SNR & ... & J1923+141 (0.03) & J1922+140 (0.10) & J1922+1403 (0.07, 0.12) \\
J1930.3+1838 & J1929.8+1832 (0.17) & unk & ... & J1930+188 (0.19) & J1930+188 (0.20) & J1929+1846u$^{\ast}$ (0.62, 0.27) \\
J1945.0+2506 & J1946.1+2436c (0.57) & unk & ... & ... & J1950+242 (1.54) & J1945+2424$^{\ast}$ (1.55, 0.72) \\
J2020.9+4030 & J2021.0+4031e (0.03) & SNR & J2021+4026 (0.12) & ... & J2020+403 (0.14) & J2020+4034 (0.08, 0.17) \\
J2252.4+5828 & J2250.6+5809 (0.41) & ... & ... & ... & ... & ... \\
\enddata
\tablecomments{Columns 2 and 3 list the associated 4FGL sources along with their classifications, as provided in the 4FGL-DR3 catalog.~The potential associations with ATNF bright pulsars, HGPS sources, 3HWC sources, and 1LHAASO sources are given in columns 4--7, respectively ($\eta$ Carinae was also considered, although it was not part of the HGPS catalog, as it was detected later by the H.E.S.S. collaboration as reported by~\citet{Abdalla20}).~The values in parentheses indicate the angular distances between the centroids of the 2FGES sources and their potential associations.~The 2FGES sources marked with a dagger ($\dag$) are linked to other 4FGL point sources with approximately the same offset:~4FGL J1408.9$-$5845c at 0.62$^{\circ}$ from 2FGES J1404.3$-$5835, 4FGL J1619.3$-$5047 at 0.25$^{\circ}$ from 2FGES J1617.8$-$5052, and 4FGL J1634.0$-$4742c at 0.09$^{\circ}$ from 2FGES J1633.5$-$4743.}
\end{deluxetable*}

\clearpage

\begin{deluxetable*}{lc}[t]
\tablecaption{Summary of 2FGES Sources Categories Based on Their Associations\label{tab:det_categories}}
\tablecolumns{2}
\tablenum{4}
{
\tablehead{[-1.ex]
\colhead{Description} & \colhead{Number of sources} 
}
\startdata
Number of previously detected 4FGL extended sources & 13 \\
Number of sources with 4FGL point source counterpart only & 6 \\
Number of sources with TeV counterpart but no bright pulsar association & 9 \\
Number of sources with bright pulsar association but no TeV counterpart & 2 \\
Number of sources with both TeV and bright pulsar associations & 4 \\
Number of sources with no counterpart at all & 1 \\
Number of dubious sources & 5 \\
\hline
Total number of 2FGES sources & 40 \\
\enddata}
\end{deluxetable*}

\subsection{2FGES Sources Superseding Previously Detected Extended Sources} \label{subsec:4fgle_srcs}

\begin{figure}[!pt]
\gridline{\fig{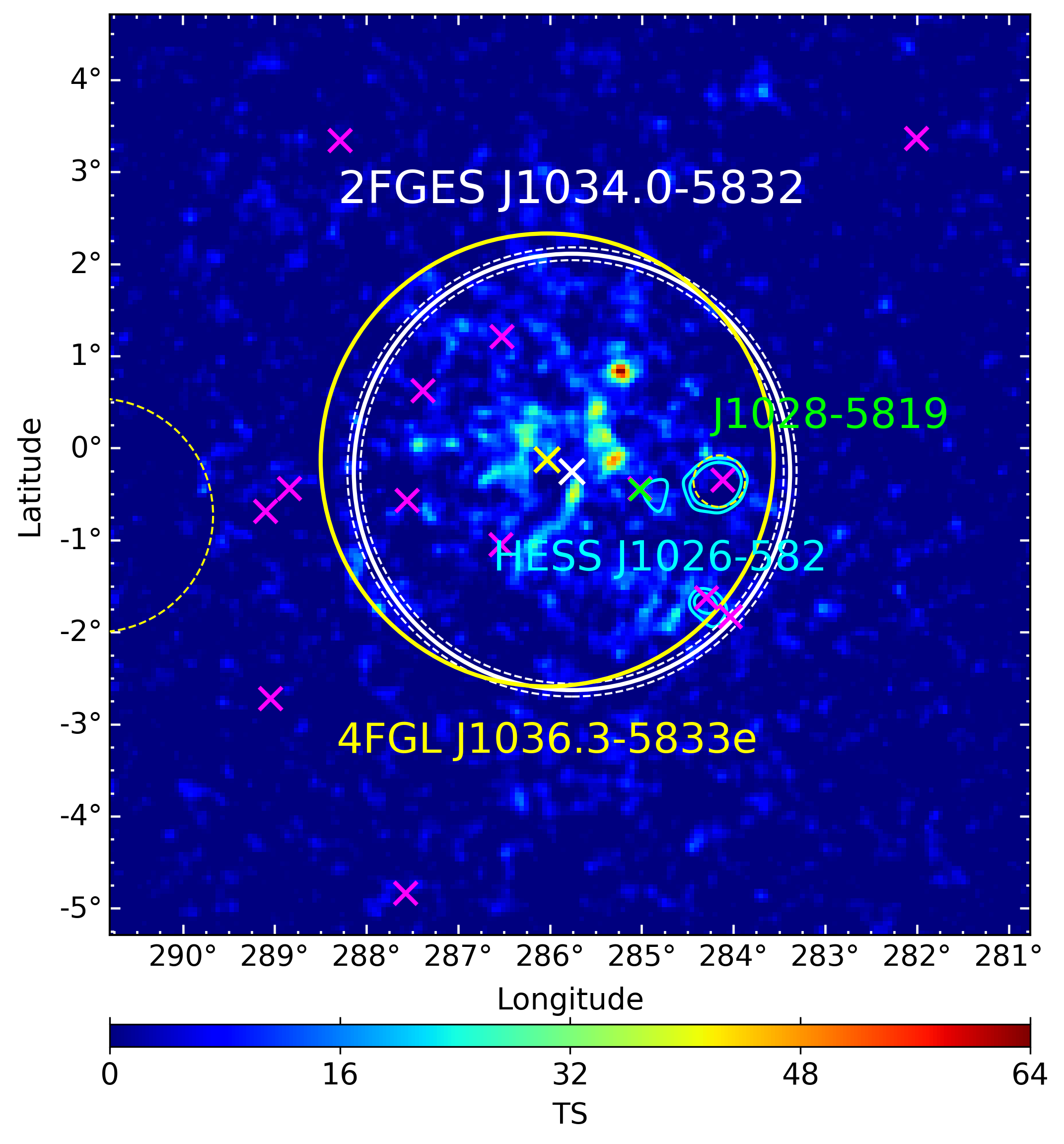}{0.35\textwidth}{}
          \fig{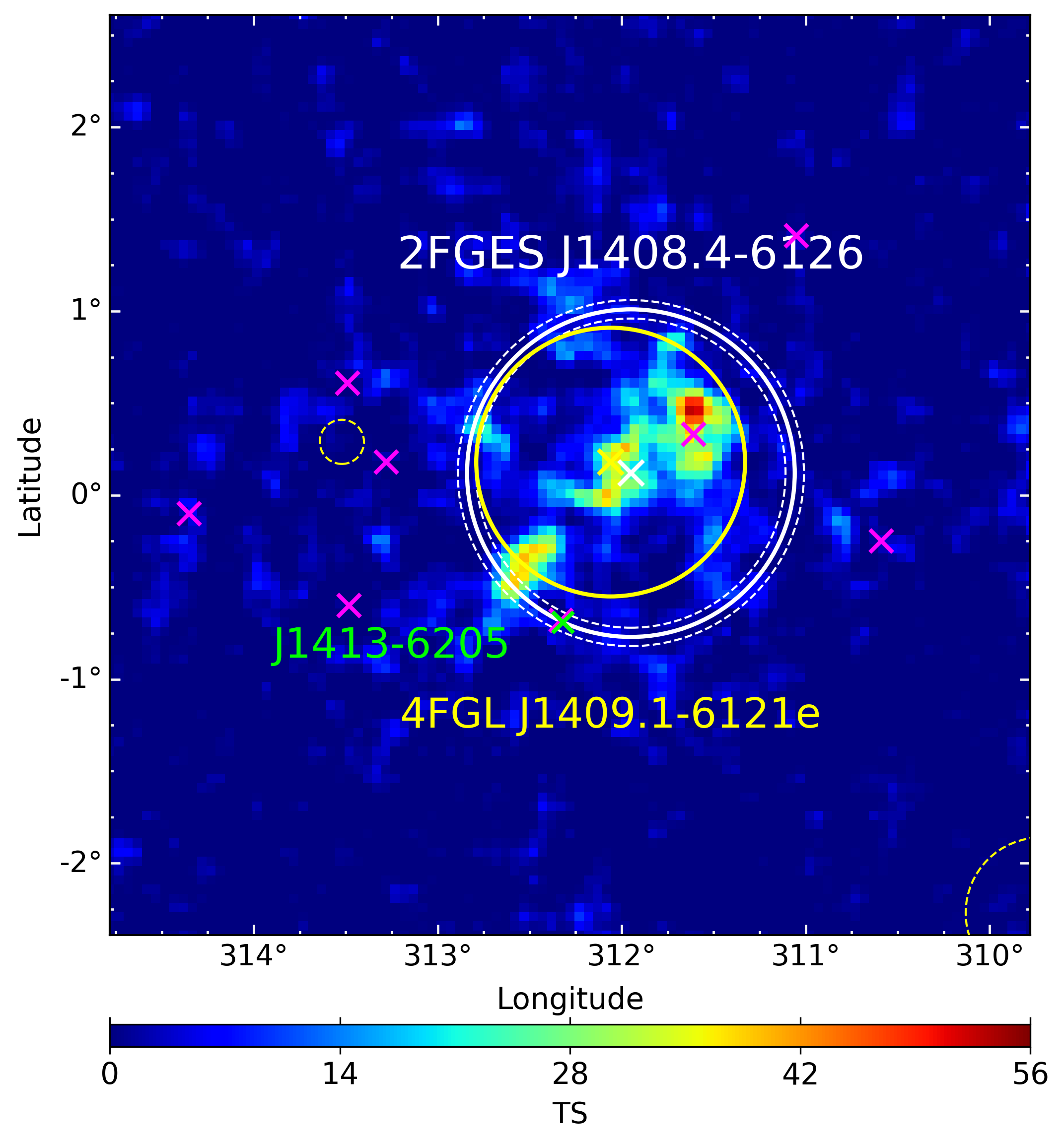}{0.35\textwidth}{}} 
\gridline{ \fig{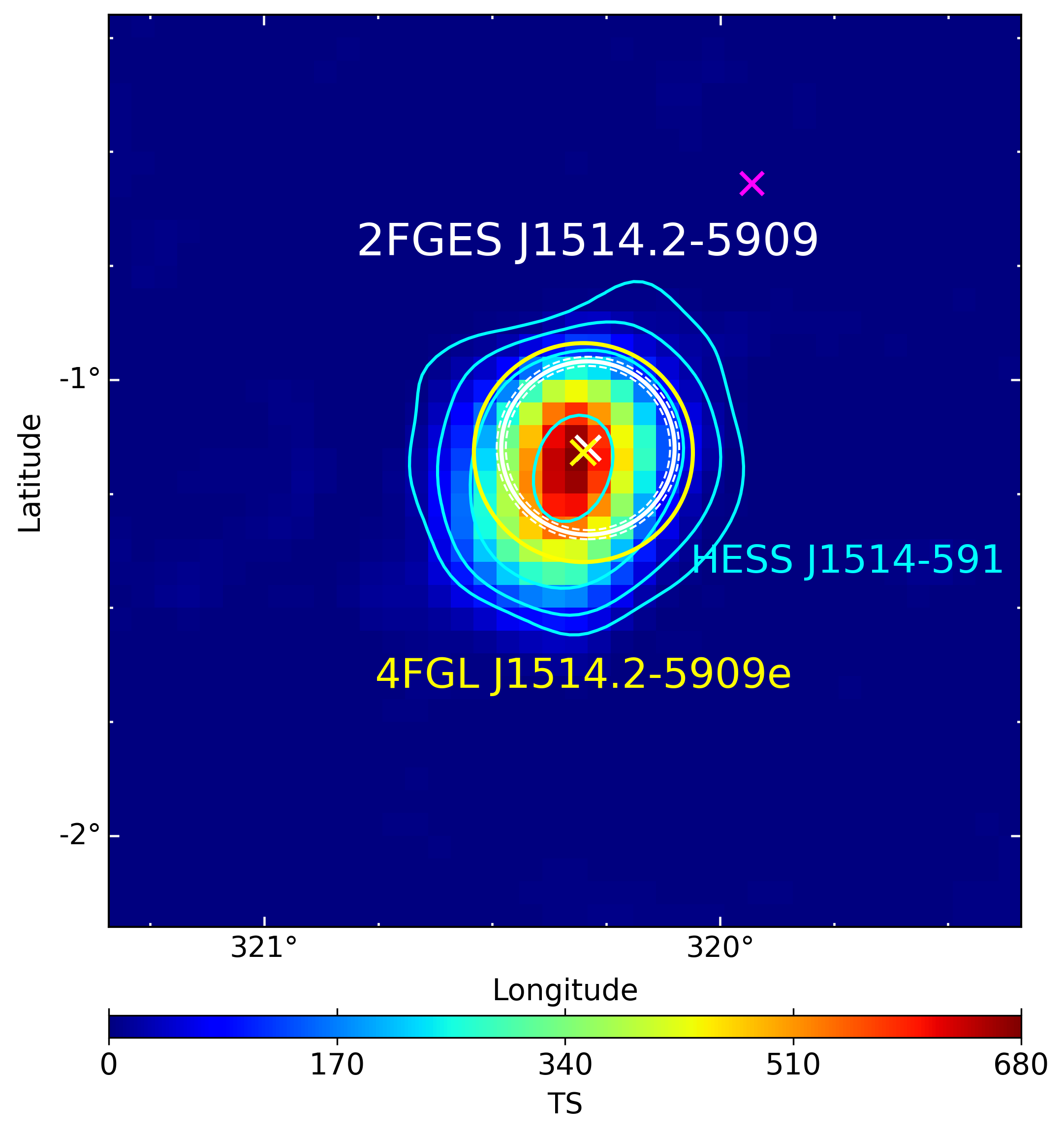}{0.35\textwidth}{}
          \fig{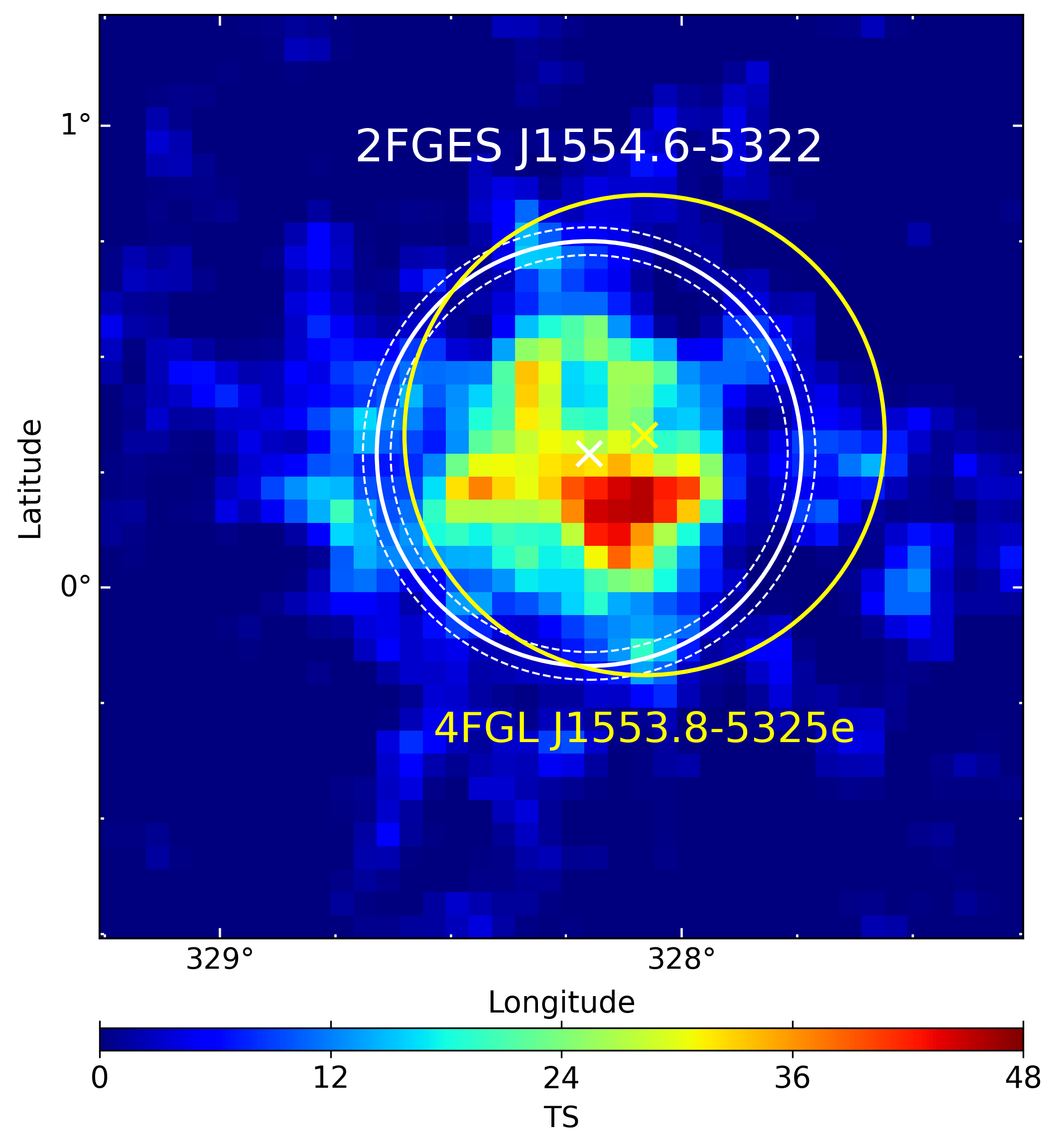}{0.35\textwidth}{}} 
\caption{TS maps of 2FGES sources superseding 4FGL extended sources.~The centroid of the 2FGES source is shown with a white cross, and its 68\% containment radius is indicated by a white circle, with the inner and outer radii representing the statistical error on the fitted extension.~The solid yellow circle indicates the extent of the previously detected 4FGL extended source superseded by the 2FGES source, while yellow dashed circles represent other fitted extended 4FGL sources in the field, some of which may overlap with the 2FGES source.~The purple circles represent the 39\% containment radii ($r_{39}$) of the associated LHAASO sources, detected by WCDA.~The magenta crosses mark surrounding 4FGL sources with TS\,$>$\,16.~The yellow, green, orange, and purple crosses depict the positions of the associated 4FGL source, bright pulsar, 3HWC source, and 1LHAASO source, respectively.~Cyan contours delineate the H.E.S.S. detection significance at 3, 5, 10, 30, and 50$\sigma$.~The corresponding 2FGES sources, from top-left to bottom-right, are 2FGES J1034.0$-$5832, 2FGES J1408.4$-$6126, 2FGES J1514.2$-$5909, and 2FGES J1554.6$-$5322.~\label{fig:4fgle}}
\end{figure}

\begin{figure}[!pt]
\gridline{ \fig{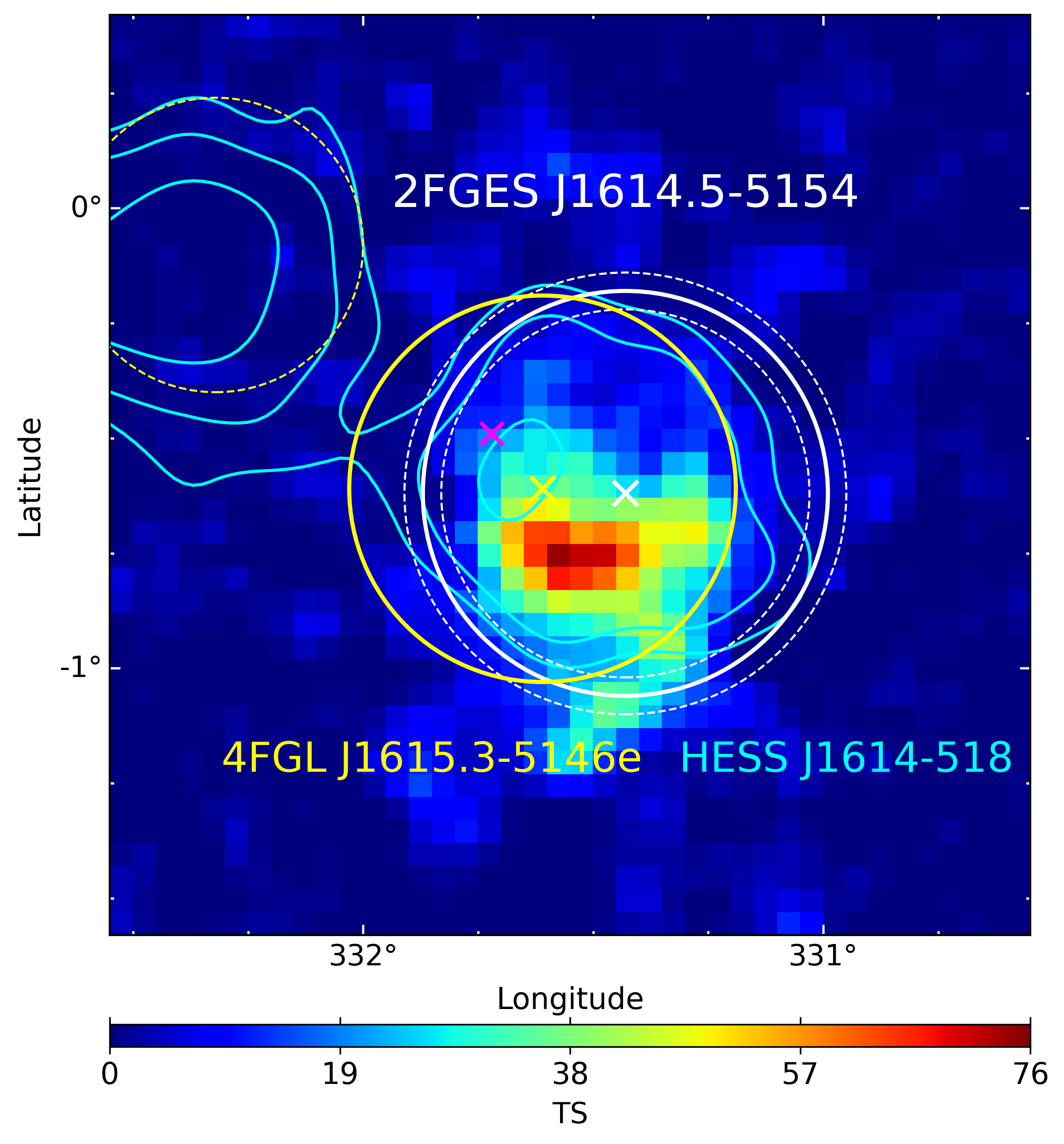}{0.35\textwidth}{}
          \fig{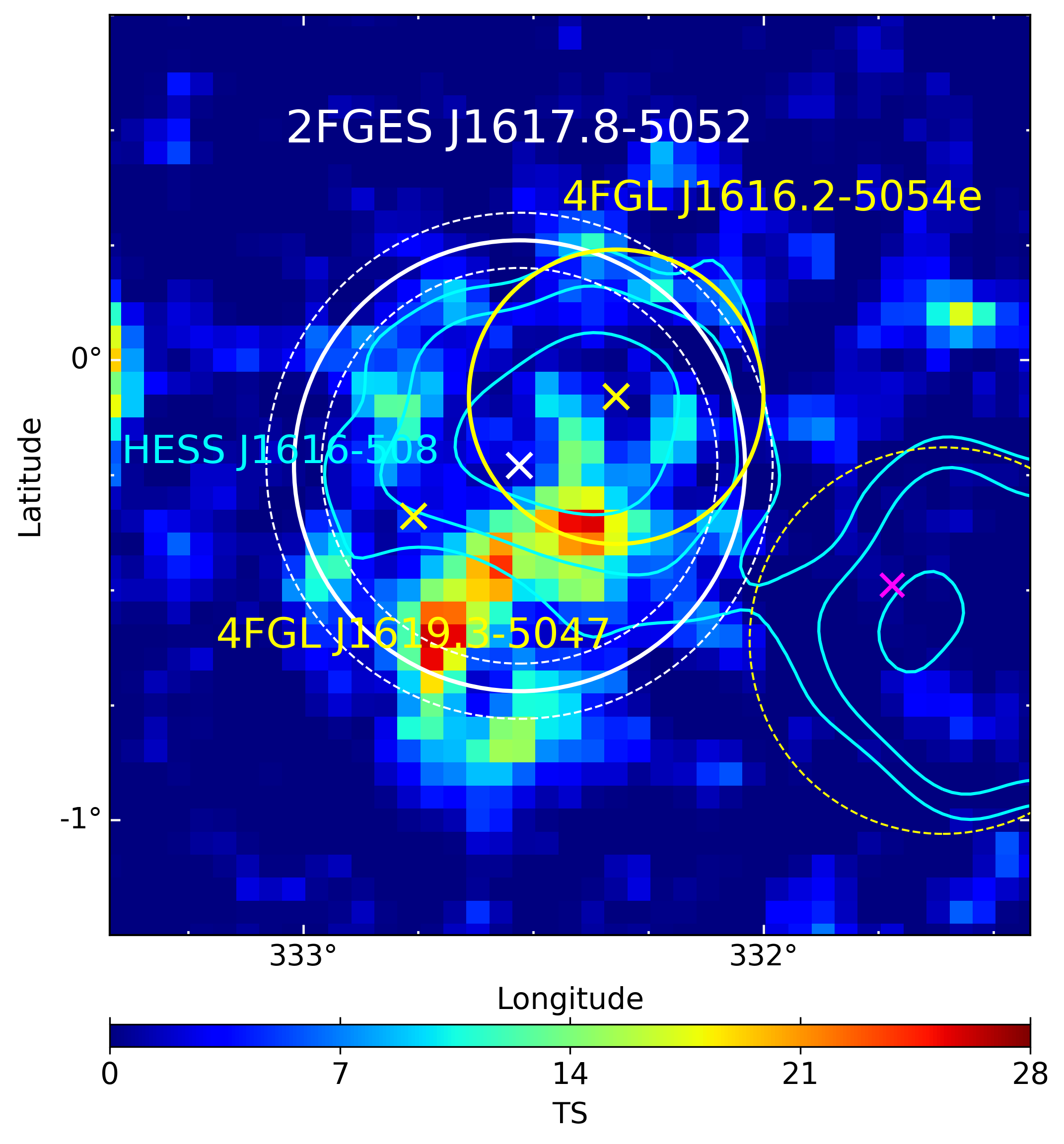}{0.35\textwidth}{}} 
 \gridline{ \fig{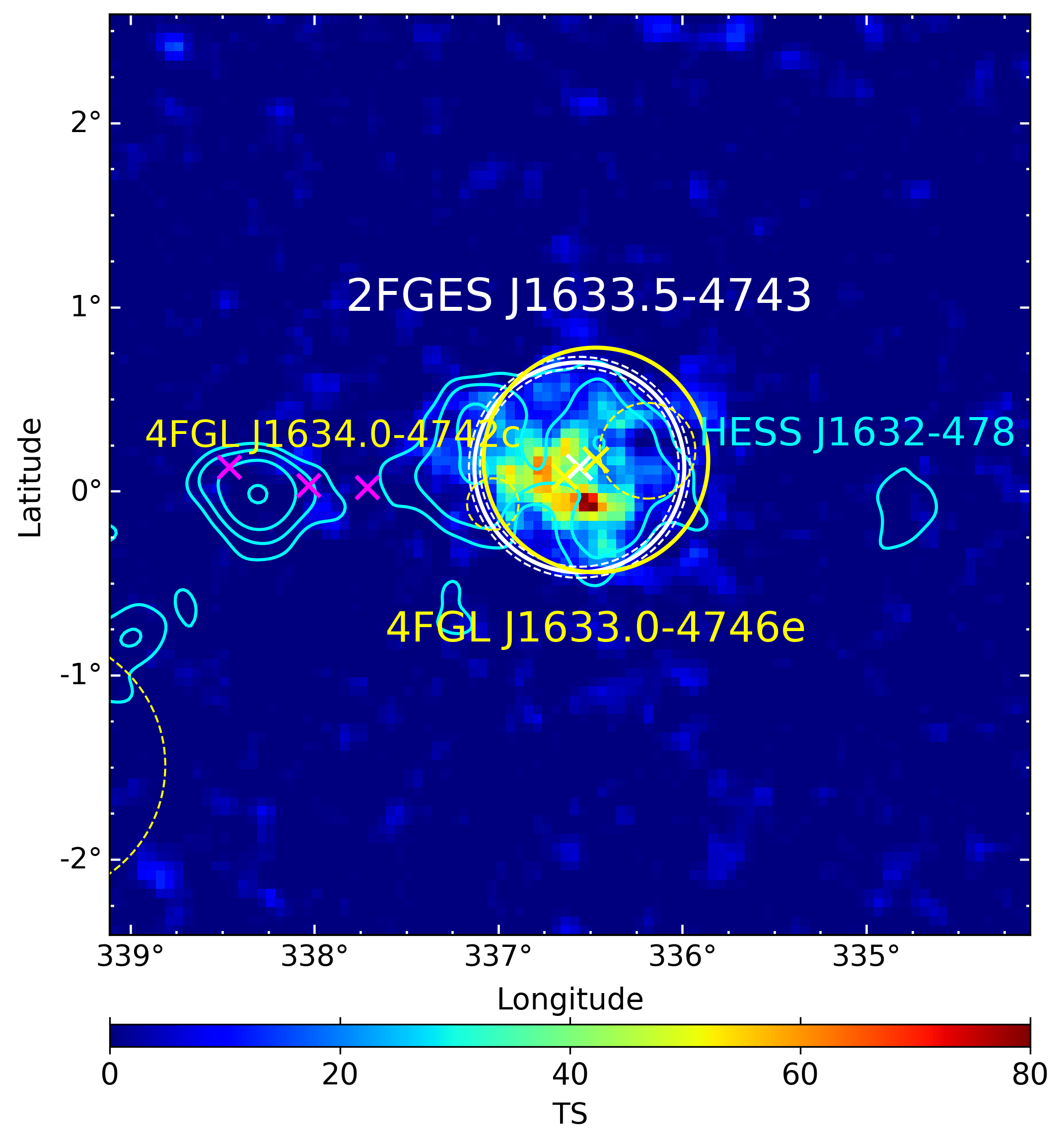}{0.35\textwidth}{}
          \fig{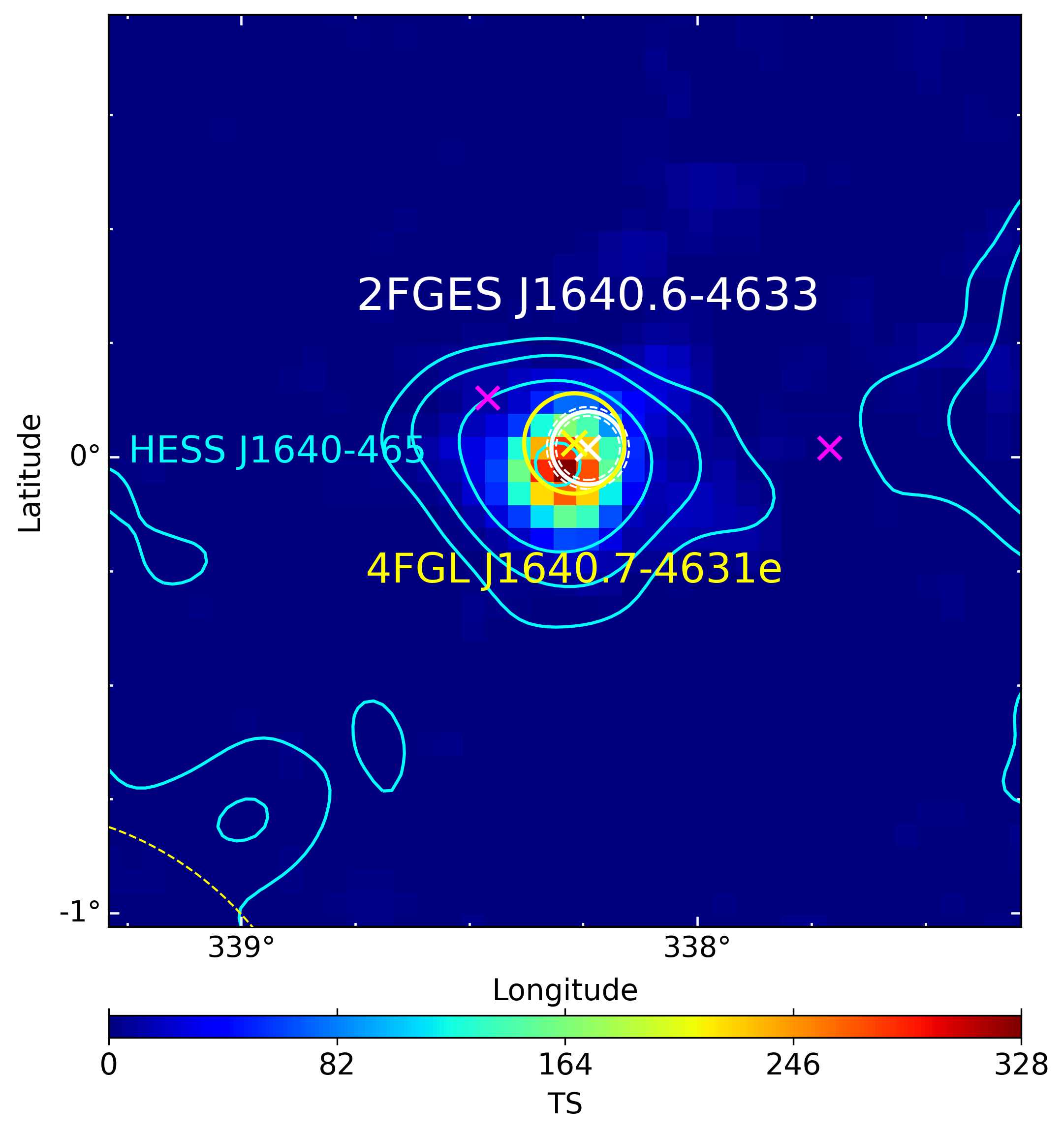}{0.35\textwidth}{}}
 \gridline{\fig{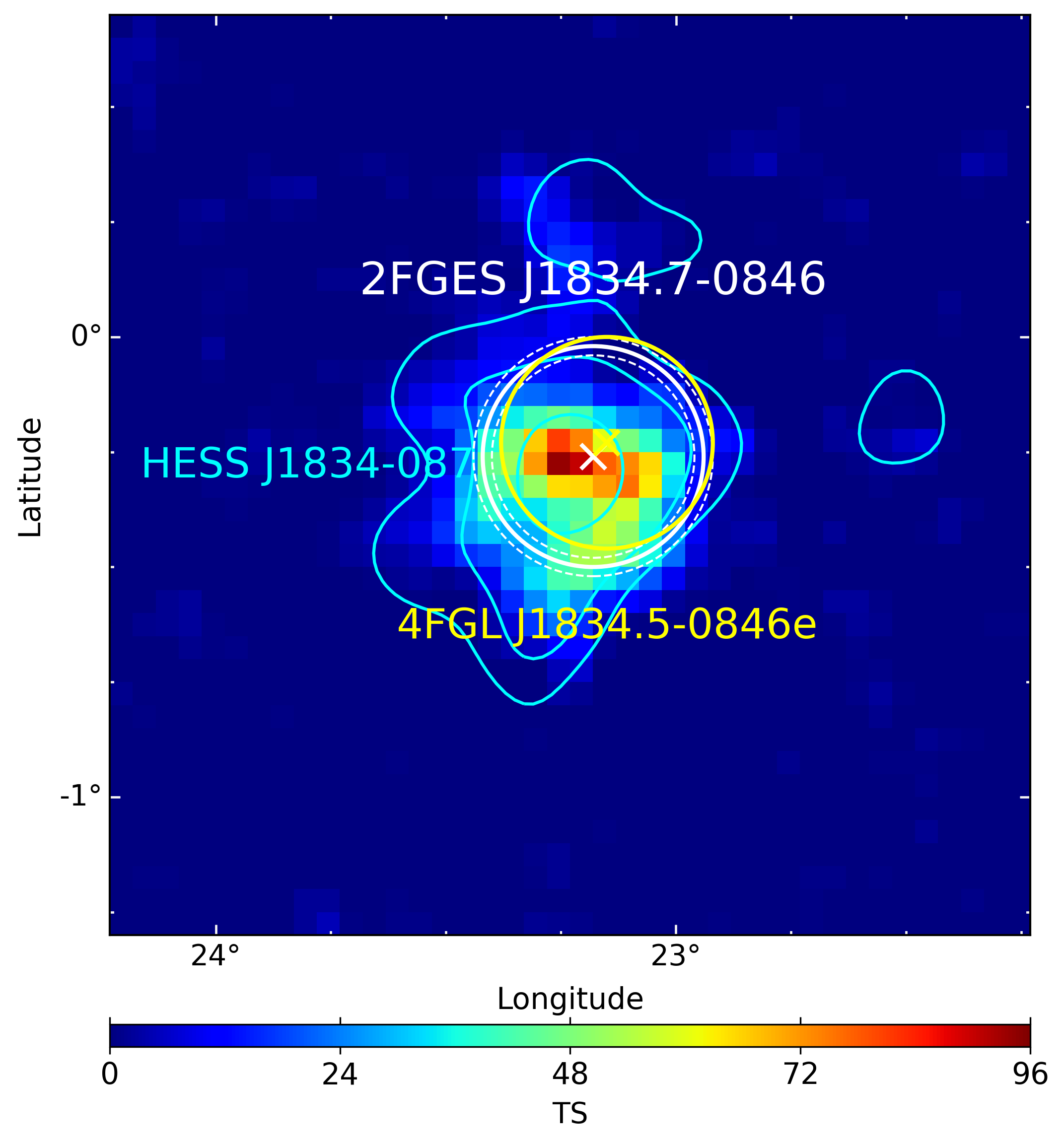}{0.35\textwidth}{}
           \fig{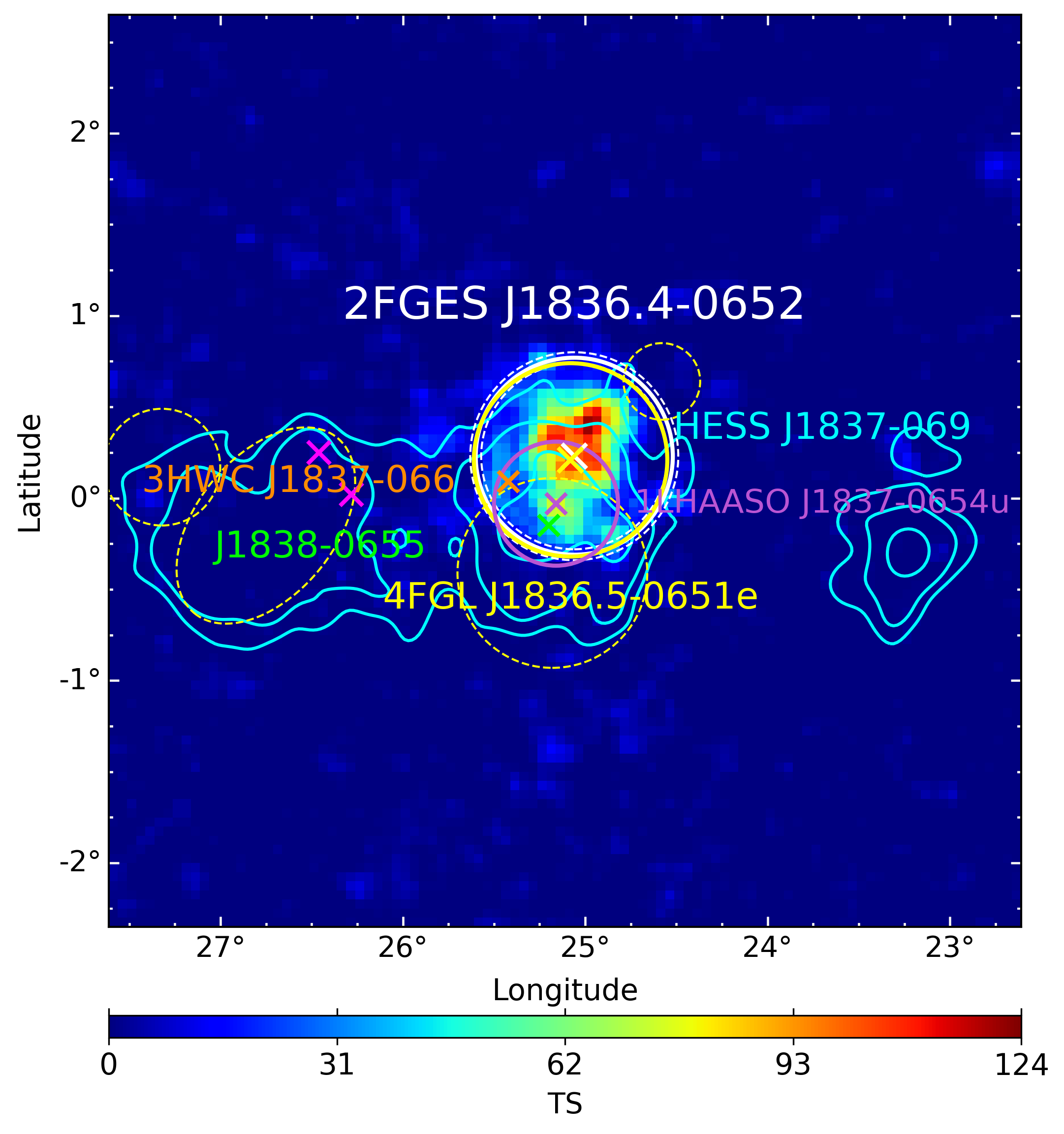}{0.35\textwidth}{}} 
\caption{TS maps of 2FGES sources superseding 4FGL extended sources, with elements as described in Fig.~\ref{fig:4fgle}.~The 2FGES sources shown here, from top-left to bottom-right, are 2FGES J1614.5$-$5154, 2FGES J1617.8$-$5052, 2FGES J1633.5$-$4743, 2FGES J1640.6$-$4633, 2FGES J1834.7$-$0846, and 2FGES J1836.4$-$0652.~\label{fig:4fgle_part2}}
\end{figure}

\begin{figure}[!pt]
\gridline{ \fig{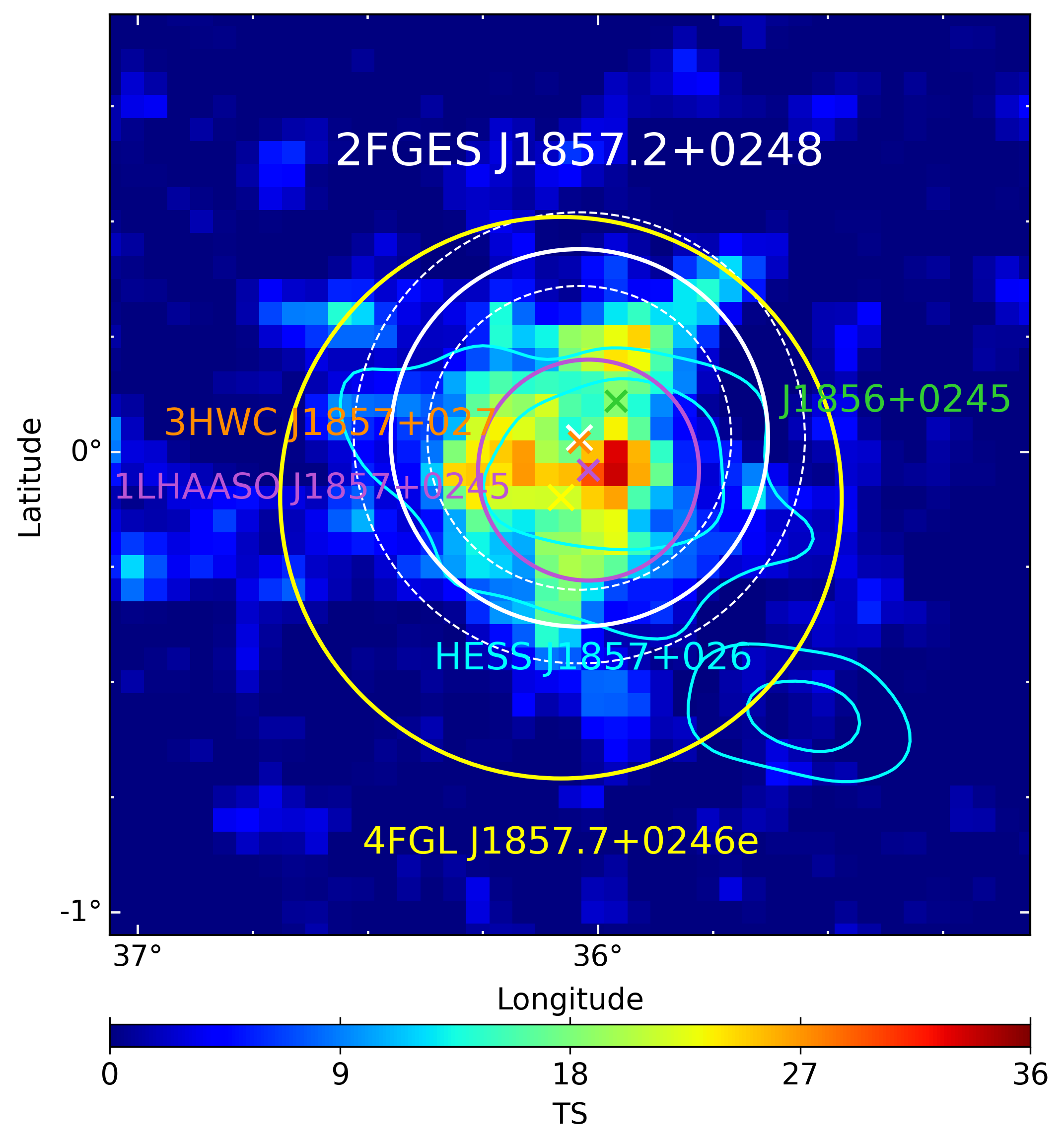}{0.35\textwidth}{}
          \fig{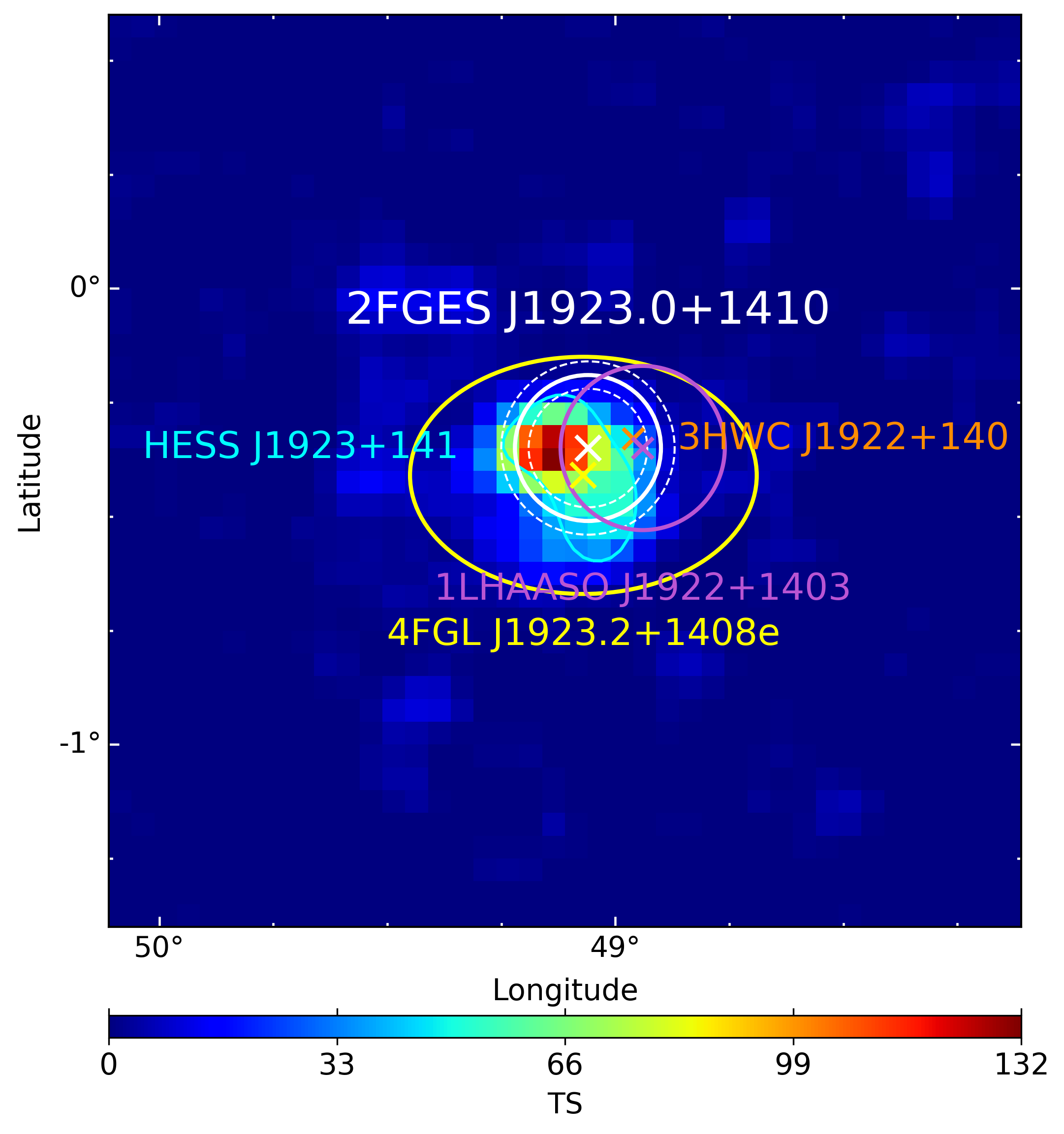}{0.35\textwidth}{}}
\gridline{ \fig{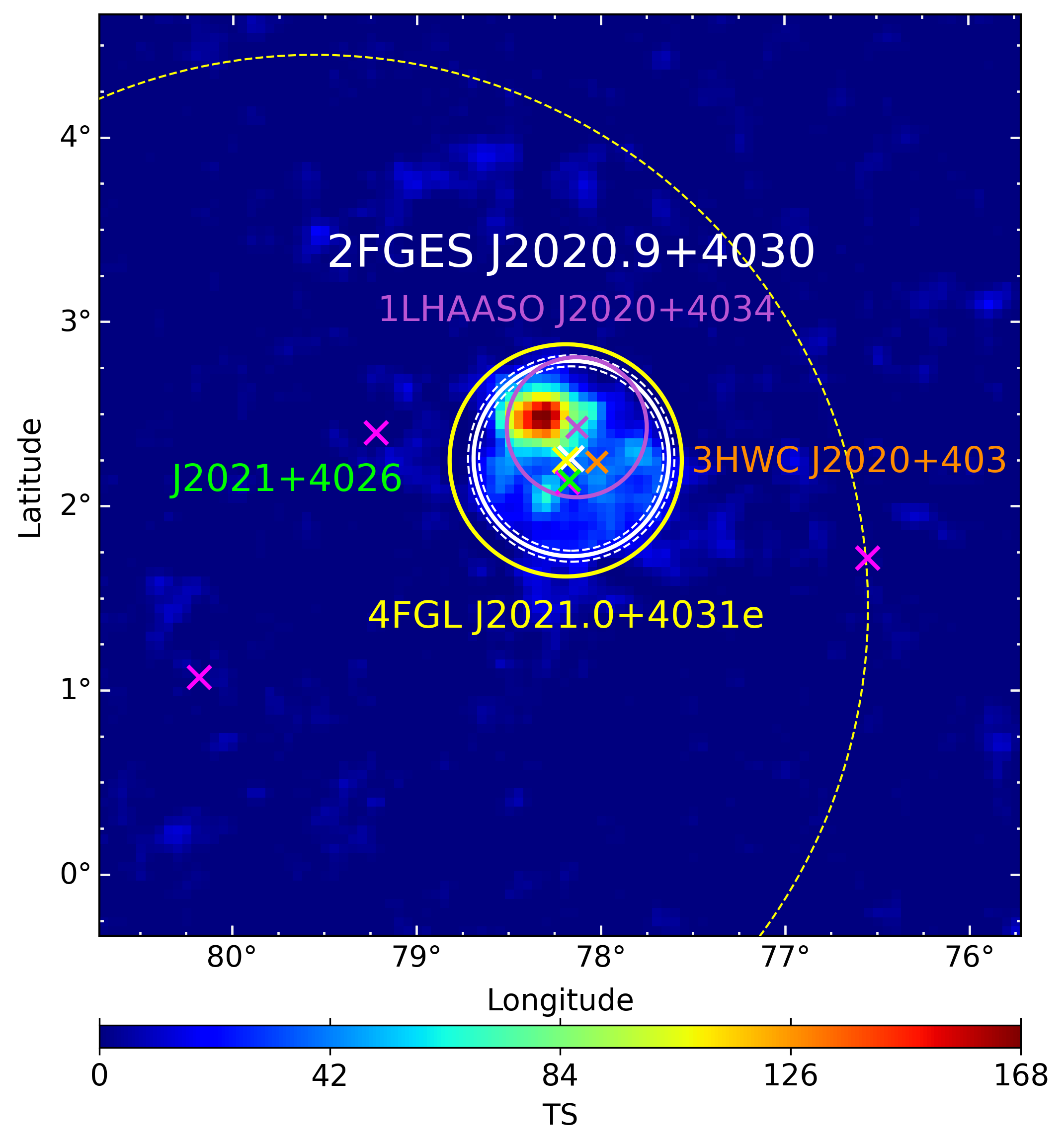}{0.35\textwidth}{}} 
\caption{TS maps of 2FGES sources superseding 4FGL extended sources, with elements as described in Fig.~\ref{fig:4fgle}.~The 2FGES sources shown here, from top-left to bottom-right, are 2FGES J1857.2+0248, 2FGES J1923.0+1410, and 2FGES J2020.9+4030.~\label{fig:4fgle_part3}}
\end{figure}

\begin{figure}[!pt]
 \gridline{ \fig{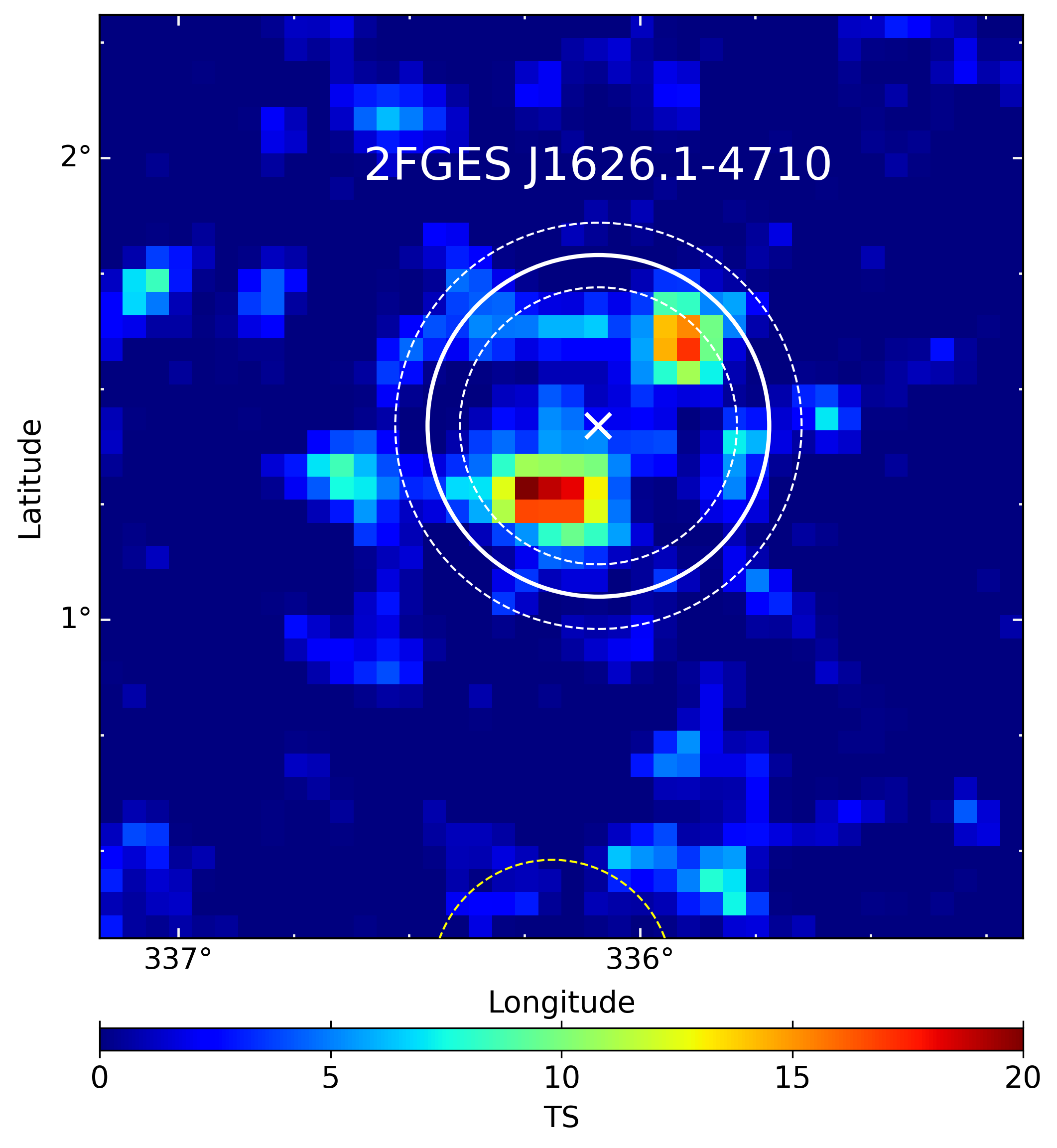}{0.35\textwidth}{}} 
\caption{TS map of 2FGES source classified as an orphan, with no associations in the scanned catalogs.~Graphical elements are defined in the caption of Fig.~\ref{fig:4fgle}.~The 2FGES source shown is 2FGES J1626.1$-$4710.~\label{fig:orphan}}
\end{figure}

Out of 40 extended sources in the catalog, 13 correspond to already known extended sources in the 4FGL-DR3 catalog.~This means that, when performing the data analysis, a given 4FGL extended source originally included in the model fitting with the morphology prescribed in the catalog was superseded by a 2FGES source (i.e., the 4FGLe source was pushed down to nearly zero in the fit, and the corresponding emission was transferred to the new 2FGES source).~This most likely happens because the catalog analysis uses one single best-fitting morphology over a broad energy range starting from 50\,MeV, while our analysis was performed above 10\,GeV.~Our interpretation is that the corresponding 2FGES sources feature a morphological description that is more suited to the highest energies than what is provided in the 4FGL catalog.~In most cases, as is illustrated in Figs.~\ref{fig:4fgle}$-$\ref{fig:4fgle_part3}, relatively minor modifications to the 4FGL morphologies are involved.~This demonstrates the ability of the analysis pipeline to recover known and documented extended emission components, several of which are unambiguously identified (as PWN or SNR, for instance).

Table~\ref{tab:previously_detected_ext_srcs} lists these previously detected extended sources along with the association proposed in the 4FGL catalog.~We note that sources 2FGES J1617.8$-$5052 and 2FGES J1633.5$-$4743 are most likely linked to 4FGL J1616.2$-$5054e and 4FGL J1633.0$-$4746e, respectively, despite also being associated with two 4FGL point sources that are slightly closer in angular distance than the corresponding 4FGL extended sources.~For the extended emission components positionally coincident with TeV sources, the associations remain consistent with those proposed in the 4FGL catalog in all but one case.~In particular, 2FGES J1034.0$-$5832 encompasses unidentified TeV source HESS J1026$-$582, but the latter is located 0.98$^{\circ}$ away from the 2FGES source's centroid, it is much smaller in size, and the GeV and TeV spectra do not connect at all, such that an association is very unlikely.~For sources 2FGES J1408.4$-$6126 and 2FGES J1554.6$-$5322, there is no association with known extended TeV sources, but we found an association of 2FGES J1408.4$-$6126 with PSR J1413$-$6205 at an angular distance of 0.89$^{\circ}$, which is equal to the 68\% containment radius of the 2FGES source.

Some 2FGES sources are detected with a smaller best-fit size compared to the morphological model used in the 4FGL catalog.~Notably, 2FGES J1857.2+0248 and 2FGES J1923.0+1410 are each associated with distinct H.E.S.S. sources and are closer in size to their respective H.E.S.S. counterparts.~This may hint at energy-dependent morphologies, holding interesting information about the underlying physics.~We emphasize, however, that size comparison is not straightforward since we used 2D Gaussian intensity distributions, while the 4FGL catalog relies on a variety of models (Gaussians, disks, maps, etc).

Lastly, while the 2FGES sources listed in Table~\ref{tab:previously_detected_ext_srcs} can be considered as replacements for the 4FGLe models in LAT data analyses above 10\,GeV, we also noted that other 2FGES sources may bear a relationship to nearby 4FGLe sources and be considered as additional components or corrections to the morphology prescribed in the 4FGL catalog.~This can be inferred from the TS maps shown in Figs.~\ref{fig:no_tev_psr_assoc}$-$\ref{fig:tev_part2}, where nearby 4FGLe sources in the field are highlighted with yellow dotted circles.~This is the case, for instance, with 2FGES J1101.7$-$6110, 2FGES J1112.6$-$6059, and 2FGES J1119.3$-$6126, all of which overlap with the much larger 4FGL 1109.4$-$6115e source, which itself is associated with the large, unknown source FGES J1109.4$-$6115.

\begin{deluxetable}{lccccc}[t] 
\tablecaption{2FGES Sources Superseding Previously Detected 4FGL Extended Sources\label{tab:previously_detected_ext_srcs}}
\tablenum{5}
\tablehead{
\colhead{2FGES Name} & \colhead{4FGL Name} & \colhead{4FGL Class} & \colhead{4FGL Association} & \colhead{4FGL Spatial Model} & \colhead{Angular Separation ($^{\circ}$)} 
}
\startdata
J1034.0$-$5832 & J1036.3$-$5833e & ... & ... & Disk & 0.30 \\
J1408.4$-$6126 & J1409.1$-$6121e & ... & ... & Disk & 0.12 \\
J1514.2$-$5909 & J1514.2$-$5909e & PWN & MSH 15$-$52 & Disk & 0.01 \\
J1554.6$-$5322 & J1553.8$-$5325e & ... & ... & Disk & 0.13 \\
J1614.5$-$5154 & J1615.3$-$5146e & SNR & HESS J1614$-$518 & Disk & 0.18 \\
J1617.8$-$5052$^{\dag}$ & J1616.2$-$5054e & PWN & HESS J1616$-$508 & Disk & 0.26 \\ 
J1633.5$-$4743$^{\dag}$ & J1633.0$-$4746e & spp & HESS J1632$-$478 & Disk & 0.10 \\ 
J1640.6$-$4633 & J1640.7$-$4631e & spp & SNR G338.3$-$0.0 & Gaussian & 0.03 \\
J1834.7$-$0846 & J1834.5$-$0846e & spp & W 41 & Gaussian & 0.04 \\
J1836.4$-$0652 & J1836.5$-$0651e & pwn & HESS J1837$-$069 & Disk & 0.03 \\
J1857.2+0248 & J1857.7+0246e & PWN & HESS J1857+026 & Disk & 0.13 \\
J1923.0+1410 & J1923.2+1408e & SNR & W 51C & Elliptical Disk & 0.06 \\
J2020.9+4030 & J2021.0+4031e & SNR & gamma Cygni & Disk & 0.03 \\
\enddata
\tablecomments{Columns 2--5 list the 4FGL extended sources recovered by the pipeline and superseded by 2FGES sources, along with their classifications, associations, and spatial models as provided in the 4FGL-DR3 catalog.~The last column represents the angular separation between the centroids of the 2FGES sources and the corresponding 4FGL extended sources.~Sources marked with a dagger ($\dag$) are each also linked to another 4FGL point source with approximately the same offset.}
\end{deluxetable}

\subsection{Extended Components Associated to 4FGL Point Sources Only} \label{subsec:isolated_ecs}

\begin{figure}[!pt]
\gridline{\fig{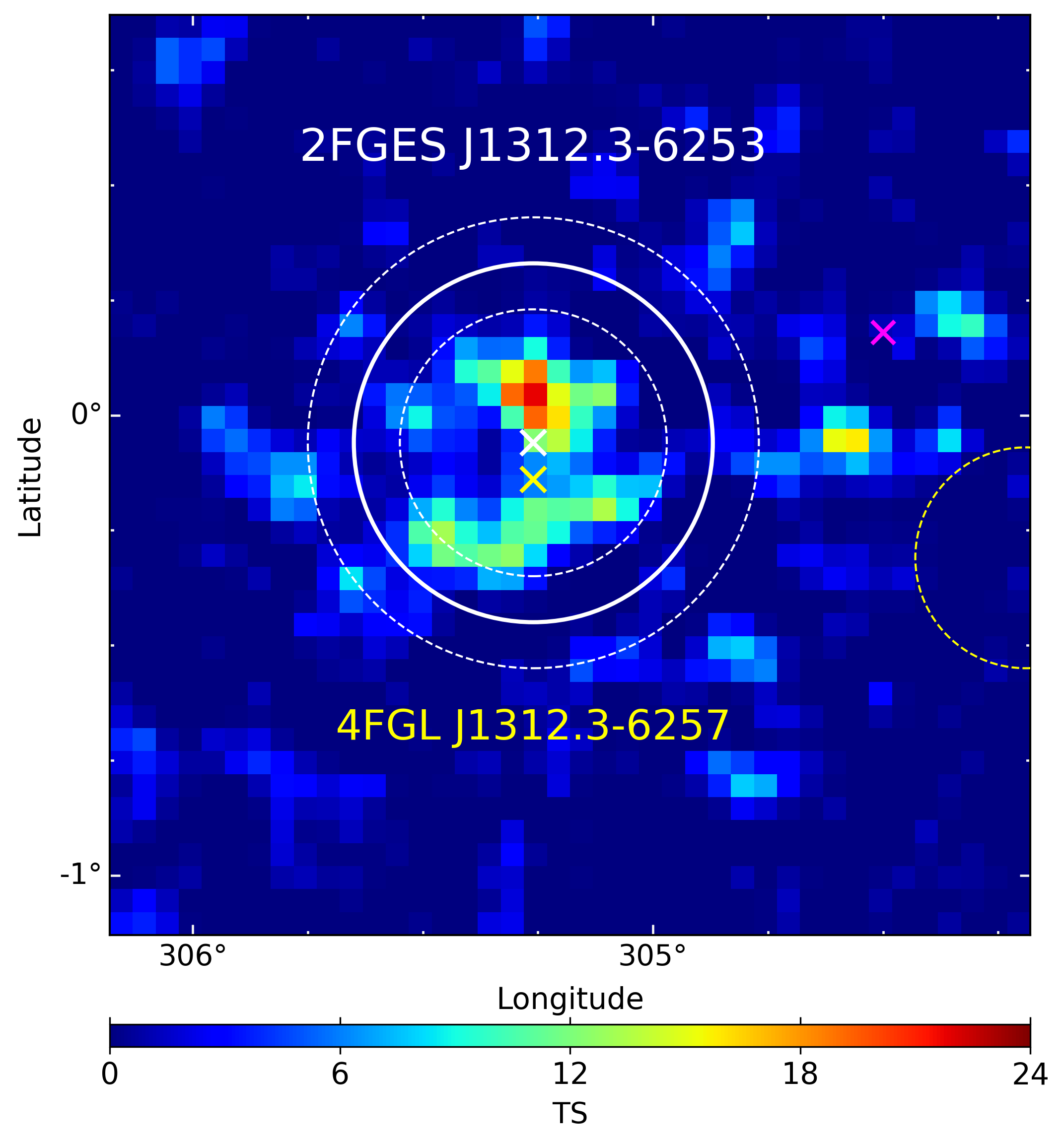}{0.35\textwidth}{}
          \fig{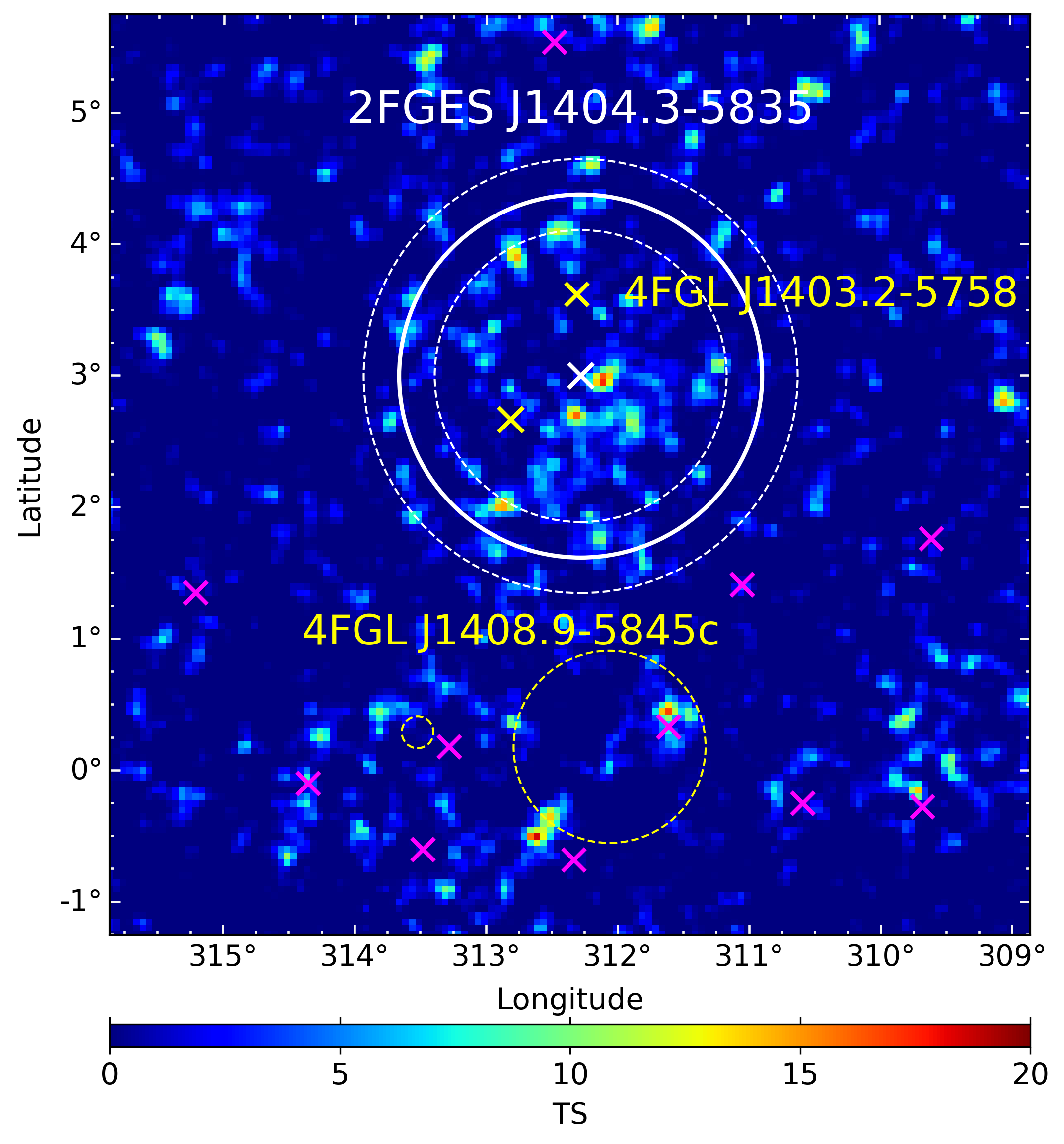}{0.35\textwidth}{}} 
\gridline{ \fig{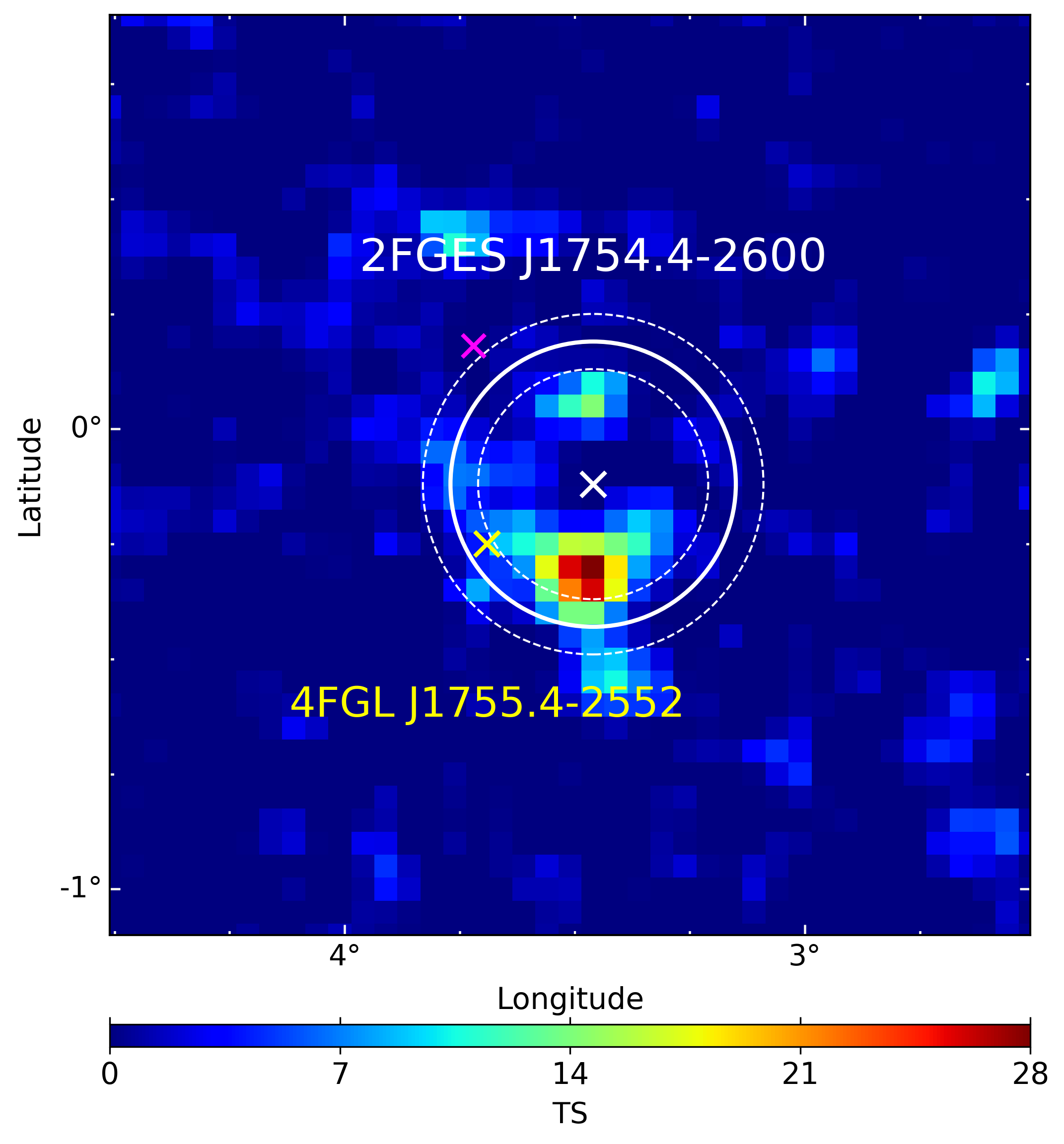}{0.35\textwidth}{}
          \fig{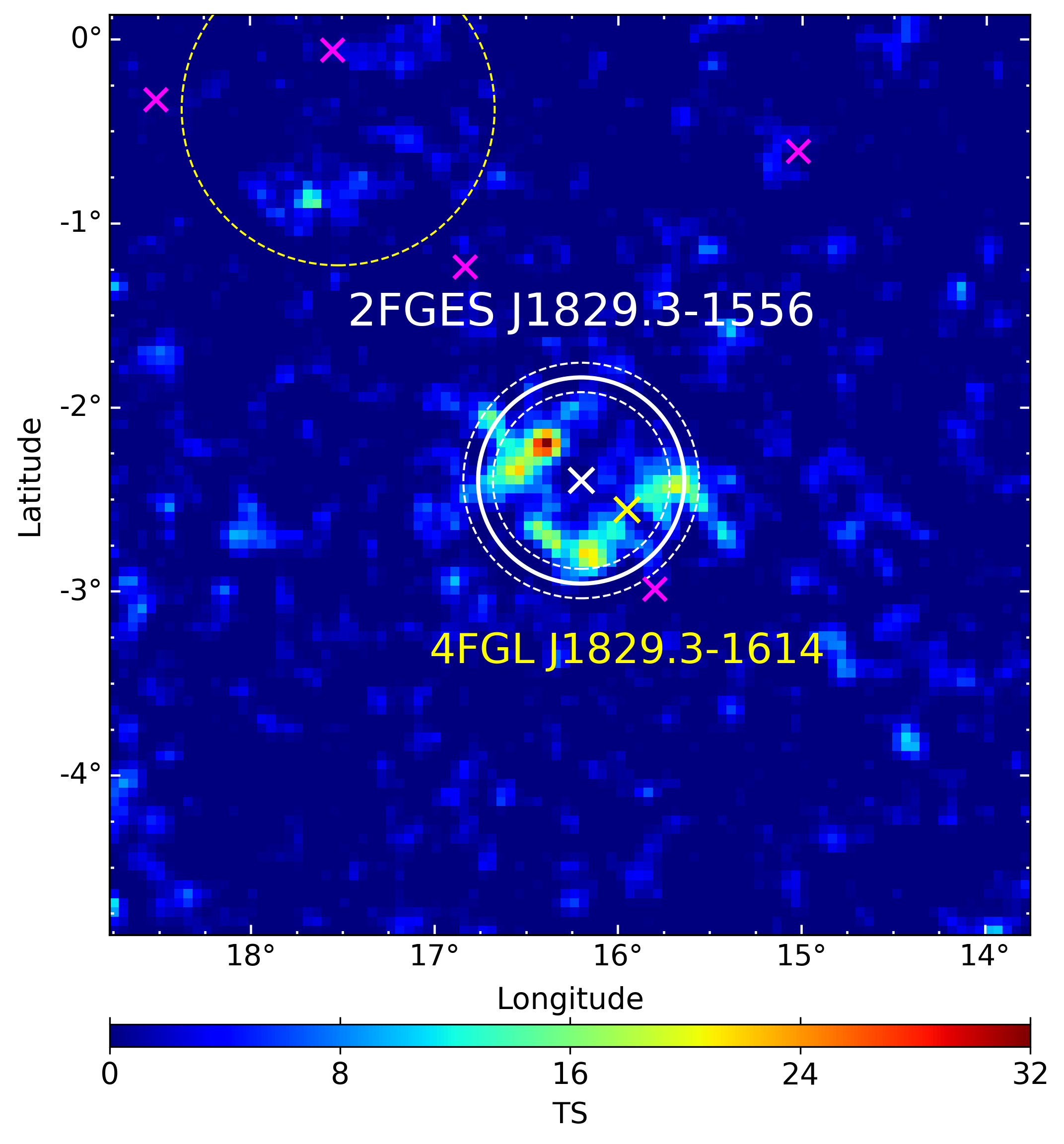}{0.35\textwidth}{}}
\gridline{ \fig{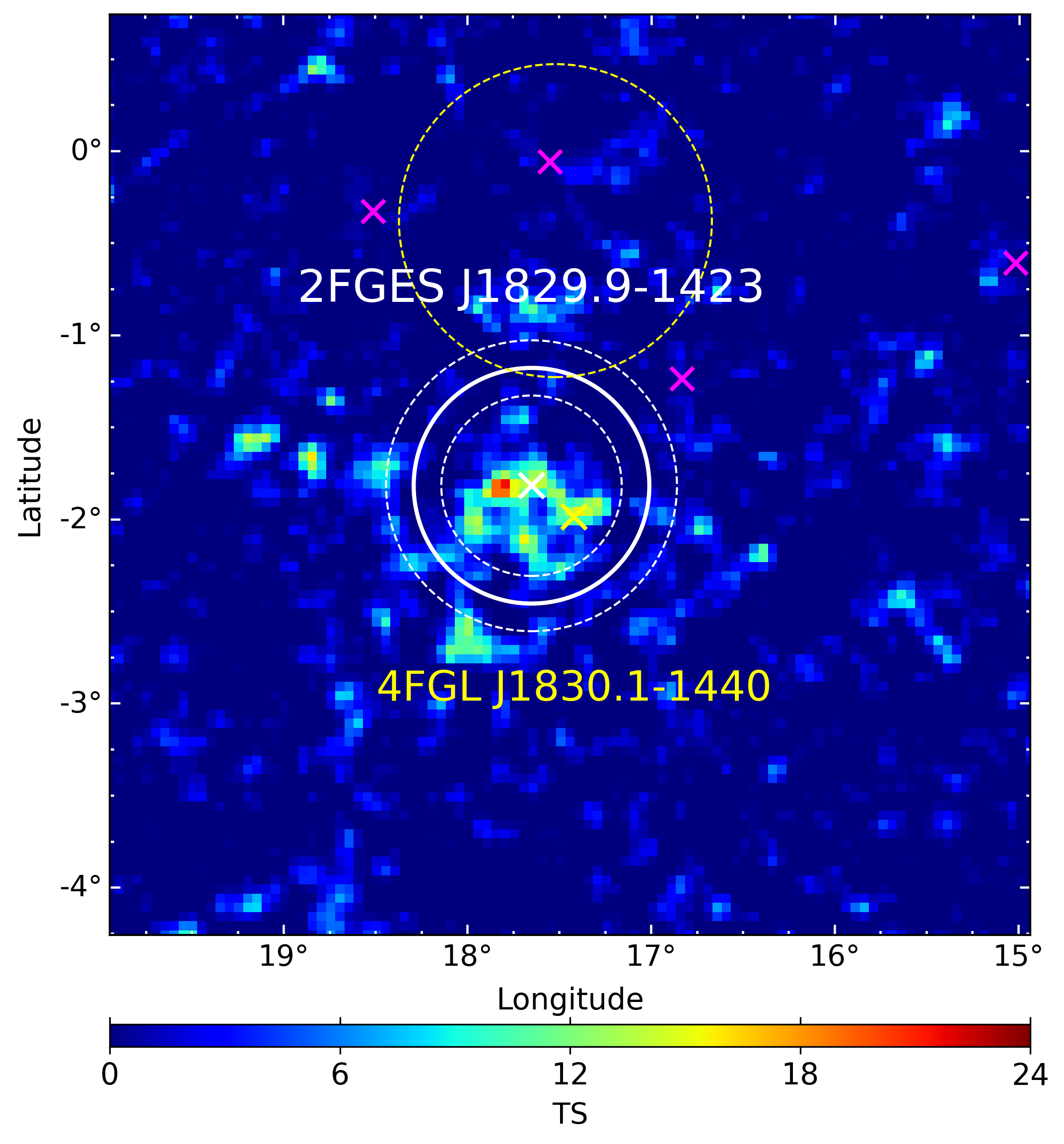}{0.35\textwidth}{}
          \fig{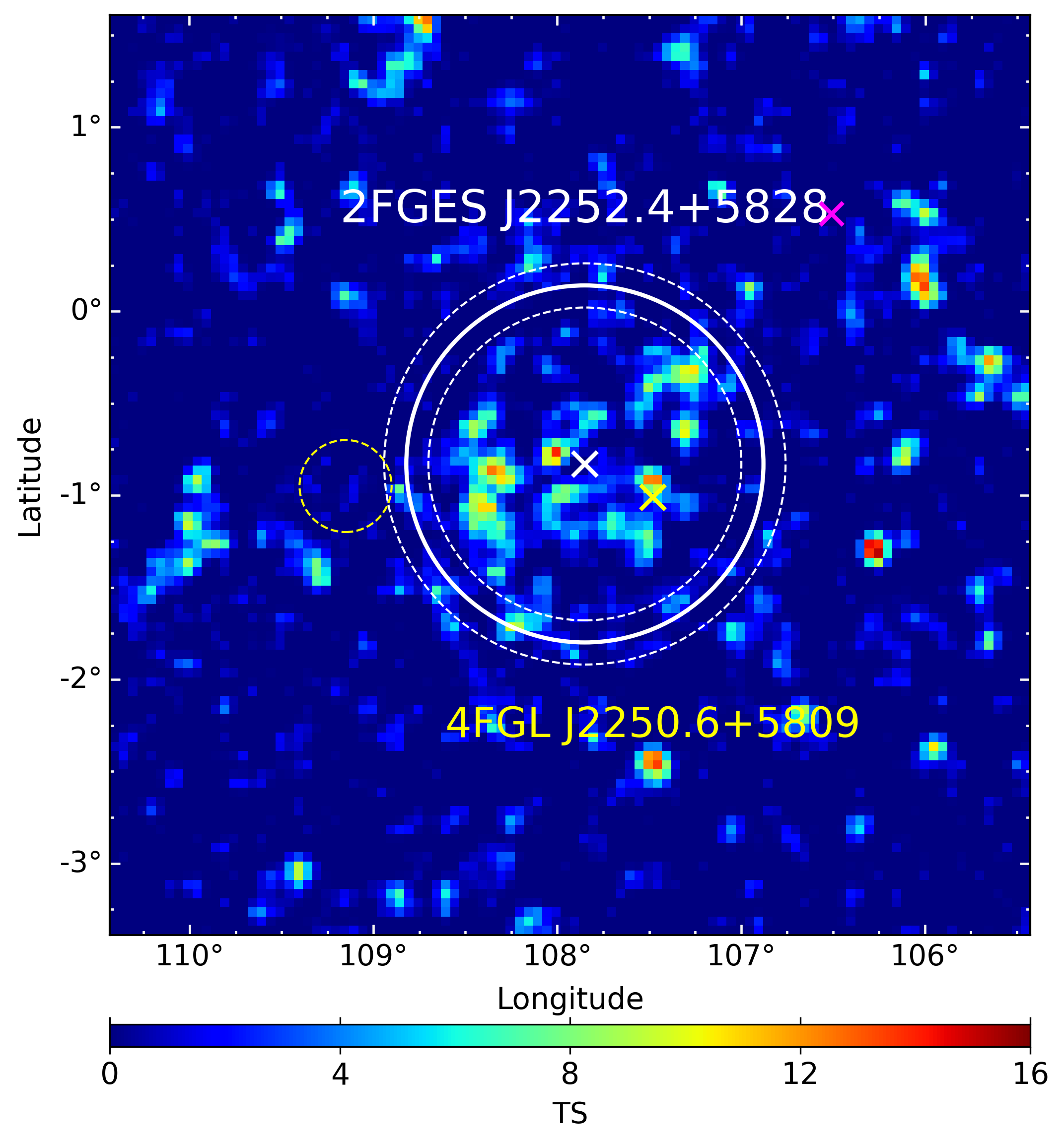}{0.35\textwidth}{}}
\caption{TS maps of 2FGES sources with only a 4FGL point source counterpart.~Graphical elements are defined in the caption of Fig.~\ref{fig:4fgle}.~The 2FGES sources, listed from top-left to bottom-right, are 2FGES J1312.3$-$6253, 2FGES J1404.3$-$5835, 2FGES J1754.4$-$2600, 2FGES J1829.3$-$1556, 2FGES J1829.9$-$1423, and 2FGES J2252.4+5828.~\label{fig:no_tev_psr_assoc}}
\end{figure}

Six 2FGES components could only be associated with 4FGL-DR3 point sources and have no TeV source or bright pulsar counterparts in the scanned catalogs.~They are listed in Table \ref{tab:4fgl_only} and the corresponding TS maps are displayed in Fig.~\ref{fig:no_tev_psr_assoc}.

2FGES J1404.3$-$5835 is significant, with a possible connection to two unassociated GeV sources, 4FGL J1403.2$-$5758 and 4FGL J1408.9$-$5845c.~The latter is just barely closer to the 2FGES source than the former.~The drop in source significance when using the alternative IEM is strong, almost 70\%, but it remains above the detection threshold of 25.~The TS map reveals scattered small-scale excesses over the field, although with a relatively coherent group of hotspots near the centroid, which warrants a dedicated analysis to confirm the reality and extent of the source.

2FGES J1754.4$-$2600 is a compact GeV source with $r_{68}\mathrel{=}0.31^{\circ}\pm0.06^{\circ}$.~The nearest GeV source, 4FGL J1755.4$-$2552, lies 0.26$^{\circ}$ away from the 2FGES source centroid and is marked as an SNR candidate due to the radio detection of the shell-type remnant G003.7$-$00.2 at the angular offset of 0.04$^{\circ}$.~The remnant was below detection threshold in~\citet{Acero16} using three years of LAT data.~Yet, the association of the 2FGES source with the radio SNR can be discarded as the former is offset and much larger.

2FGES J1829.3$-$1556 is a significant emission component with TS = 99.10, and it is associated with 4FGL J1829.3$-$1614, an SPP candidate\footnote{Sources flagged with \enquote{SPP} in capital letters in the 4FGL-DR3 catalog are of unknown nature, with potential associations with either a PWN or an SNR, and those flagged with \enquote{spp} in lowercase letters are SPP candidates.}, which has a potential connection to the shell-type remnant SNR G016.2$-$02.7.~The remnant was below detection threshold in~\citet{Acero16}.~Its association with the 2FGES source is however unlikely because the latter is 3$-$4 times as large, and the emission layout does not match the radio intensity distribution.

2FGES J1829.9$-$1423 is a highly significant extended emission detected near 2FGES J1829.3$-$1556.~The nearby shell-type remnant SNR G017.4$-$02.3 with a radio size of 0.2$^{\circ}$ and angular distance of 0.53$^{\circ}$ is too small and too far away to be a plausible counterpart for the GeV emission.

\begin{deluxetable*}{lccccc}[t]
\tablecaption{2FGES Sources With a 4FGL Point Source Association Only\label{tab:4fgl_only}}
\tablenum{6}
\tablehead{
\multirow{2}{*}{2FGES Name} & \multirow{2}{*}{$r_{68}$ ($^{\circ}$)} & \multirow{2}{*}{$\Gamma$} & \multirow{2}{*}{4FGL Name} & \multirow{2}{*}{4FGL Class} \\ 
 &  &  &  & 
}
\startdata
J1312.3$-$6253 & 0.39 & 2.70 & J1312.3$-$6257 & ... \\
J1404.3$-$5835$^{\dag}$ & 1.38 & 2.24 & J1403.2$-$5758 & ... \\  
J1754.4$-$2600 & 0.31 & 2.00 & J1755.4$-$2552 & snr \\
J1829.3$-$1556 & 0.56 & 1.86 & J1829.3$-$1614 & spp \\
J1829.9$-$1423 & 0.64 & 1.85 & J1830.1$-$1440 & ... \\
J2252.4+5828 & 0.97 & 1.96 & J2250.6+5809 & ... \\
\enddata
\tablecomments{Columns 2 and 3 provide the spatial extension and the photon index of the 2FGES sources, respectively.~The 4FGL association and related classification are listed in columns 4 and 5, respectively.~The source marked with a dagger ($\dag$) is linked to an additional unassociated source, 4FGL J1408.9$-$5845c, which is located at the same angular offset of 0.62$^{\circ}$ from it.} 
\end{deluxetable*}

\subsection{Extended Components Associated to Bright Pulsars} \label{subsec:ecs_associated_bright_psrs}

Our search for associations between 2FGES sources and bright pulsars resulted in six newly detected extended sources, each with an angular separation of $<1^{\circ}$ from a pulsar, as listed in Table~\ref{tab:ext_comps_assoc_bright_psrs} (see the TS maps in Figs.~\ref{fig:psr}$-$\ref{fig:tev_psr}).~These pulsars have $\dot{E}/d^{2}\,\geq\,10^{34}$ erg s$^{-1}$ kpc$^{-2}$, characteristic ages of $\sim$10--170 kyr, and are located up to 7\,kpc from Earth.~Among the 2FGES sources possibly associated with bright pulsars, four are also coincident with TeV sources, of which two are yet unidentified and two are PWNe.~This group of six sources includes three 2FGES sources where the associated 4FGL sources and the bright pulsars correspond to the same objects.

2FGES J0006.0+7319, detected around an energetic $\gamma$-ray pulsar, shows a hard GeV spectrum and is potentially linked to the recently discovered TeV source 1LHAASO J0007+7303u, identified as a PWN in TeVCat.\footnote{See http://tevcat.uchicago.edu and~\citet{Wakely.Horan08}.}~Similarly, 2FGES J1417.5$-$6057 also exhibits a hard spectrum and is positionally in perfect agreement with the unassociated GeV source 4FGL J1417.7$-$6057 and PWN HESS J1418$-$609 (also known as the Rabbit nebula).~Notably, two of the associated pulsars in this list (J0007+7303 and J1913+1011) have recently been introduced as TeV halo candidates due to their associations with 3HWC and/or 1LHAASO sources~\citep{Albert20,Cao24}.~This subset of 2FGES sources is therefore of particular interest to us as potential GeV pulsar halo candidates, with further discussion provided in Section~\ref{sec:Discussion}.

We note that the TS map for 2FGES J1101.7$-$6110 reveals several small-scale hotspots scattered over nearly 2$^\circ$, which does not a priori support its association with a relatively distant pulsar located at 7\,kpc from us (this will be discussed more quantitatively in Section~\ref{sec:Discussion}).

\begin{longrotatetable}
\tabletypesize{\footnotesize}
\begin{deluxetable}{lcccclccc}
\tablecaption{2FGES Sources With a Bright Pulsar Association\label{tab:ext_comps_assoc_bright_psrs}}
\tablenum{7}
\tablehead{[-1.ex]
\colhead{2FGES Name} & \colhead{PSR Name} & \colhead{$d$} & \colhead{$\tau_{\rm{c}}$} & \colhead{$\dot E$} & \colhead{4FGL Name (Class)} & \colhead{TeV Name (Class)} & \colhead{$r_{68}$} & \colhead{$\Gamma$} \\
\colhead{} & \colhead{} & \colhead{(kpc)} & \colhead{(kyr)} & \colhead{(erg~s$^{-1}$)} & \colhead{} & \colhead{} & \colhead{($^{\circ}$)} & \colhead{} 
}
\startdata
J0006.0+7319 & J0007+7303 (0.28) & 1.4 & 13.90 & 4.51e+35 & J0007.0+7303 (PSR) & 1LHAASO J0007+7303u (PWN) & 0.85 & 1.85 \\ 
J1101.7$-$6110 & J1101$-$6101 (0.15) & 7.0 & 116.00 & 1.36e+36 & J1102.0$-$6054 (bcu) & ... & 0.98 & 2.79 \\
J1112.6$-$6059 & J1112$-$6103 (0.07) & 4.5 & 32.70 & 4.53e+36 & J1112.2$-$6055 (...) & ... & 0.39 & 1.95 \\
J1417.5$-$6057 & J1418$-$6058 (0.15) & 1.9 & 10.30 & 4.95e+36 & J1417.7$-$6057 (...) & HESS J1418$-$609 (PWN) & 0.17 & 1.68 \\
J1647.6$-$4551 & J1648$-$4611 (0.34) & 4.5 & 110.00 & 2.09e+35 & J1648.4$-$4611 (PSR) & HESS J1646$-$458 (Unid) & 0.54 & 2.15 \\
J1913.5+1024 & J1913+1011 (0.22) & 4.6 & 169.00 & 2.87e+36 & J1913.3+1019 (PSR) & HESS J1912+101, 3HWC J1912+103, 1LHAASO J1912+1014u (Unid) & 0.77 & 2.27 \\ 
\enddata
\tablecomments{Column 2 lists the bright pulsars associated with the 2FGES sources, along with the angular separations between their centroids shown in parentheses.~The properties of the bright pulsars, including distance, characteristic age, and spin-down luminosity, are given in columns 3--5, respectively.~The potential GeV and TeV associations, along with their classifications in parentheses, are given in columns 6 and 7.~The last two columns provide the spatial extension and spectral index of the 2FGES sources.}
\end{deluxetable}
\end{longrotatetable}

\subsection{Extended Components Associated to TeV Sources} \label{subsec:ecs_associated_tev_srcs}

\begin{figure}[!pt]
 \gridline{ \fig{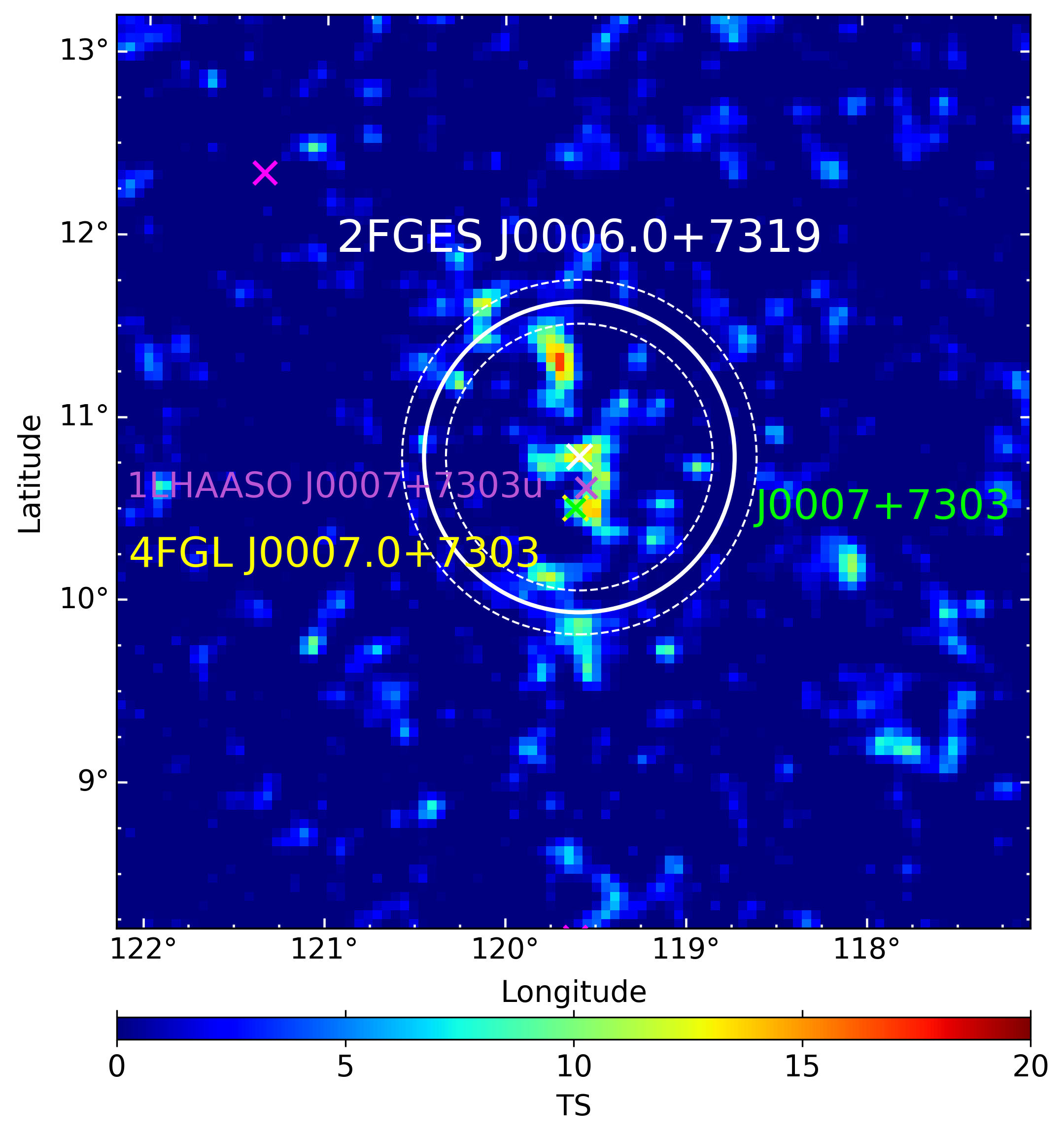}{0.35\textwidth}{}
          \fig{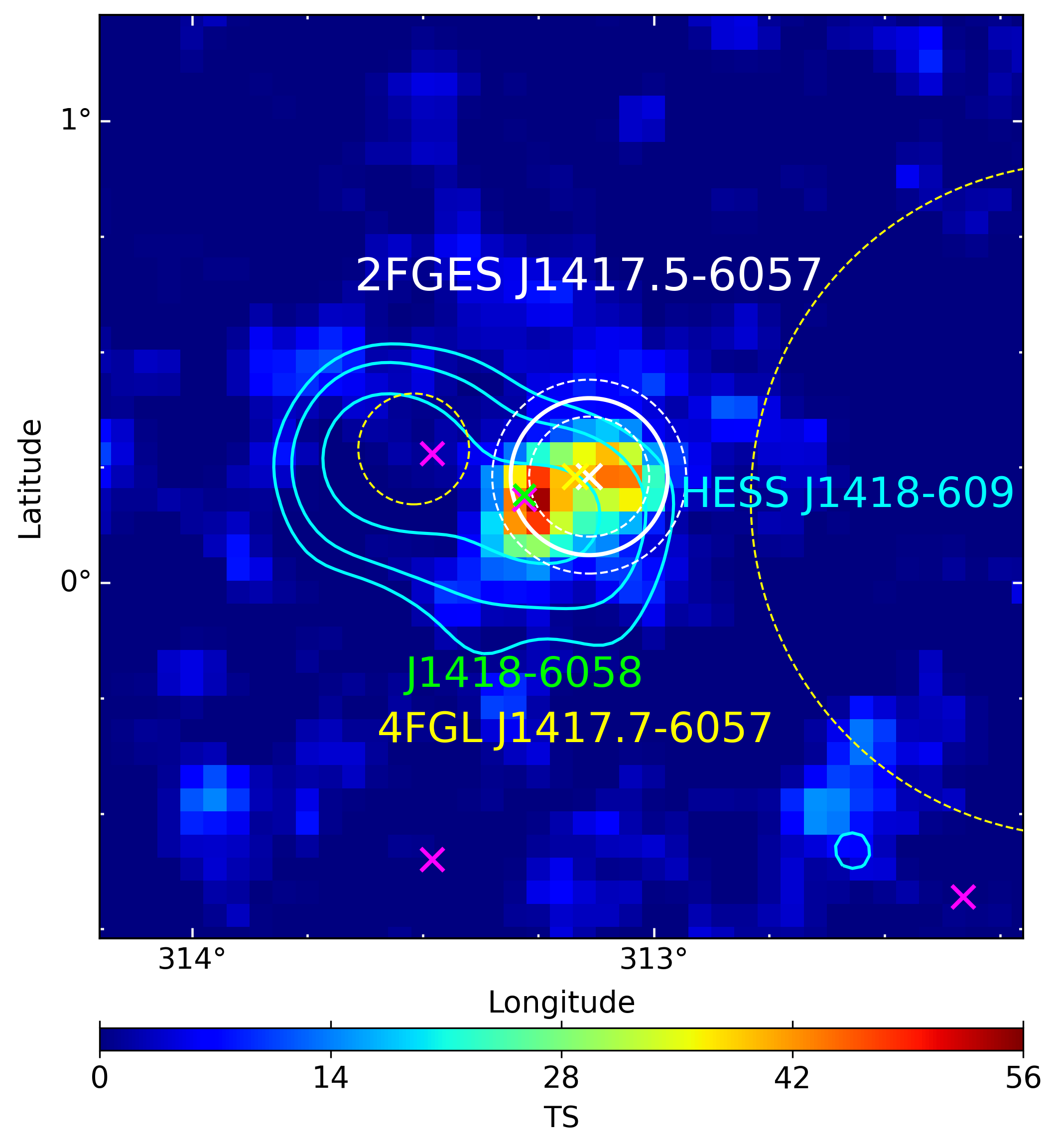}{0.35\textwidth}{}} 
 \gridline{ \fig{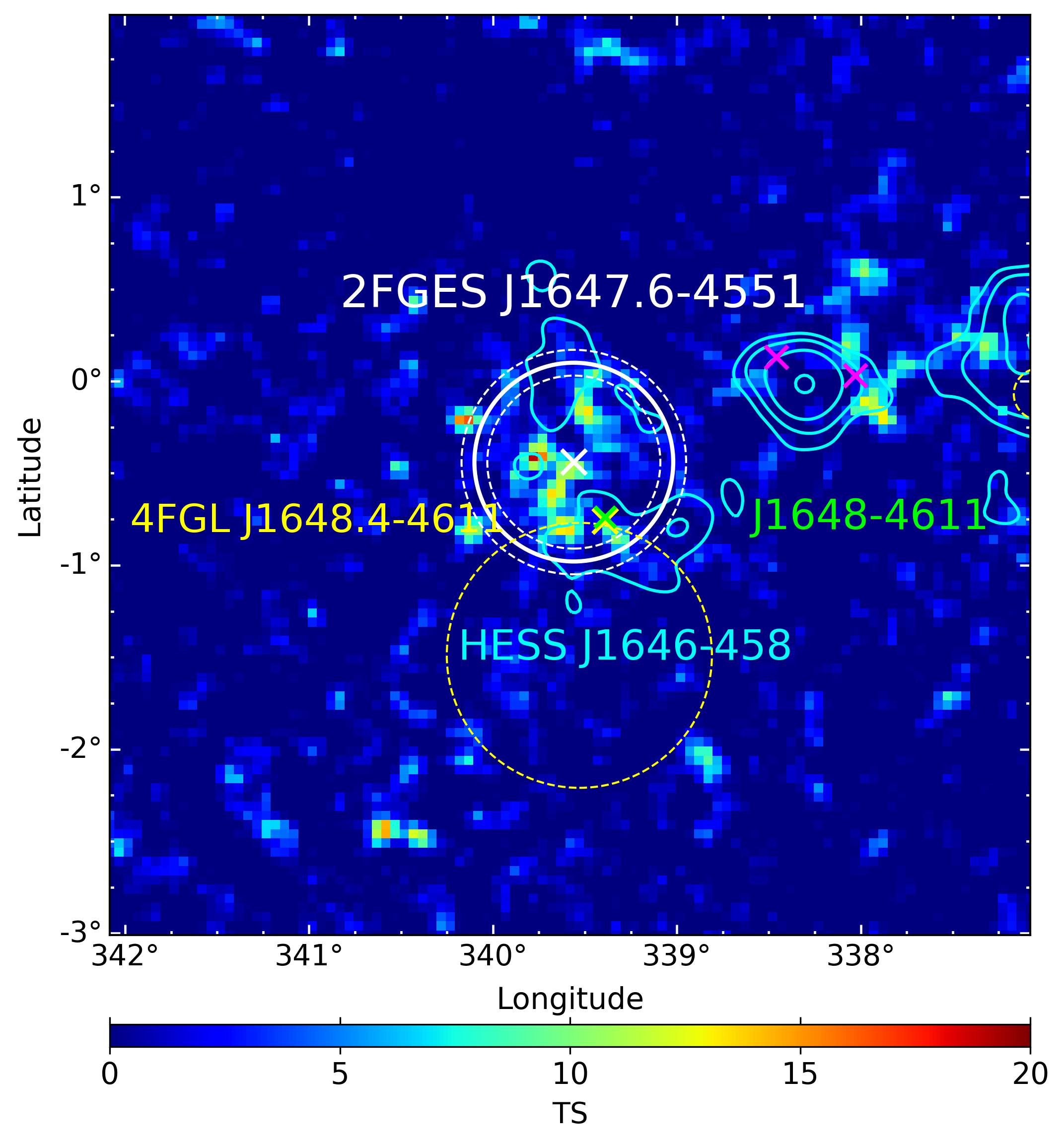}{0.35\textwidth}{}
          \fig{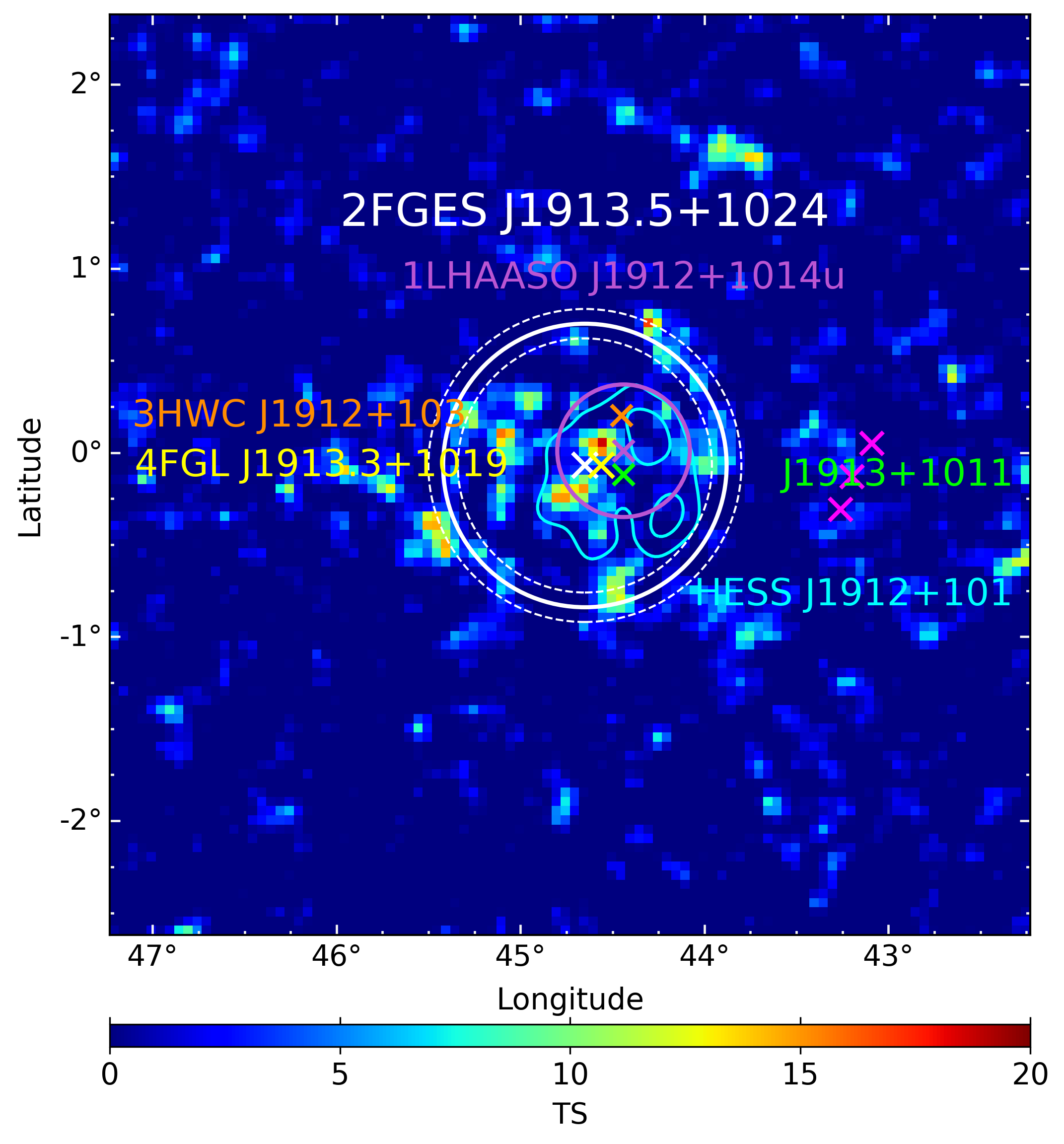}{0.35\textwidth}{}}
\caption{TS maps of 2FGES sources associated with both a TeV source and a bright pulsar.~Graphical elements are defined in the caption of Fig.~\ref{fig:4fgle}.~The 2FGES sources, shown from top-left to bottom-right, are 2FGES J0006.0+7319, 2FGES J1417.5$-$6057, 2FGES J1647.6$-$4551, and 2FGES J1913.5+1024.~\label{fig:tev_psr}}
\end{figure}

\begin{figure}[!pt]
 \gridline{ \fig{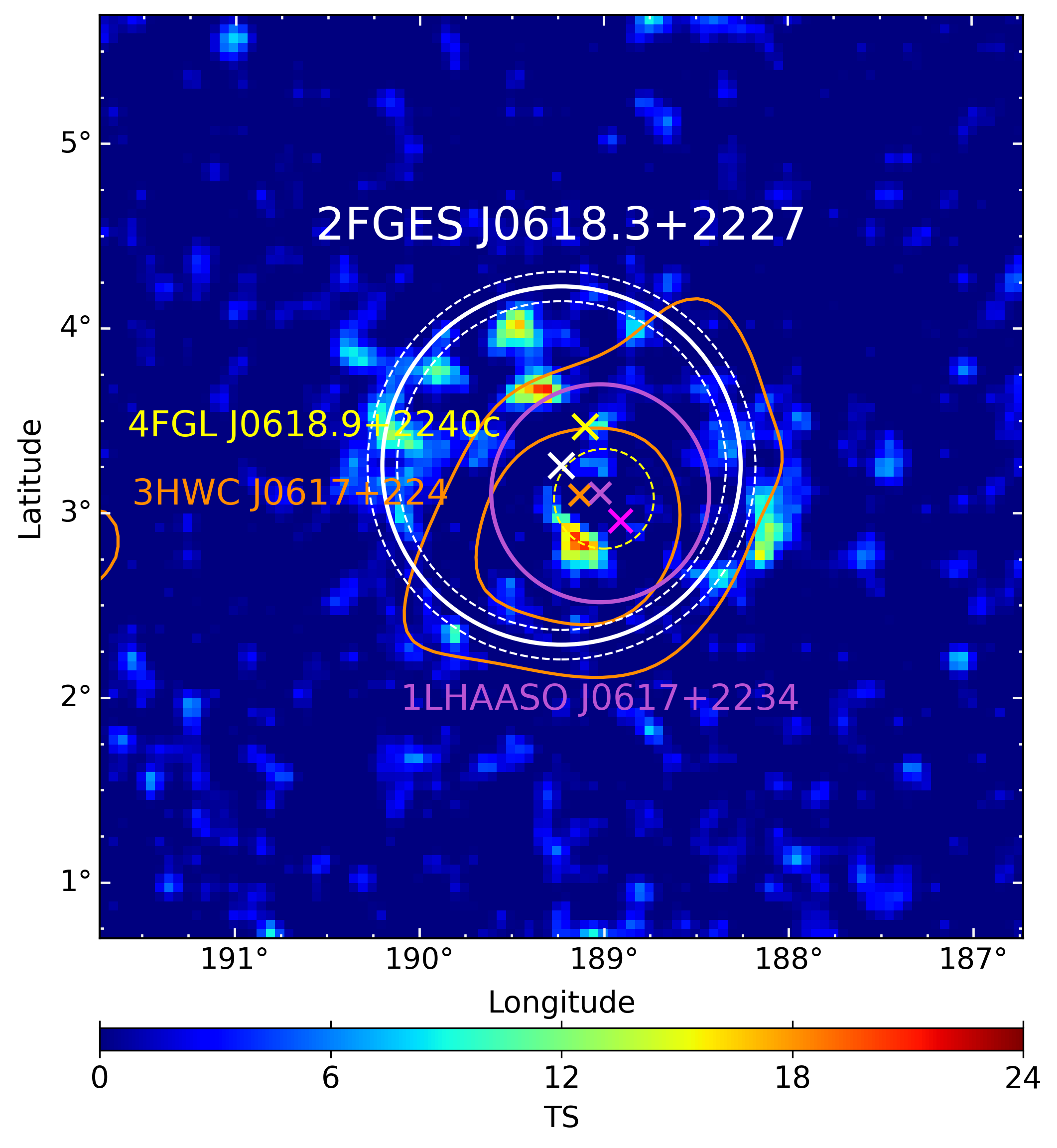}{0.35\textwidth}{}
          \fig{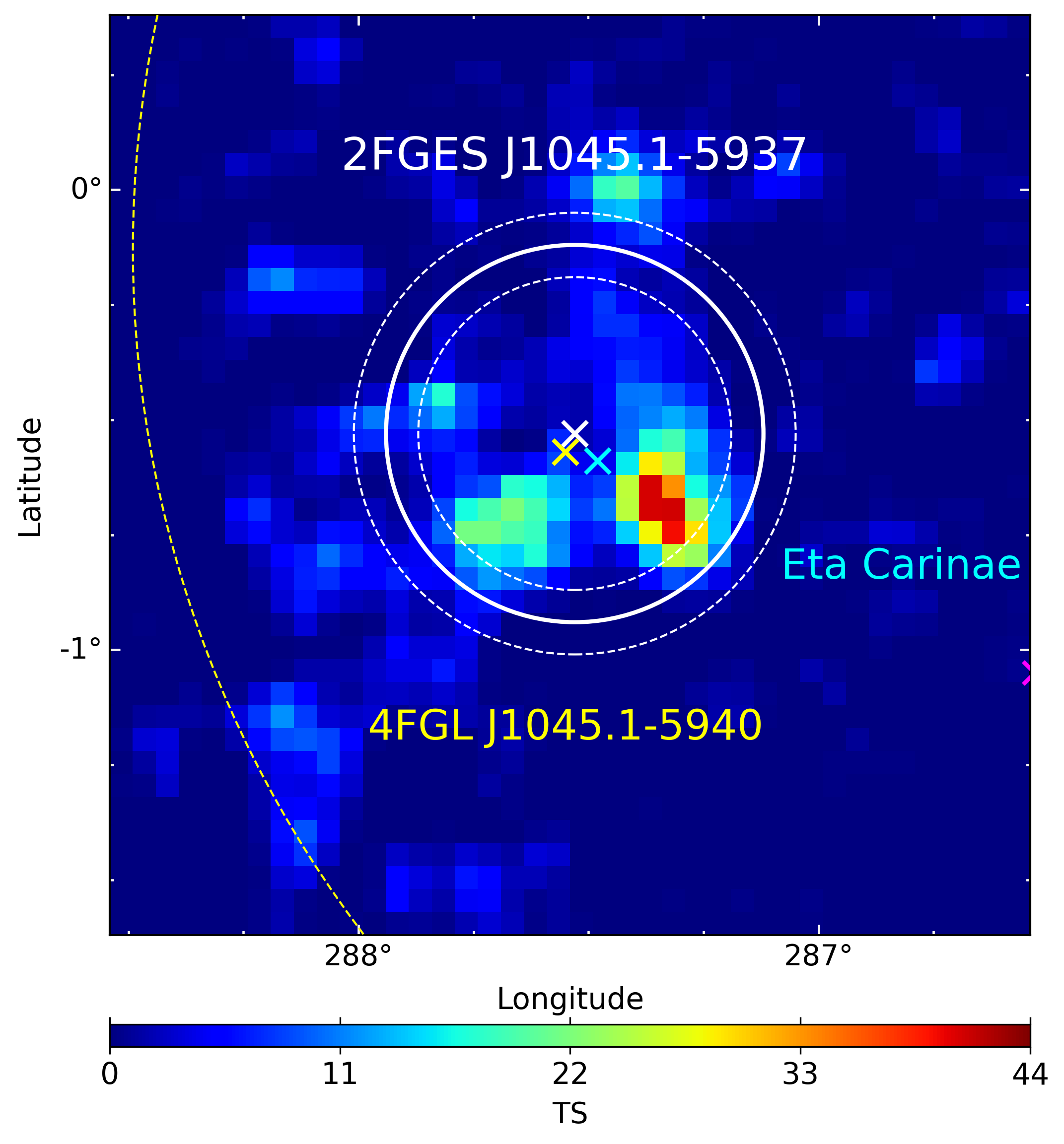}{0.35\textwidth}{}}
 \gridline{ \fig{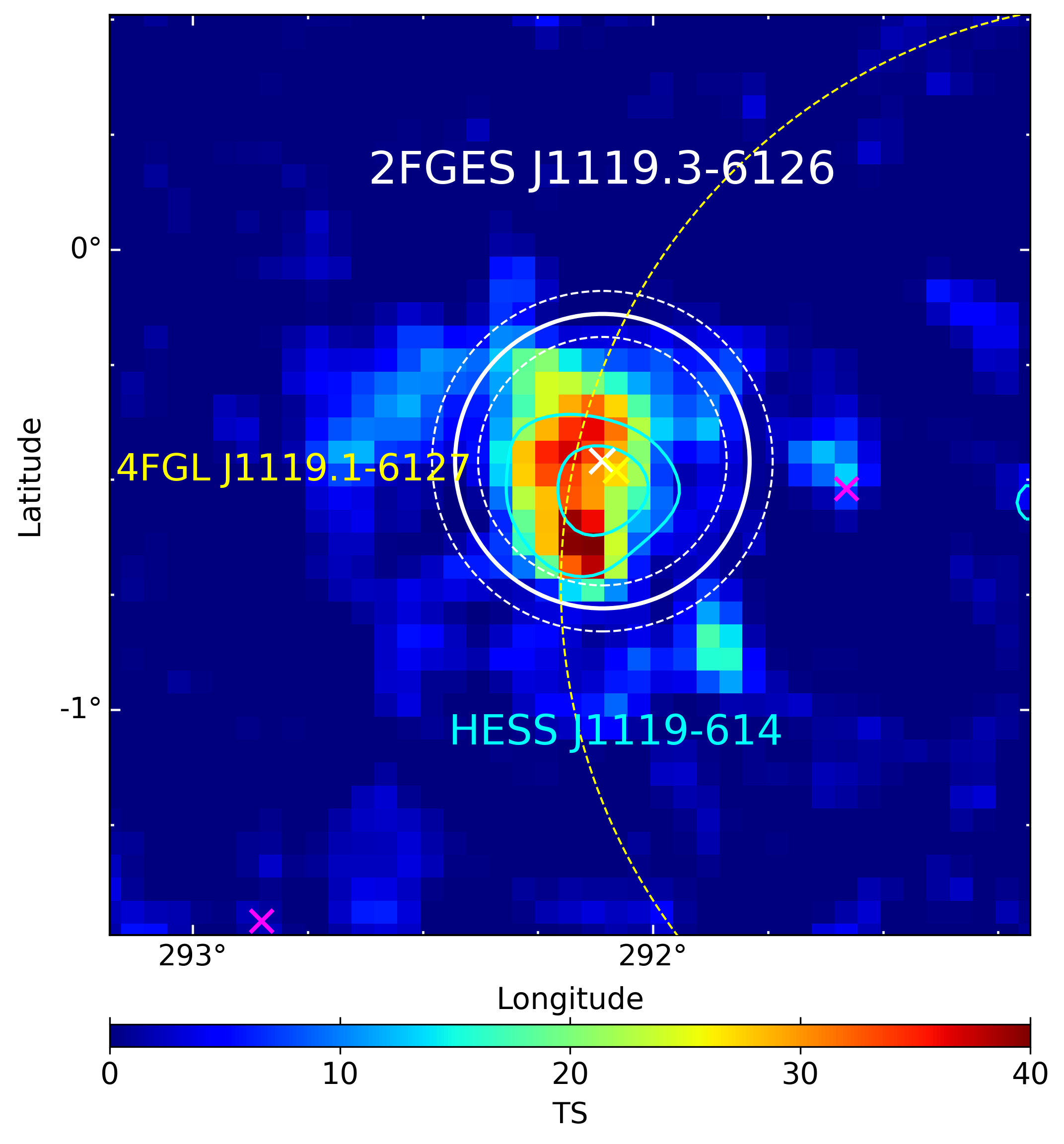}{0.35\textwidth}{}
          \fig{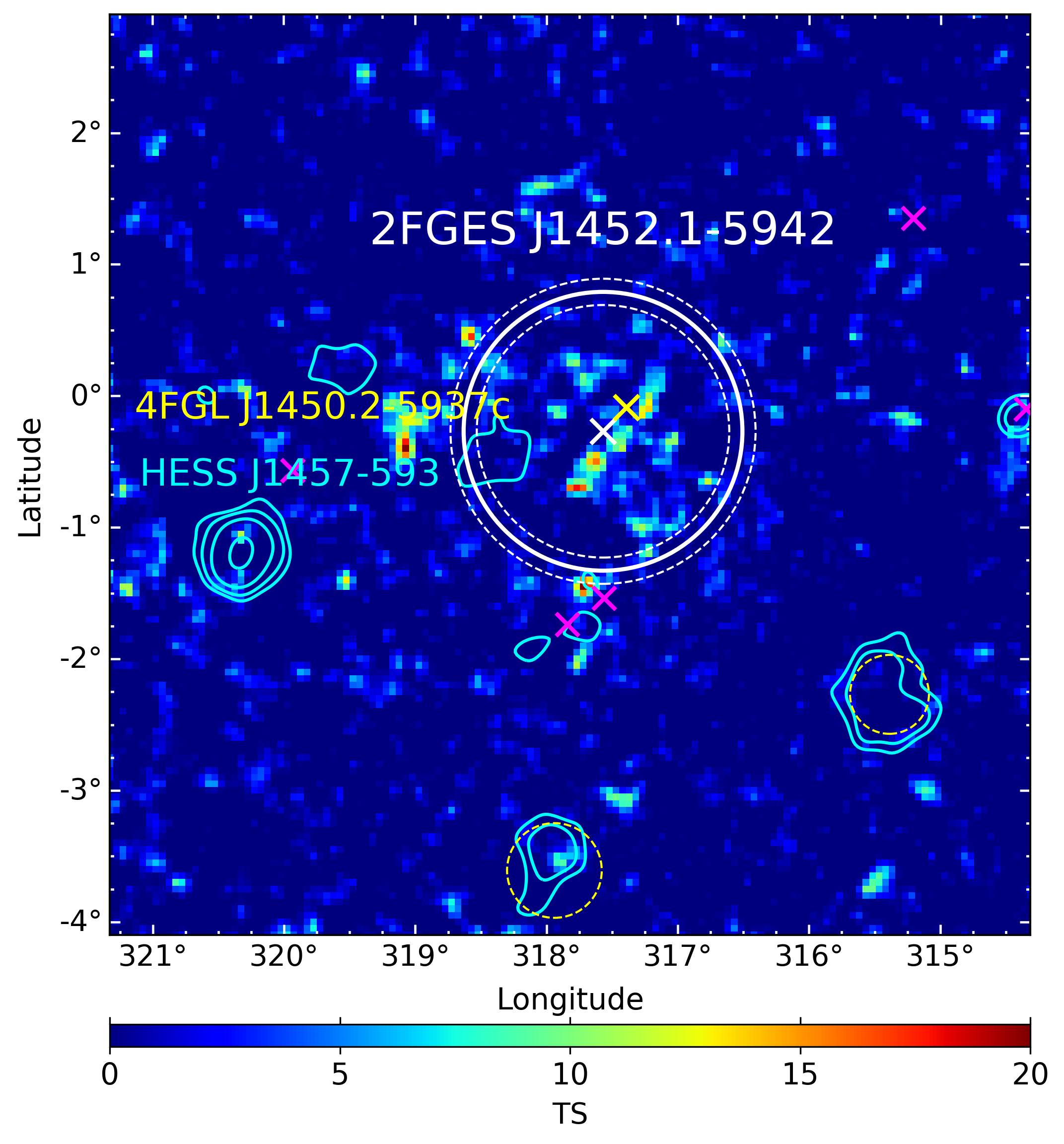}{0.35\textwidth}{}} 
\gridline{ \fig{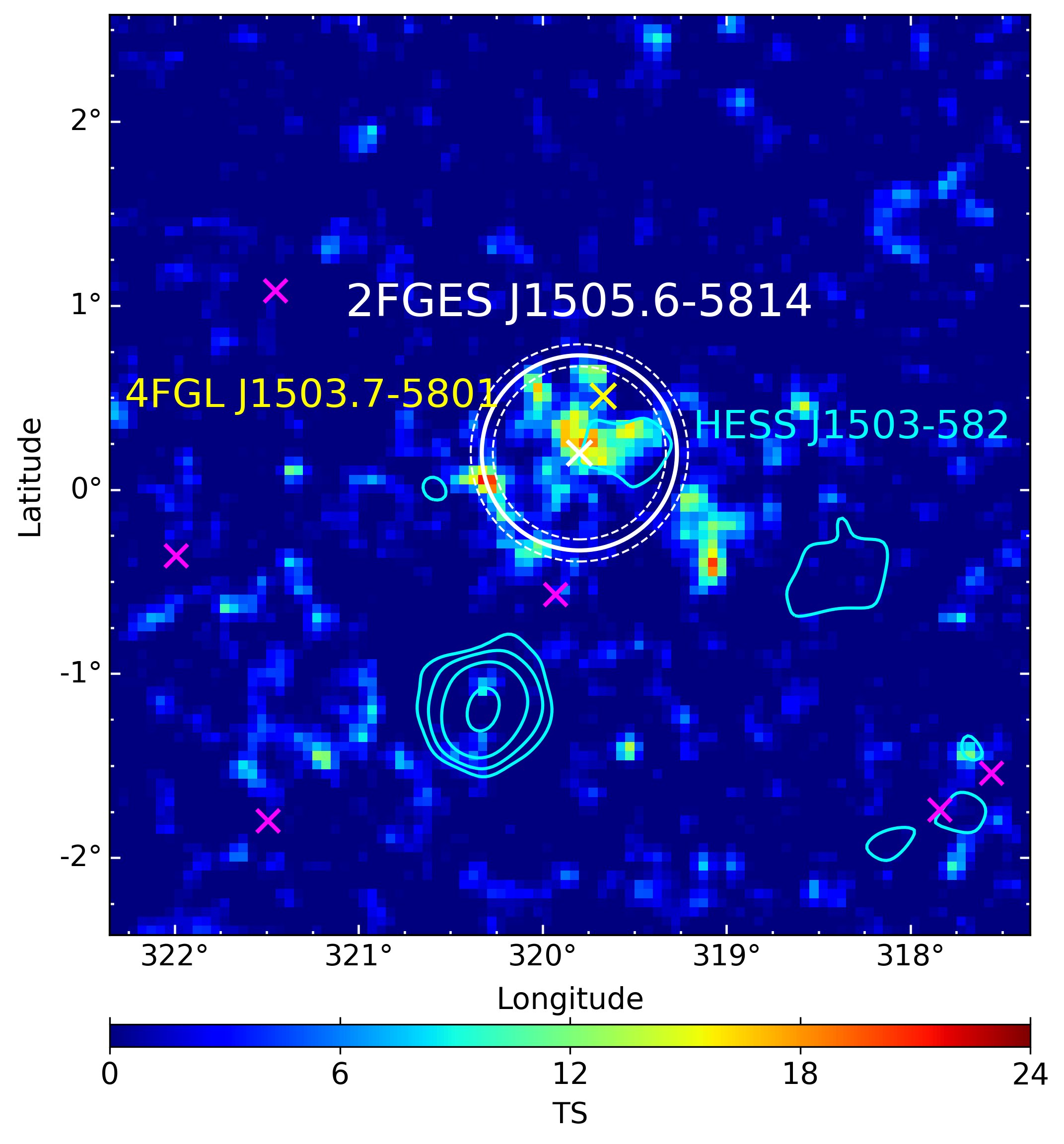}{0.35\textwidth}{}
          \fig{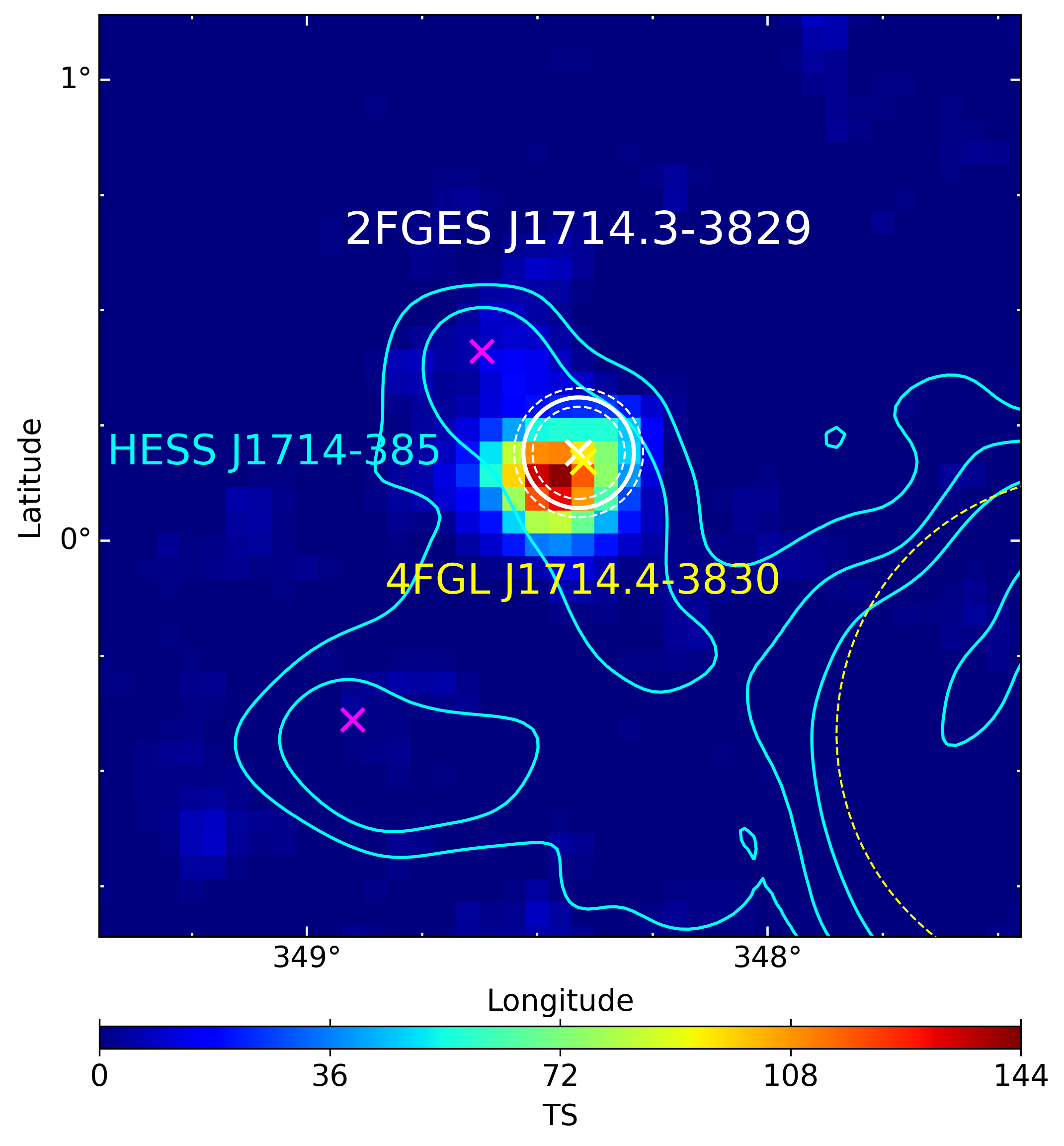}{0.35\textwidth}{}} 
\caption{TS maps of 2FGES sources associated with a TeV source but not with any bright pulsar.~Orange contours delineate the HAWC detection significance at 3, 5, 10, 30, and 50$\sigma$.~Other graphical elements are defined in the caption of Fig.~\ref{fig:4fgle}.~The 2FGES sources, shown from top-left to bottom-right, are 2FGES J0618.3+2227, 2FGES J1045.1$-$5937, 2FGES J1119.3$-$6126, 2FGES J1452.1$-$5942, 2FGES J1505.6$-$5814, and 2FGES J1714.3$-$3829.~\label{fig:tev}}
\end{figure}

\begin{figure}[!pt]
\gridline{ \fig{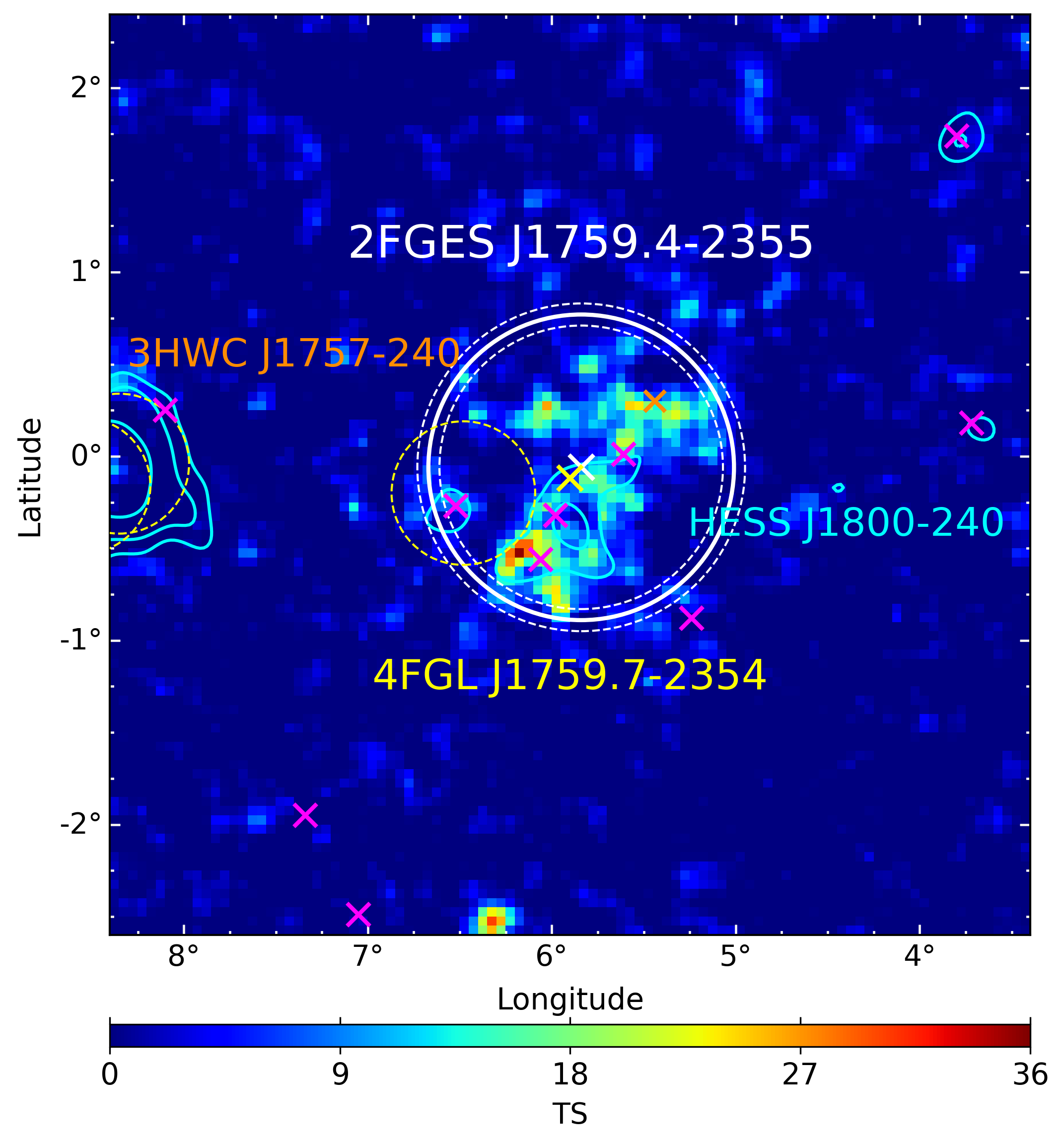}{0.35\textwidth}{}
          \fig{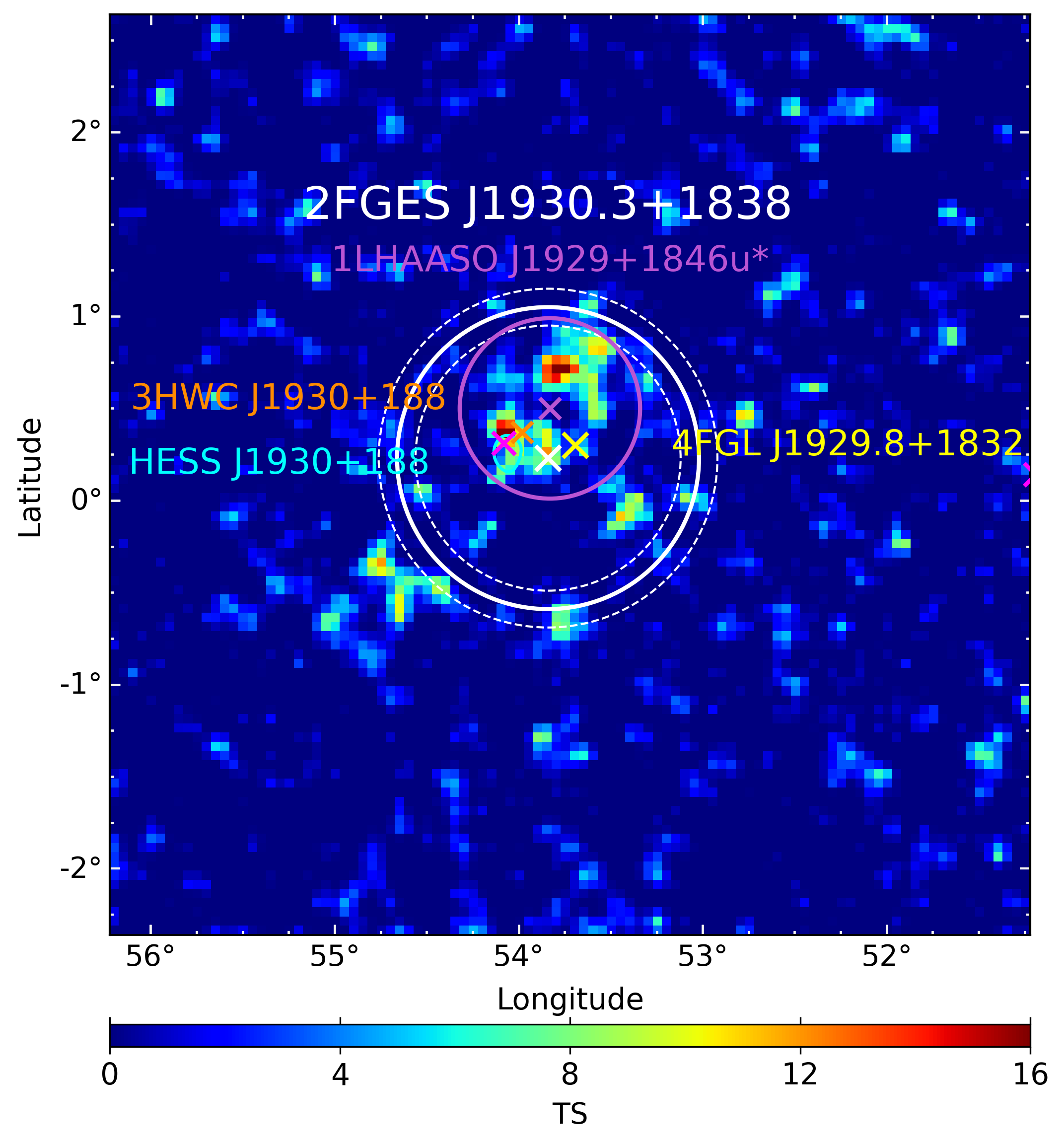}{0.35\textwidth}{}} 
 \gridline{ \fig{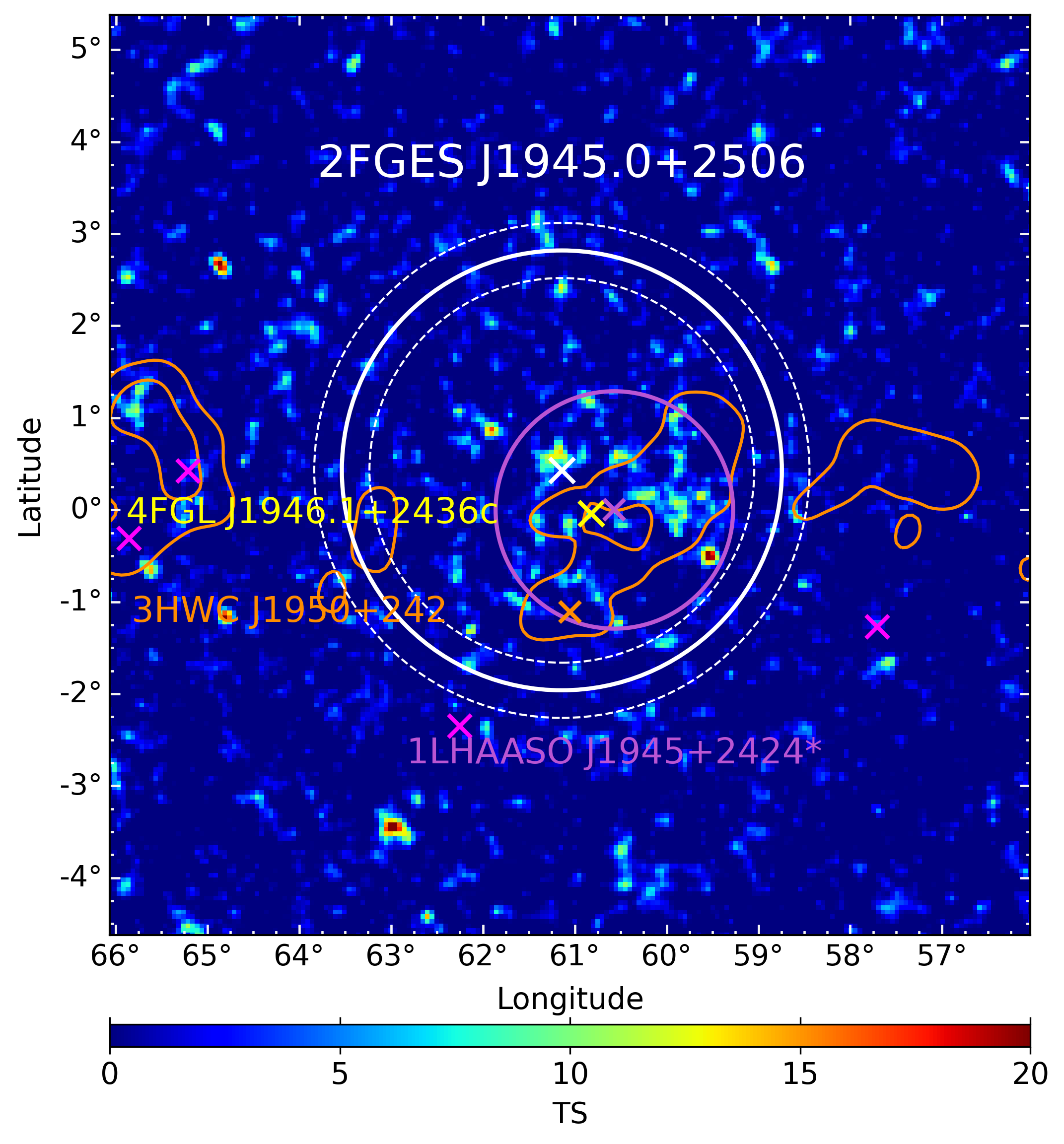}{0.35\textwidth}{}}
\caption{TS maps of 2FGES sources associated with a TeV source but not with any bright pulsar, as described in Fig.~\ref{fig:tev}.~The 2FGES sources, shown from top-left to bottom-right, are 2FGES J1759.4$-$2355, 2FGES J1930.3+1838, and 2FGES J1945.0+2506.~\label{fig:tev_part2}}
\end{figure}

Table~\ref{tab:tev_assoc_ext_srcs} lists the 2FGES sources that are associated with TeV sources, totaling 13 out of 40 objects.~The corresponding TS maps are displayed in Figs.~\ref{fig:tev_psr}$-$\ref{fig:tev_part2}.~Six of the possible TeV counterparts are unidentified objects, and most 2FGES sources are new GeV detections, which may help pinpoint the nature of these sources.\footnote{Some of the 2FGES sources discussed in this section were previously reported, although maybe not with the exact same properties.~For instance, 2FGES J1119.3$-$6126, associated with pulsar 4FGL J1119.1$-$6127 and TeV source HESS J1119$-$614 shares similarities with 4FGL J1119.0$-$6127e included in the fourth release of the 4FGL catalog~(4FGL-DR4;~\citealt{Ballet23}), which was made public during the preparation of this paper.}

\begin{figure}[!pt]
\gridline{\fig{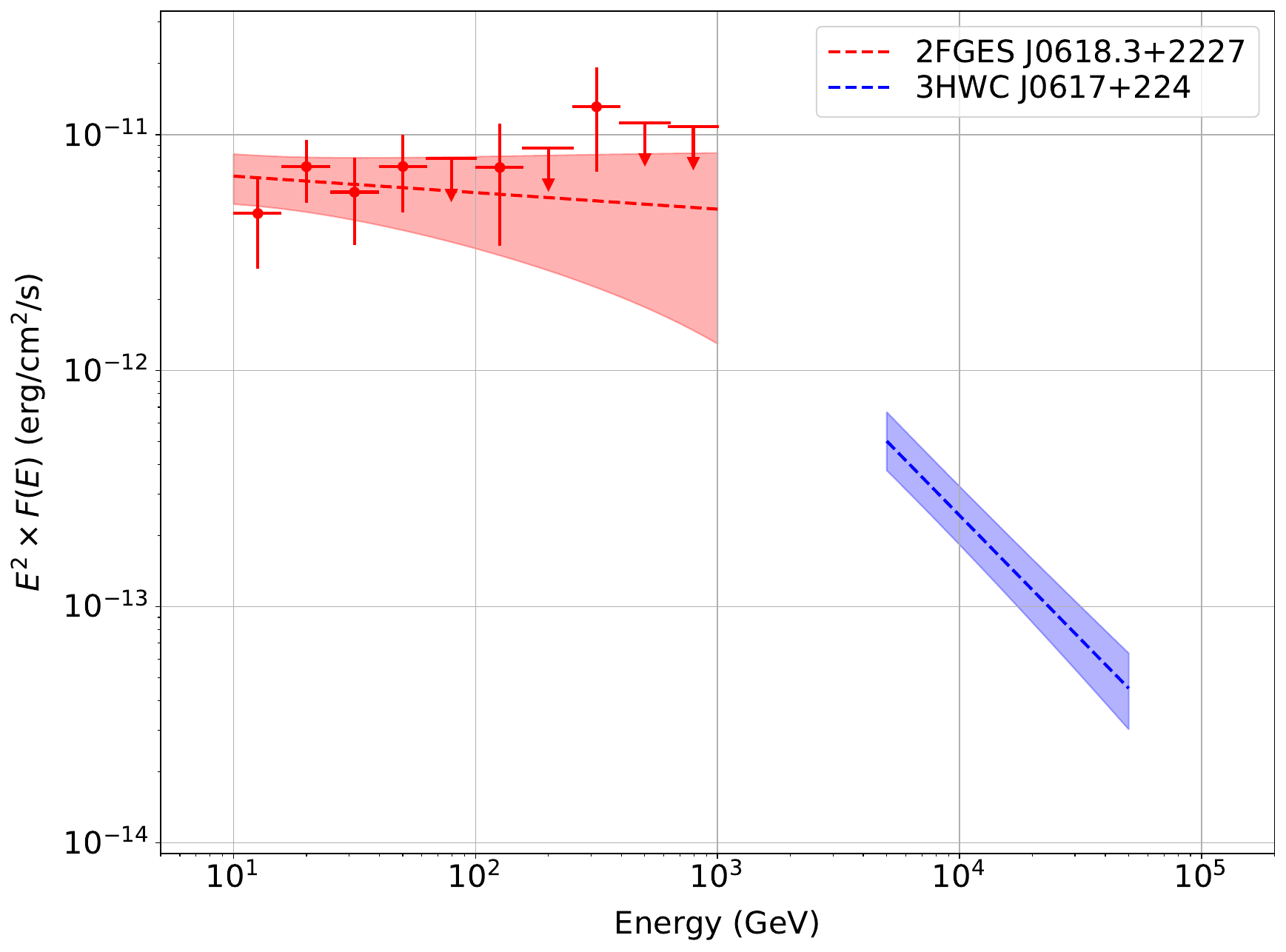}{0.45\textwidth}{}
          \fig{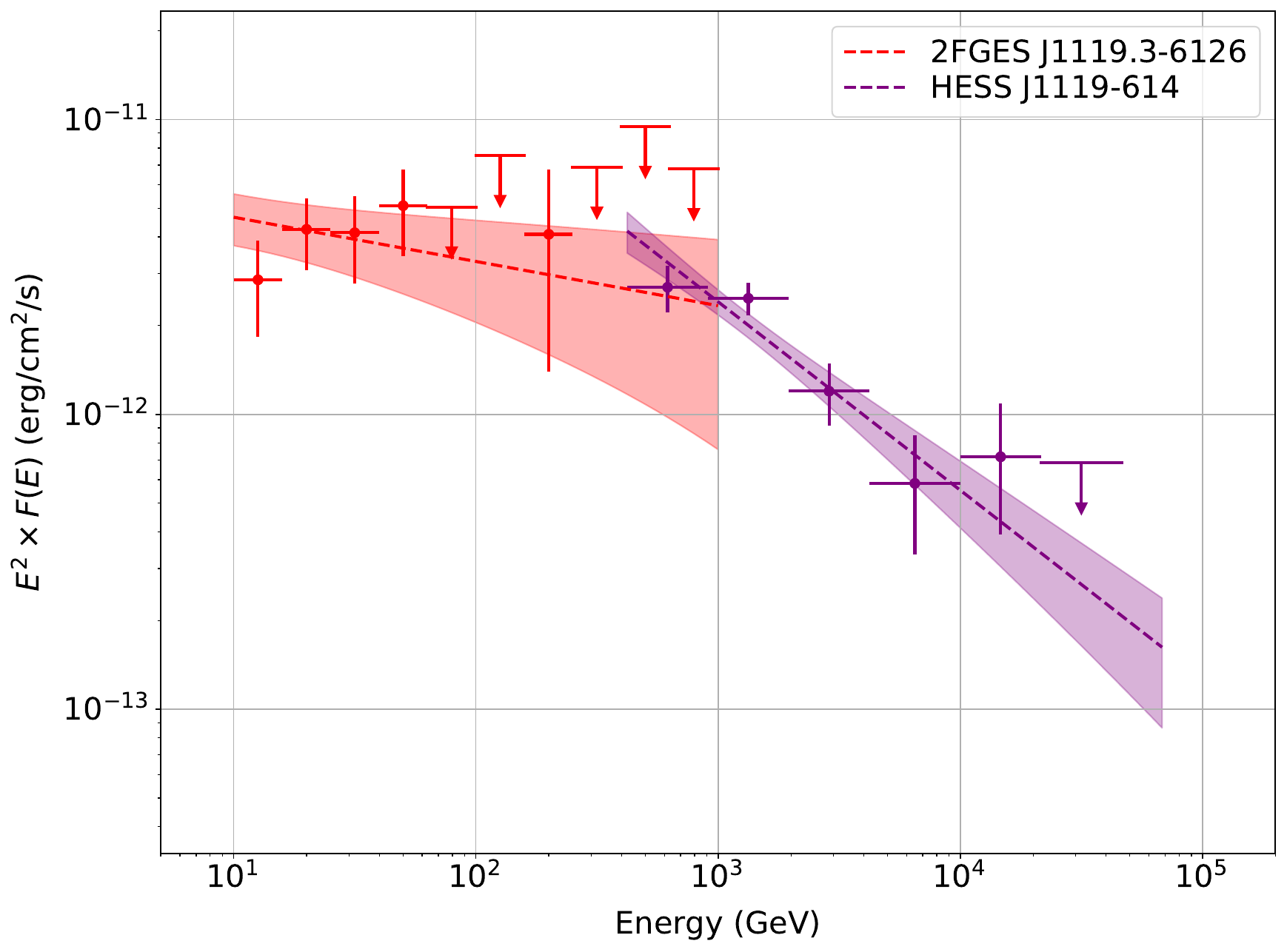}{0.45\textwidth}{}} 
\gridline{ \fig{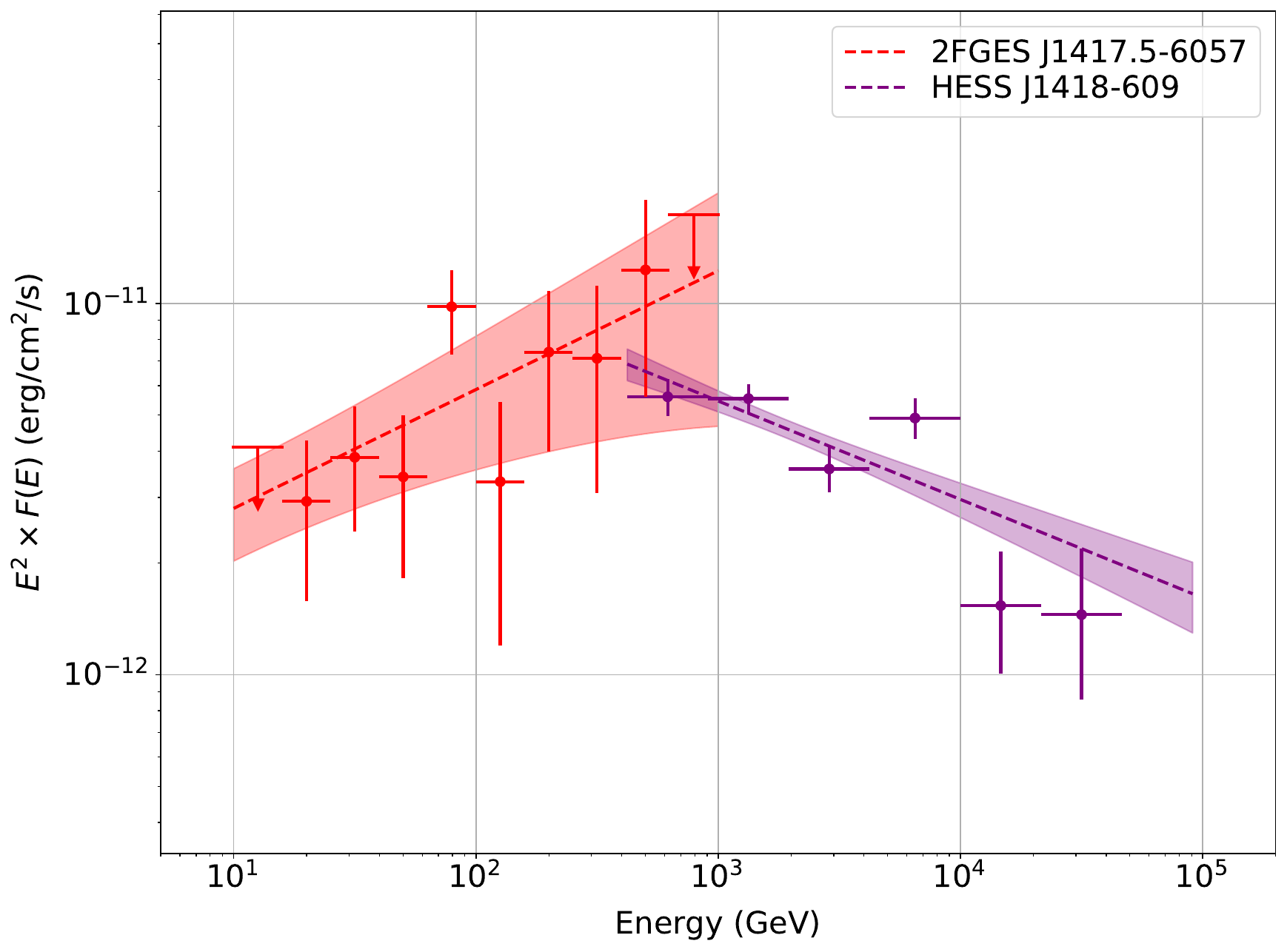}{0.45\textwidth}{}
          \fig{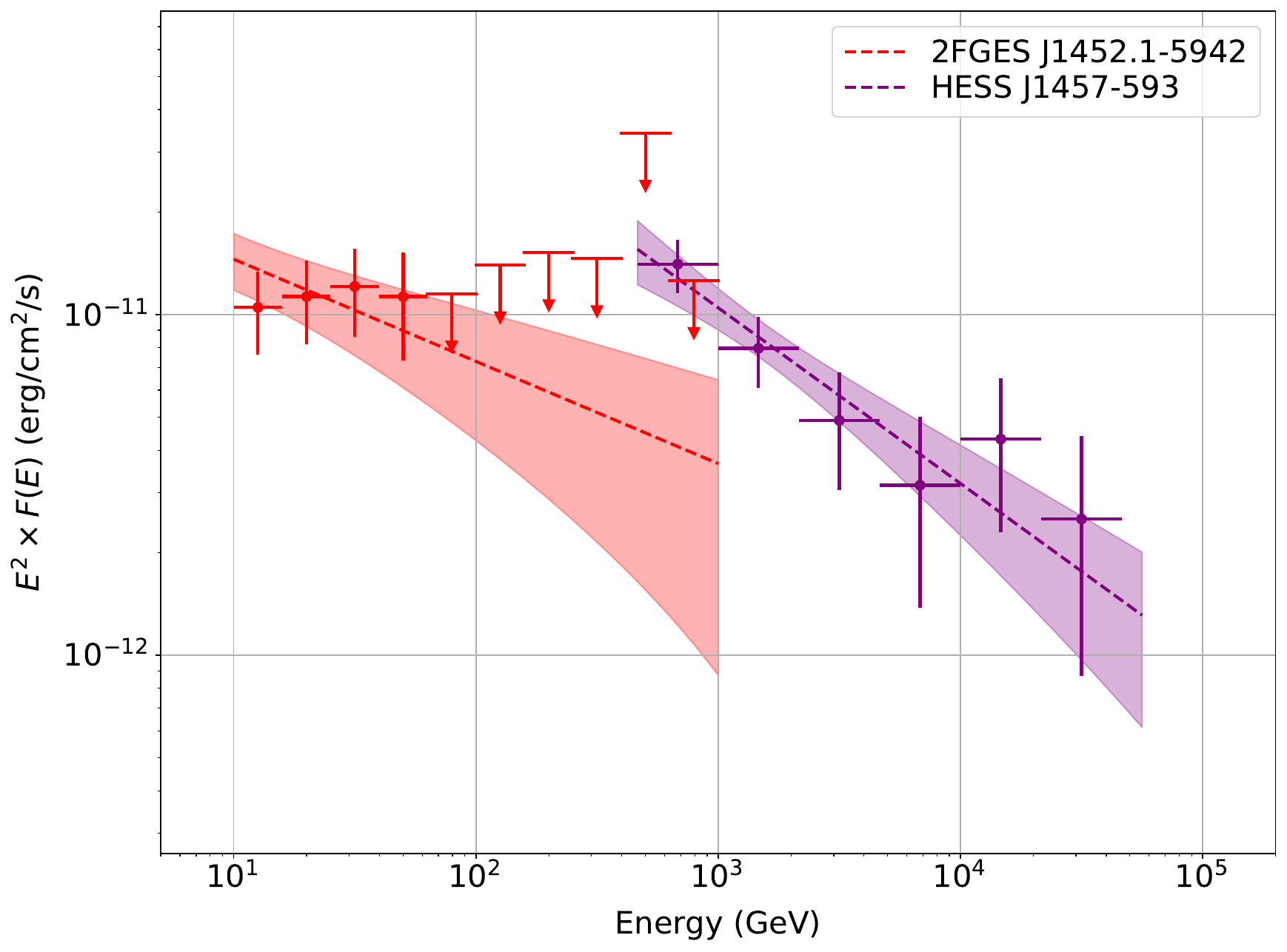}{0.45\textwidth}{}} 
\gridline{ \fig{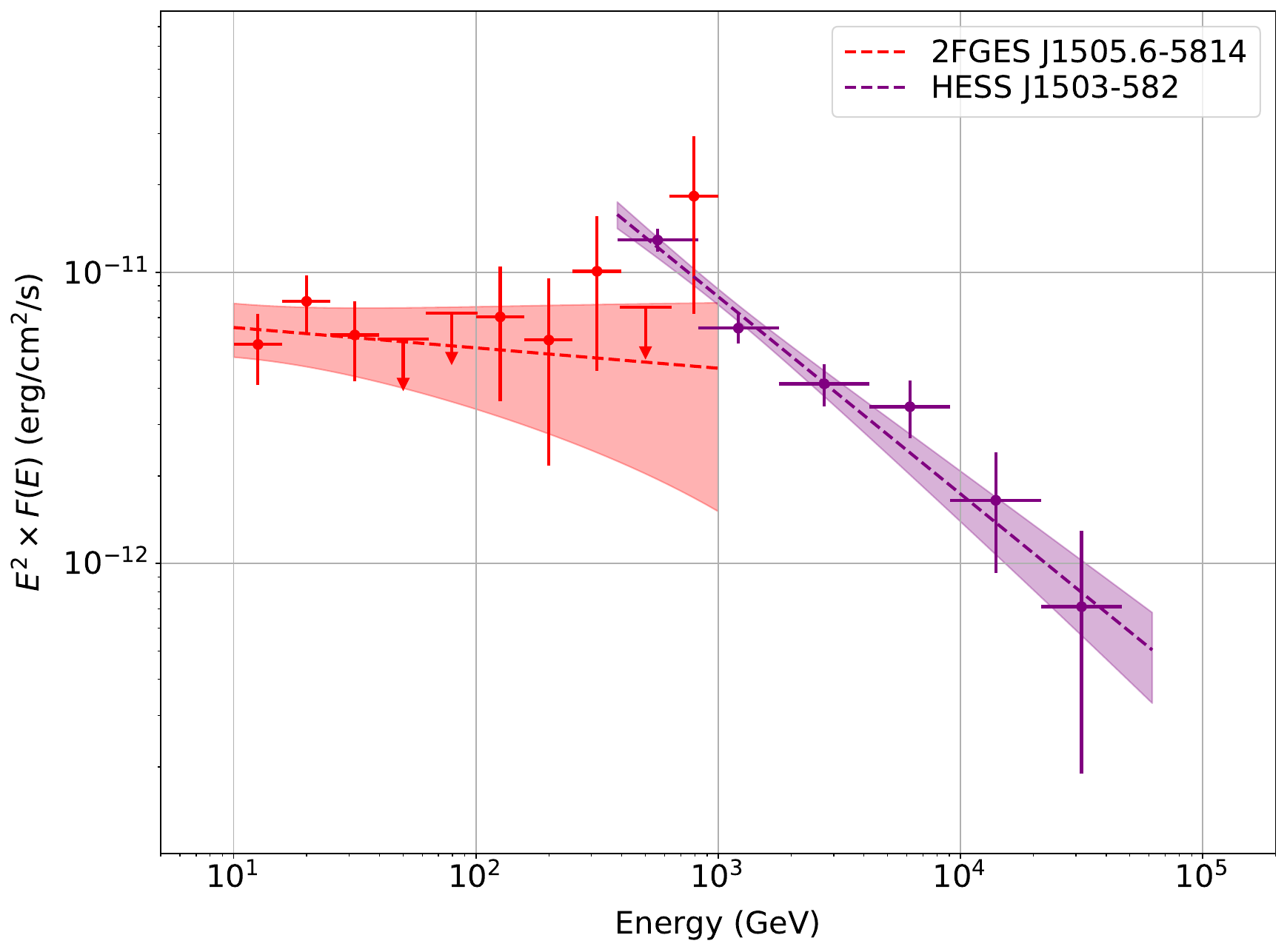}{0.45\textwidth}{}
          \fig{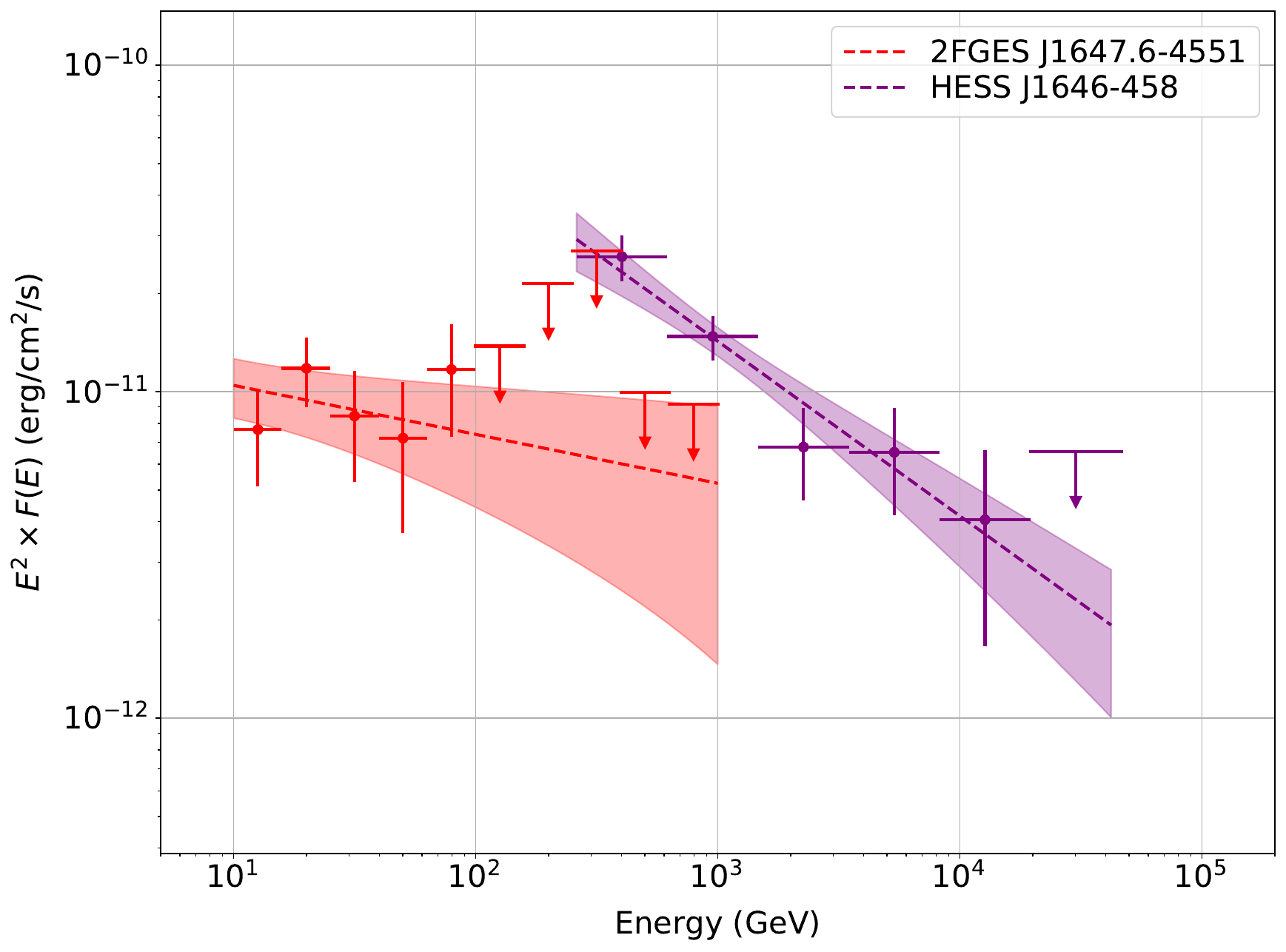}{0.45\textwidth}{}}          
\caption{Spectral energy distributions for the 2FGES sources associated with TeV sources.~The spectra of 2FGES sources are shown in red, while the data for the associated 3HWC and HGPS sources are shown in blue and purple, respectively.~The upper limits are given at 95\% confidence level.~\label{fig:gev_tev_sed}}
\end{figure}

\begin{figure}[!pt] 
 \gridline{ \fig{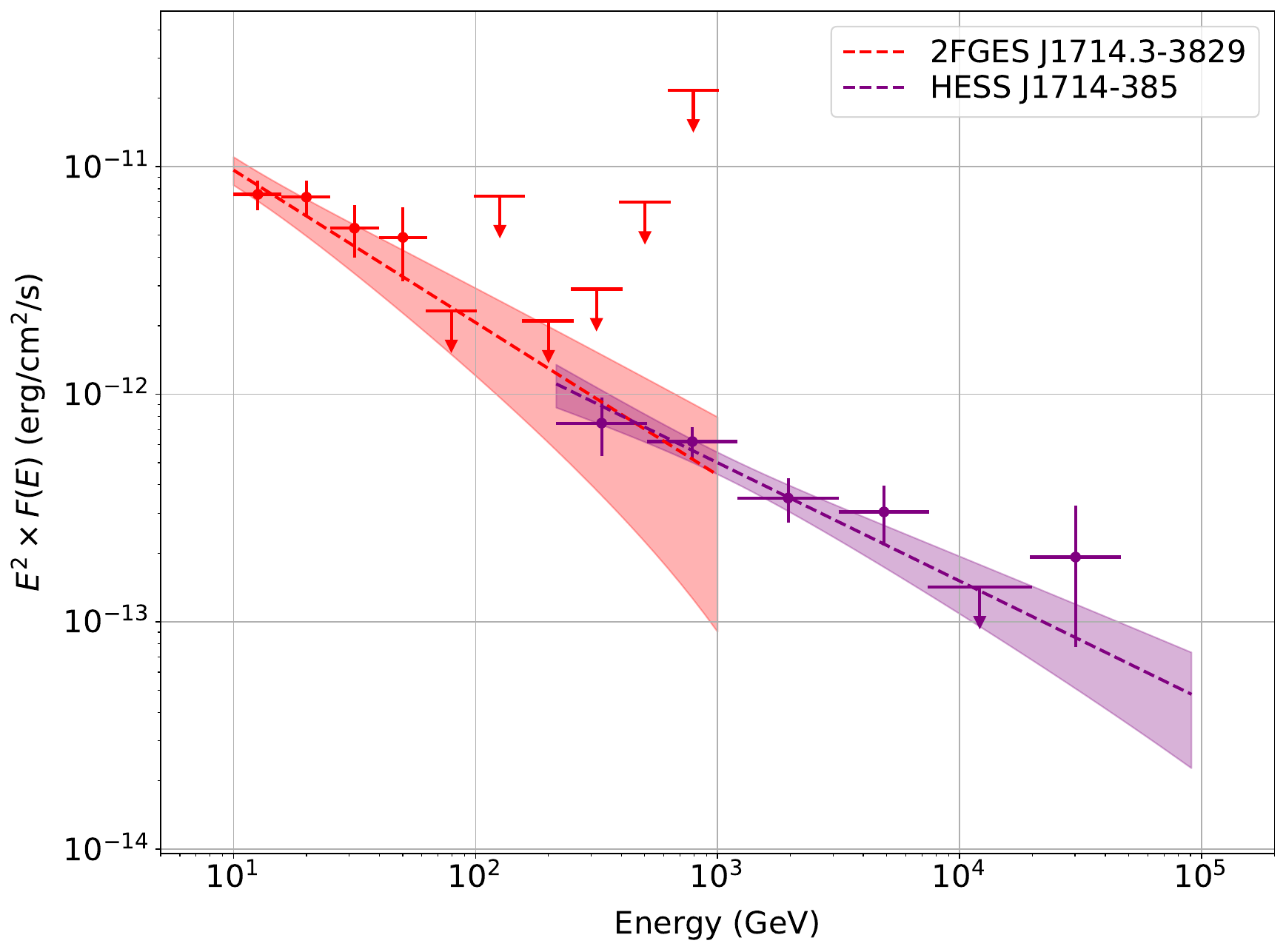}{0.45\textwidth}{}
          \fig{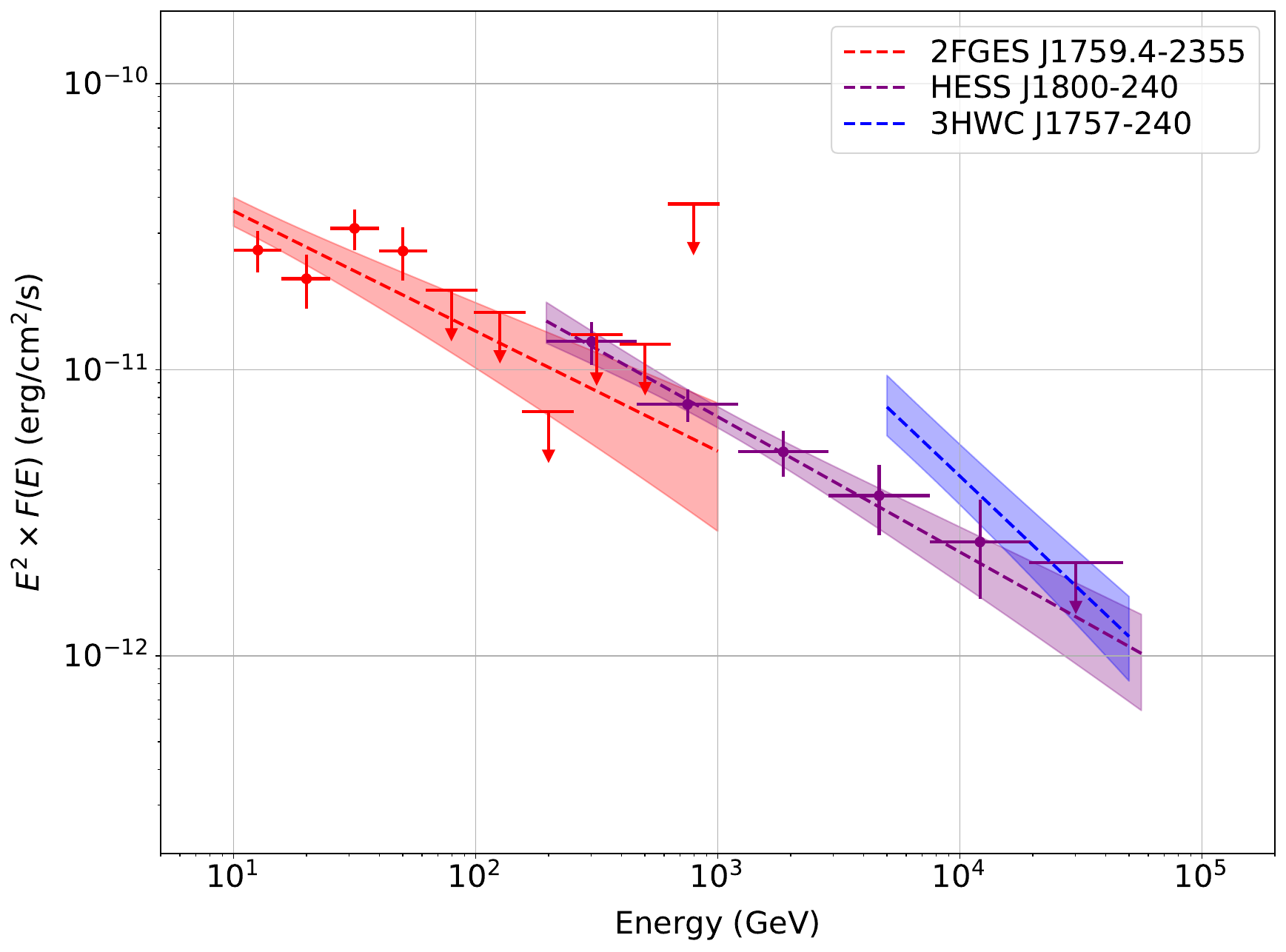}{0.45\textwidth}{}} 
 \gridline{ \fig{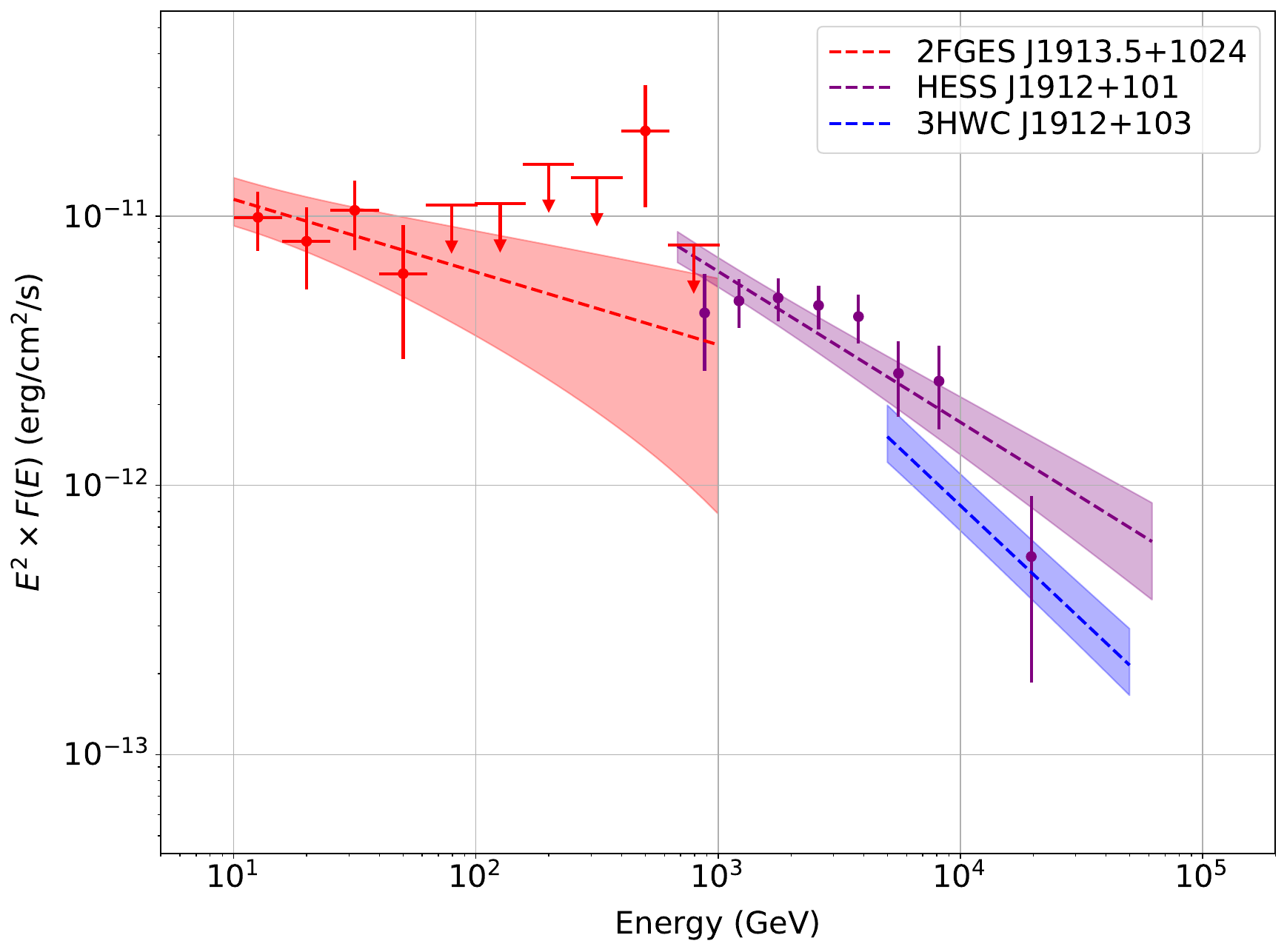}{0.45\textwidth}{}
          \fig{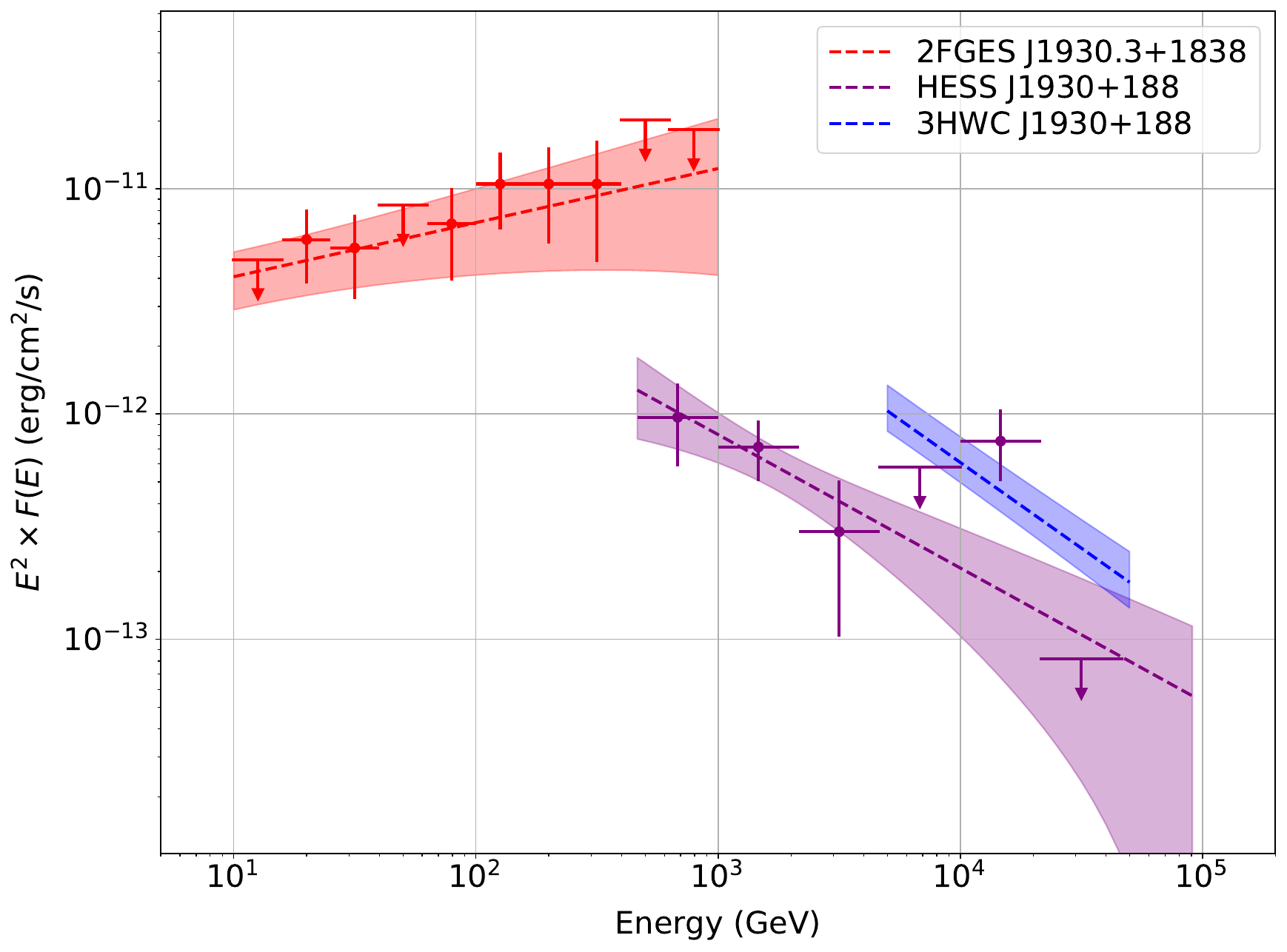}{0.45\textwidth}{}}
\gridline{ \fig{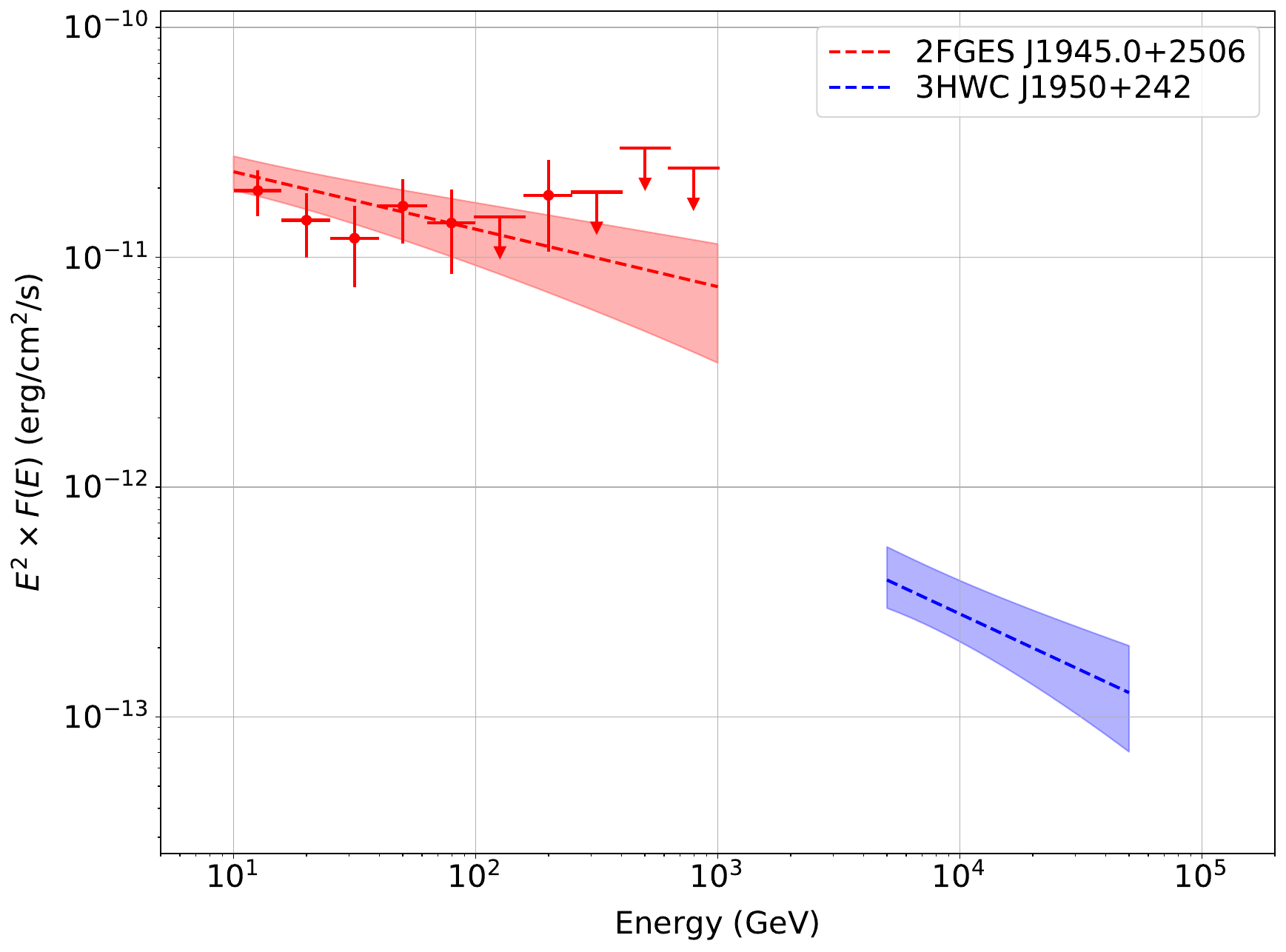}{0.45\textwidth}{}}           
\caption{Same as Fig.~\ref{fig:gev_tev_sed}, for the rest of the 2FGES sources associated with TeV sources.~\label{fig:gev_tev_sed_part2}}
\end{figure}

To assess the likelihood of a connection between the 2FGES and TeV sources, we examined the correlation with the contours of the detection significance obtained from the HGPS and 3HWC data on the TS maps.~We also compared our best-fit power-law spectra to the spectral data provided for the HGPS and 3HWC sources to check whether the GeV-TeV spectra connect well, which would support the possibility that a 2FGES and TeV source are one and the same object or region.\footnote{The TeV spectral information was extracted from the HGPS and 3HWC catalog FITS files.~The GeV-TeV spectra are provided for only 11 out of 13 sources, as $\eta$ Carinae was not included in the HGPS catalog, and one 2FGES source has only a 1LHAASO counterpart, for which spectral information is not public.}~The corresponding plots are displayed in Figs.~\ref{fig:gev_tev_sed}$-$\ref{fig:gev_tev_sed_part2}, showing that the GeV-TeV spectra connect well for the majority of 2FGES sources, except in five cases discussed below.~Interestingly, there are discussions in \citet{Abdalla18} about data analysis caveats concerning the measured TeV spectra of HESS J1503$-$582, HESS J1646$-$458, and HESS J1930+188, suggesting that the actual level of the emission may need to be shifted upwards or downwards in a way that would support the spectral GeV-TeV connection.

For 2FGES J1452.1$-$5942, 2FGES J1505.6$-$5814, and 2FGES J1647.6$-$4551, the level of GeV emission appears significantly below a plausible extrapolation of the TeV spectra to lower energies, even taking into account a possible break or curvature in the broadband spectrum.~The TS map for 2FGES J1452.1$-$5942, for instance, shows that this source is much larger than the coincident HESS J1457$-$593, with its centroid quite offset from the TeV source.~A possible interpretation of the new HGPS source consisted of a molecular cloud~\citep{Hofverberg10} illuminated by escaping cosmic rays from a nearby SNR.~However, the radio SNR is much smaller and located at the edge of the 2FGES source, making a physical connection seem unlikely.~HESS J1503$-$582 is also a new HGPS source without a compelling association, but the connection with 2FGES J1505.6$-$5814 seems more likely, as their morphologies are more similar and the spectral mismatch is less severe.~The 4FGL J1503.7$-$5801 source, associated with 2FGES J1505.6$-$5814 and identified as a blazar candidate (WISE J150334.23$-$580119.2 = FRC J1503$-$5801;~\citealt{Schinzel17}), is located near the top edge of the 2FGES emission region, making it unlikely to account for the bulk of the GeV extended emission.~2FGES J1647.6$-$4551 lies in the complex Westerlund 1 region, and its morphology shares some similarities with the layout of the TeV emission revealed in a recent study~\citep{Aharonian22} both in terms of position and extent, with some overlap with two out of three identified TeV emission substructures.~Notably, there are many 4FGL point-like sources clustering in this region, and the very large extended source 4FGL J1652.2$-$4633e partially overlaps with the 2FGES source, suggesting that the observed spectral mismatch could potentially be influenced by an inadequate description of this complicated region.

For 2FGES J1930.3+1838 and maybe also 2FGES J1945.0+2506, the level of GeV emission seems significantly above that measured in the TeV range, suggesting that these sources are capturing emission beyond the boundaries of the TeV-emitting objects.~And indeed, the TS maps seem to confirm that interpretation.~In the case of 2FGES J1930.3+1838, the smaller coincident LHAASO source seems to encompass the main cluster of excess emission on small scales and it may be that the large extent of the 2FGES source is driven by residual emission further away in the field.~The same situation holds for 2FGES J1945.0+2506, with most hotspots in the TS maps being found within the extent of the HAWC and LHAASO sources.

2FGES J1045.1$-$5937 has a soft spectrum with an index of $\Gamma\mathrel{=}2.76\pm0.26$ and is coincident with 4FGL J1045.1$-$5940, which is associated with $\eta$ Carinae.~It is located inside the spatially large extended source 4FGL J1036.3$-$5833e, which is identified in our work as one of the 2FGES sources (see Table~\ref{tab:previously_detected_ext_srcs}).~Recently,~\citet{Ge22} reported the detection of two extended components above 0.5\,GeV in the star-forming region of the Carina Nebula complex, one of which is perfectly consistent with the 2FGES source, albeit with a harder spectrum with an index of $\Gamma\mathrel{=}2.36\pm0.01$.~\citet{MartiReimer21} described the spectrum of the 4FGL source as best fit by a smoothly broken PL in the 80\,MeV--500\,GeV energy range, with a high-energy index of $\Gamma_{2}\mathrel{=}2.64\pm0.09$ above a spectral break at $E_{b}\mathrel{=}500~\rm{MeV}$, which is in good agreement with our results.~\citet{HESS20} reported the detection of very-high-energy point-like $\gamma$-ray emission towards the colliding-wind binary $\eta$ Carinae, which is located 0.08$^{\circ}$ away from the centroid of 2FGES J1045.1$-$5937.~The bright pulsar PSR J1048$-$5832 is located at 1.16$^{\circ}$ away from the centroid of 2FGES source, making any contamination unlikely.~Moreover,~\cite{Steinmassl23} provide further insights into the mechanisms of cosmic ray escape and $\gamma$-ray emission in the vicinity of $\eta$ Carinae, supporting the interpretation of the newly detected extended source as potentially connected to the binary system $\eta$ Carinae, possibly in the form of a halo of radiation from particles escaping the colliding-wind region.

For all other 2FGES sources, those for which the GeV and TeV spectra seem to connect smoothly, the correlation of the 2FGES source with the layout of TeV emission is pretty good.~Some 2FGES-TeV correlations are very convincing, like 2FGES J1119.3$-$6126, 2FGES J1417.5$-$6057, or 2FGES J1714.3$-$3829, while other connections like 2FGES J1759.4$-$2355 or 2FGES J1913.5+1024, probably deserve a more dedicated modeling of the corresponding regions.

\begin{longrotatetable}
\tabletypesize{\footnotesize}
\begin{deluxetable}{lcccccc}
\tablecaption{2FGES Sources Associated With TeV Sources\label{tab:tev_assoc_ext_srcs}}
\tablenum{8}
\tablehead{
\colhead{2FGES Name} & \colhead{TeV Name ($^{\circ}$)} & \colhead{TeV Class} & \colhead{4FGL Name} & \colhead{PSR Name} & \colhead{$r_{68}$ ($^{\circ}$)} & \colhead{$\Gamma$} 
}
\startdata
 J0006.0+7319$^{\dag}$ & 1LHAASO J0007+7303u (0.28, 0.17) & PWN & J0007.0+7303 & J0007+7303 & 0.85 & 1.85 \\ 
J0618.3+2227 & 3HWC J0617+224 (0.19), 1LHAASO J0617+2234 (..., 0.25) & Shell & J0618.9+2240c & ... & 0.97 & 2.07 \\ 
J1045.1$-$5937 & $\eta$ Carinae (0.08) & Binary & J1045.1$-$5940 & ... & 0.41 & 2.76 \\
J1119.3$-$6126 & HESS J1119$-$614 (0.04) & Composite & J1119.1$-$6127 & ... & 0.32 & 2.15 \\
J1417.5$-$6057 & HESS J1418$-$609 (0.06) & PWN & J1417.7$-$6057 & J1418$-$6058 & 0.17 & 1.68 \\
J1452.1$-$5942 & HESS J1457$-$593 (0.74) & Unid & J1450.2$-$5937c & ... & 1.06 & 2.30 \\
J1505.6$-$5814 & HESS J1503$-$582 (0.31) & Unid & J1503.7$-$5801 & ... & 0.53 & 2.07 \\
J1647.6$-$4551 & HESS J1646$-$458 (0.40) & Unid & J1648.4$-$4611 & J1648$-$4611 & 0.54 & 2.15 \\
J1714.3$-$3829 & HESS J1714$-$385 (0.04) & Composite & J1714.4$-$3830 & ... & 0.12 & 2.67 \\
J1759.4$-$2355 & HESS J1800$-$240 (0.32), 3HWC J1757$-$240 (0.54) & Unid & J1759.7$-$2354 & ... & 0.83 & 2.42 \\
J1913.5+1024$^{\dag}$ & HESS J1912+101 (0.24), 3HWC J1912+103 (0.34), 1LHAASO J1912+1014u (0.10, 0.23) & Unid & J1913.3+1019 & J1913+1011 & 0.77 & 2.27 \\ 
J1930.3+1838 & HESS J1930+188 (0.19), 3HWC J1930+188 (0.20), 1LHAASO J1929+1846u$^{\ast}$ (0.62, 0.27) & Composite & J1929.8+1832 & ... & 0.82 & 1.76 \\
J1945.0+2506 & 3HWC J1950+242 (1.54) & Unid & J1946.1+2436c & ... & 2.39 & 2.25 \\
\enddata
\tablecomments{Column 2 lists the TeV sources that are positionally coincident with 2FGES sources, along with the angular separations between their centroids shown in parentheses (for each LHAASO source, a doublet is provided corresponding to the KM2A and WCDA detections of the source, respectively).~In column 3, the classification of TeV sources is taken from the HGPS catalog if the source exists; otherwise, TeVCat~\citep{Wakely.Horan08} is used.~The associations with 4FGL sources and energetic PSRs are given in columns 4 and 5, respectively.~The last two columns provide the size and spectral index of the 2FGES sources, respectively.~The 2FGES sources flagged with a dagger ($\dag$) are potentially associated with TeV halo candidates.} 
\end{deluxetable}
\end{longrotatetable}

\subsection{Dubious 2FGES Sources} \label{subsec:dubious_srcs}

\begin{figure}[!pt]
 \gridline{ \fig{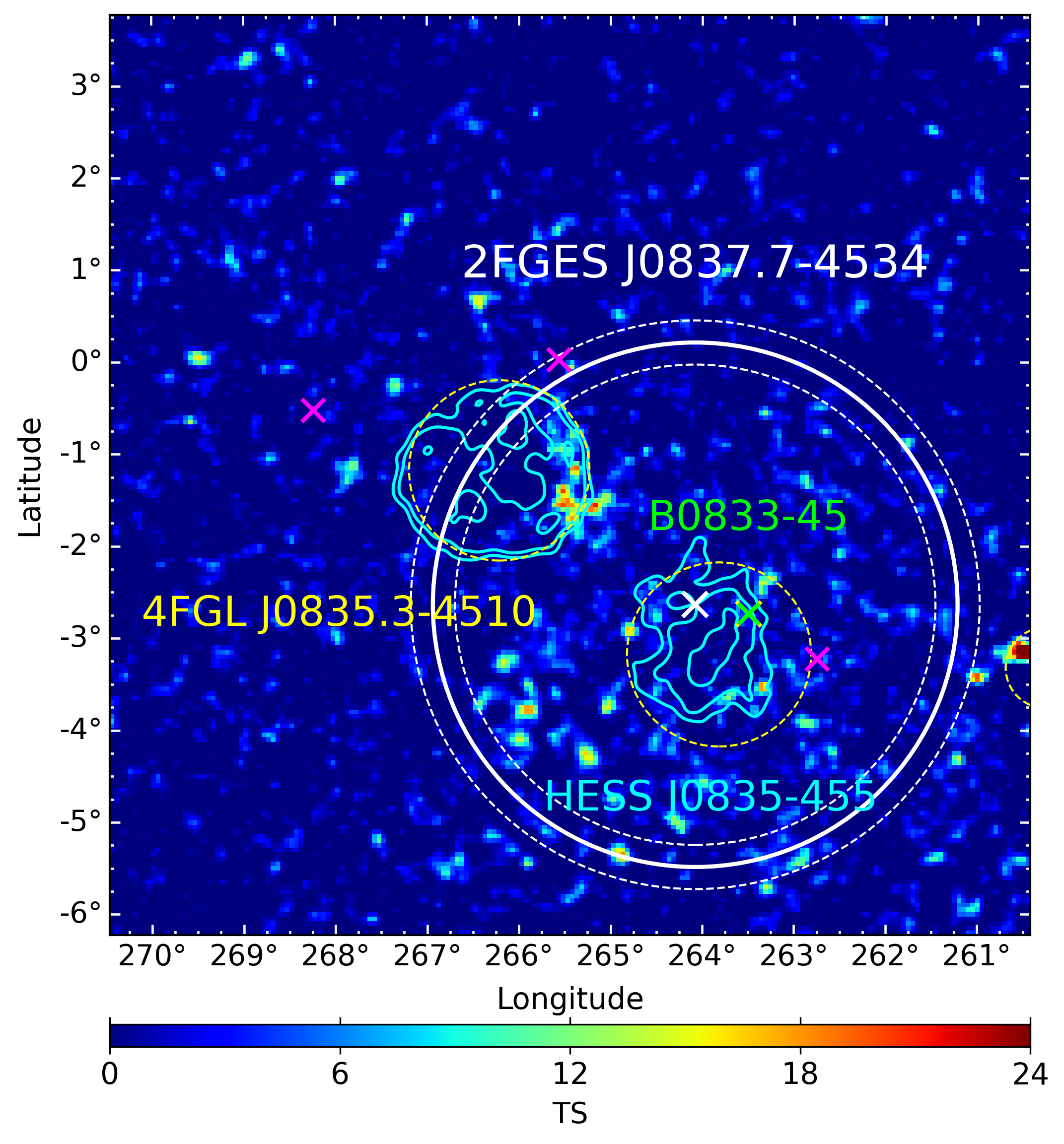}{0.35\textwidth}{}
          \fig{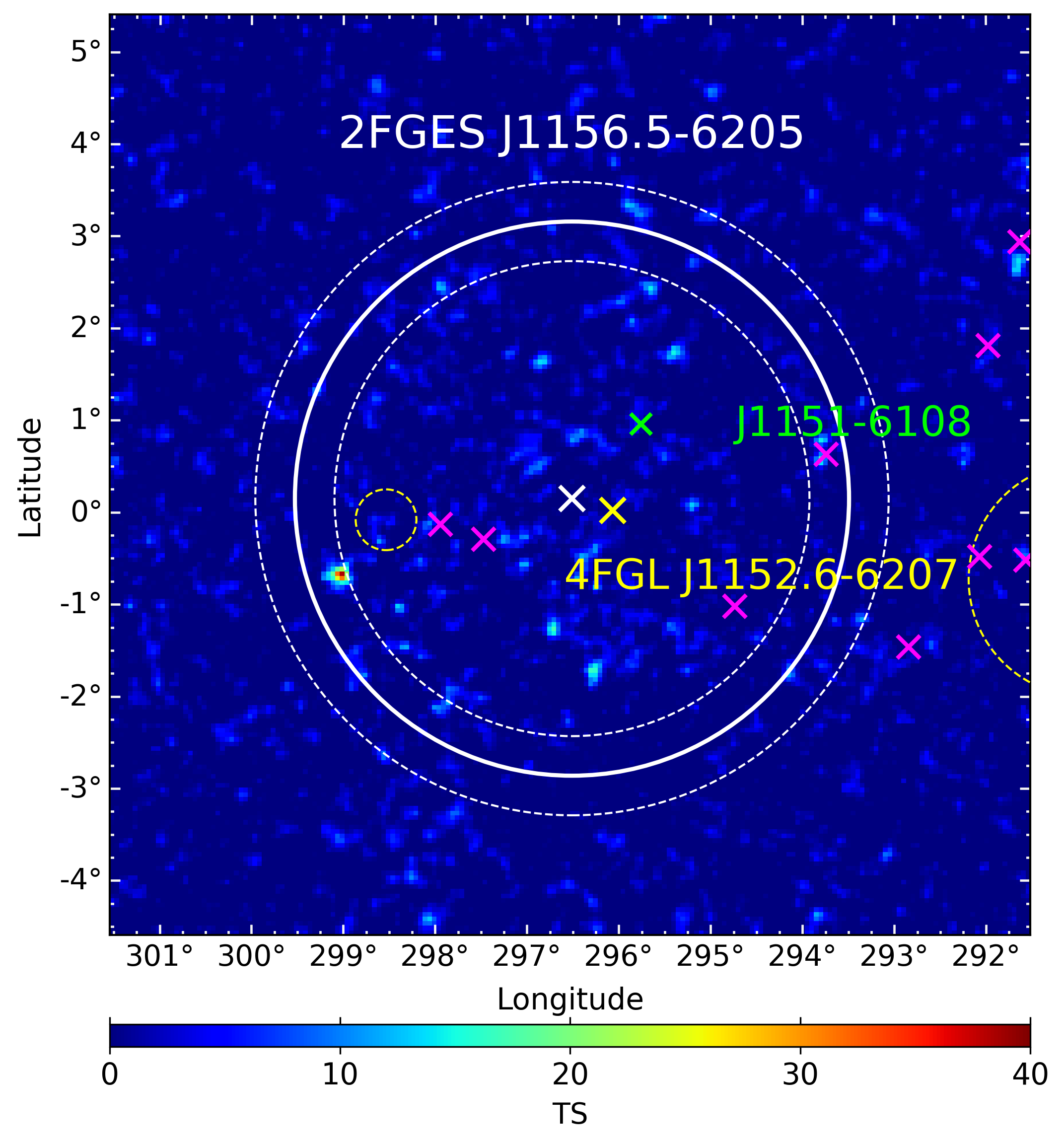}{0.35\textwidth}{}} 
 \gridline{ \fig{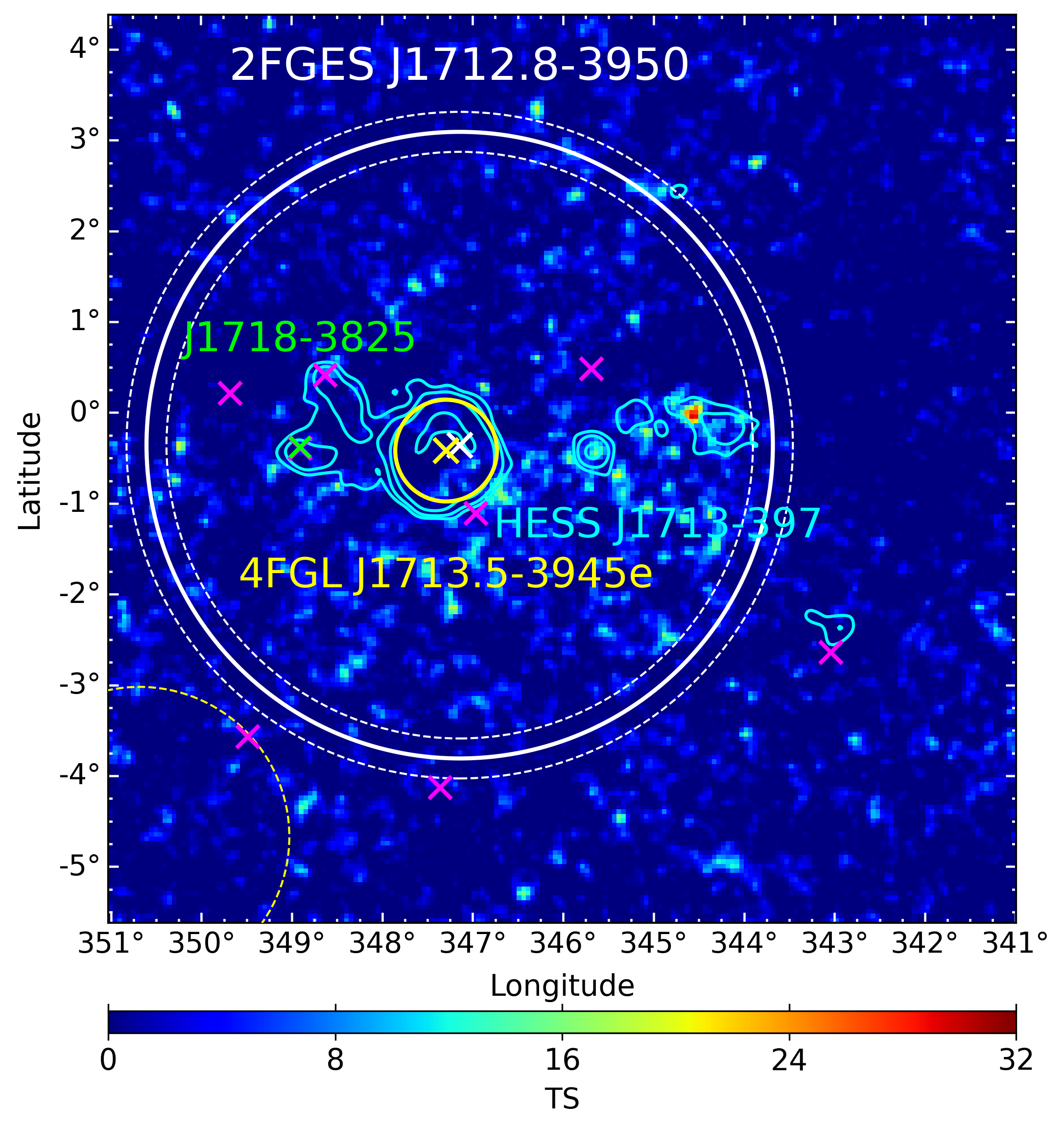}{0.35\textwidth}{}
          \fig{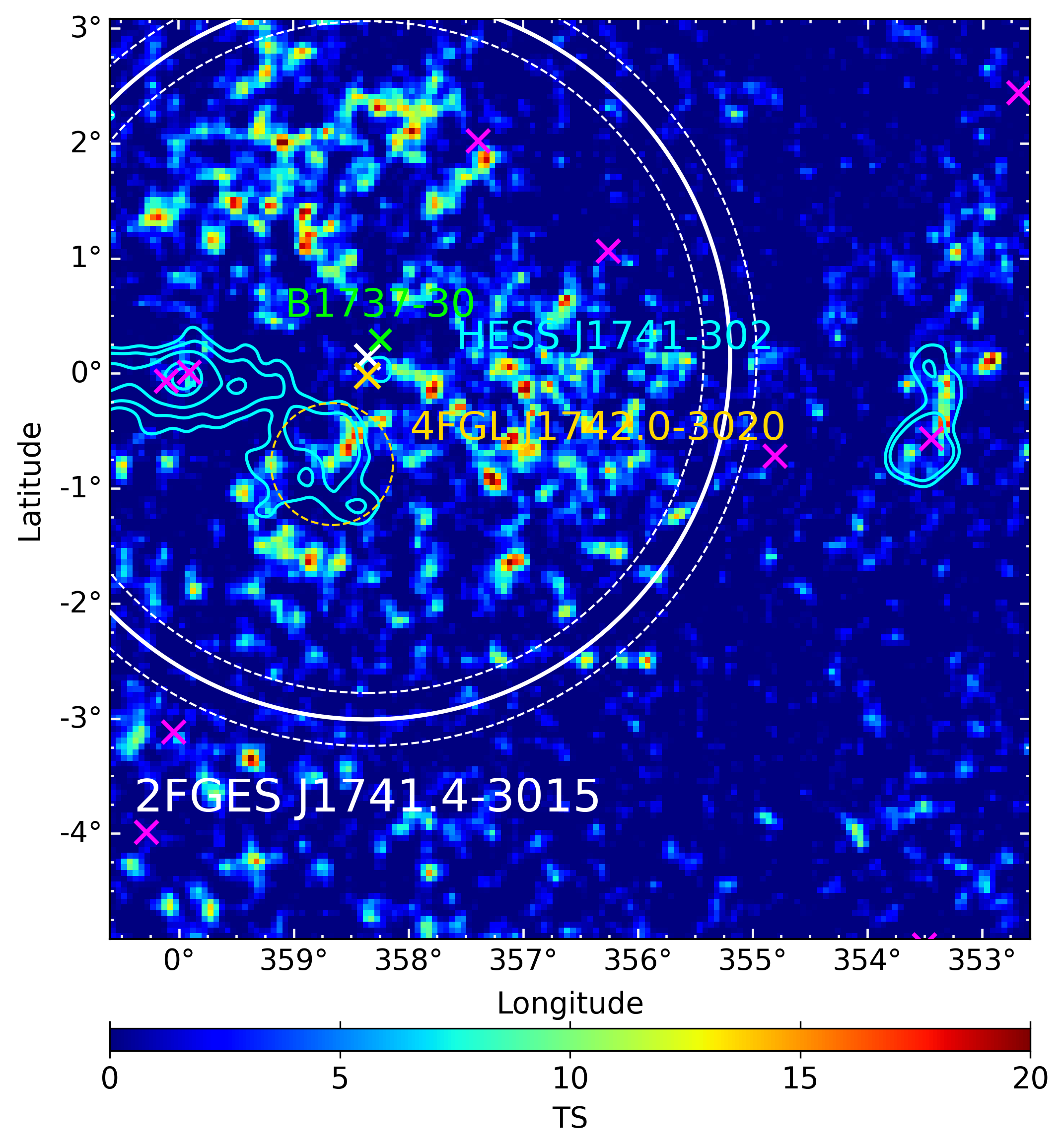}{0.35\textwidth}{}} 
\gridline{ \fig{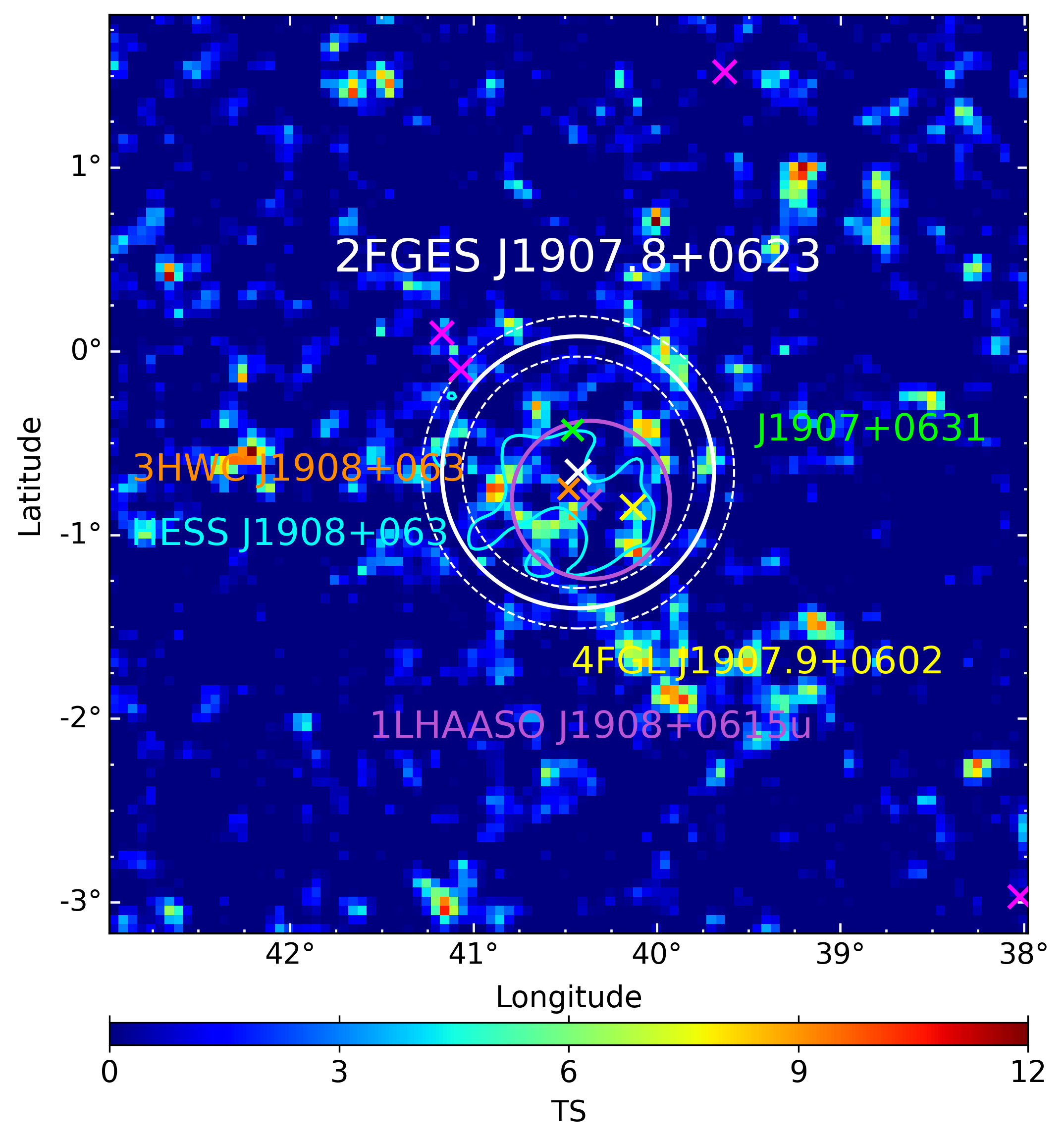}{0.35\textwidth}{}} 
\caption{TS maps of 2FGES sources classified as dubious.~Graphical elements are defined in the caption of Fig.~\ref{fig:4fgle}.~The 2FGES sources, shown from top-left to bottom-right, are 2FGES J0837.7$-$4534, 2FGES J1156.5$-$6205, 2FGES J1712.8$-$3950, 2FGES J1741.4$-$3015, and 2FGES J1907.8+0623.~\label{fig:dub}}
\end{figure}

Table~\ref{tab:dubious_ext_srcs} lists five 2FGES sources that we have classified as dubious detections.~This classification is based on a visual inspection of the TS maps (see Fig.~\ref{fig:dub}) and relies on two main qualitative criteria: (i) the distribution of the excess emission, supposedly captured by the 2FGES source, does not follow the assumed (2D Gaussian) morphology; and (ii) the field exhibits significant residual emission on small or intermediate angular scales, beyond the typical extent of the source and with a comparable magnitude.~Although the classification is somewhat subjective, the final list of dubious sources reveals systematic trends: all but one of the dubious sources have large sizes of more than 2$^\circ$ and show a notable decrease in their significance when using the alternative IEM.

It is important to note that classifying a source as dubious does not imply that the signal captured by the 2FGES source is not (partly or entirely) real; rather, it indicates that the semi-automated analysis performed here did not produce a sufficiently satisfactory description of the field, leading to cautious interpretation.~A given analysis may fulfill criterion (ii) if the IEM is inadequate for the region, compromising reliable extraction of an extended signal.~This is particularly relevant in regions near the Galactic center, such as 2FGES J1712.8$-$3950 and 2FGES J1741.4$-$3015.~Alternatively, criterion (i) may be fulfilled if an actual extended source has a morphology that cannot be accurately represented by our assumed 2D Gaussian intensity distribution.~This may also apply to the same two sources or to 2FGES J0837.7$-$4534 in the complex Vela region.

\begin{deluxetable}{lcc|lcc}[t] 
\tablecaption{2FGES Sources Classified as Dubious\label{tab:dubious_ext_srcs}}
\tablenum{9}
\tablehead{
\colhead{2FGES Name} & \colhead{$r_{68}$ ($^{\circ}$)} & \colhead{TS$_{\rm{mod.}}$ (\%)} & \colhead{2FGES Name} & \colhead{$r_{68}$ ($^{\circ}$)} & \colhead{TS$_{\rm{mod.}}$ (\%)}
}
\startdata
J0837.7$-$4534 & 2.85 & 39 & J1741.4$-$3015 & 3.15 & 87 \\ 
J1156.5$-$6205 & 3.01 & 40 & J1907.8+0623 & 0.74 & 15 \\ 
J1712.8$-$3950 & 3.45 & 58 & ... & ... & ... \\
\enddata
\tablecomments{Columns 2 and 5 list the sizes of the dubious 2FGES sources.~The modulation in the TS (as a percentage) due to changes in the IEM is provided in columns 3 and 6, derived as $\nicefrac{(X_{\rm{std.}} - X_{\rm{alt.}})}{X_{\rm{std.}}}$, where $X_{\rm{std.}}$ and $X_{\rm{alt.}}$ represent the measured TS values using the standard and alternative IEMs, respectively.}
\end{deluxetable}

\clearpage

\clearpage

\clearpage
\begin{figure}[!pt]
 \gridline{ \fig{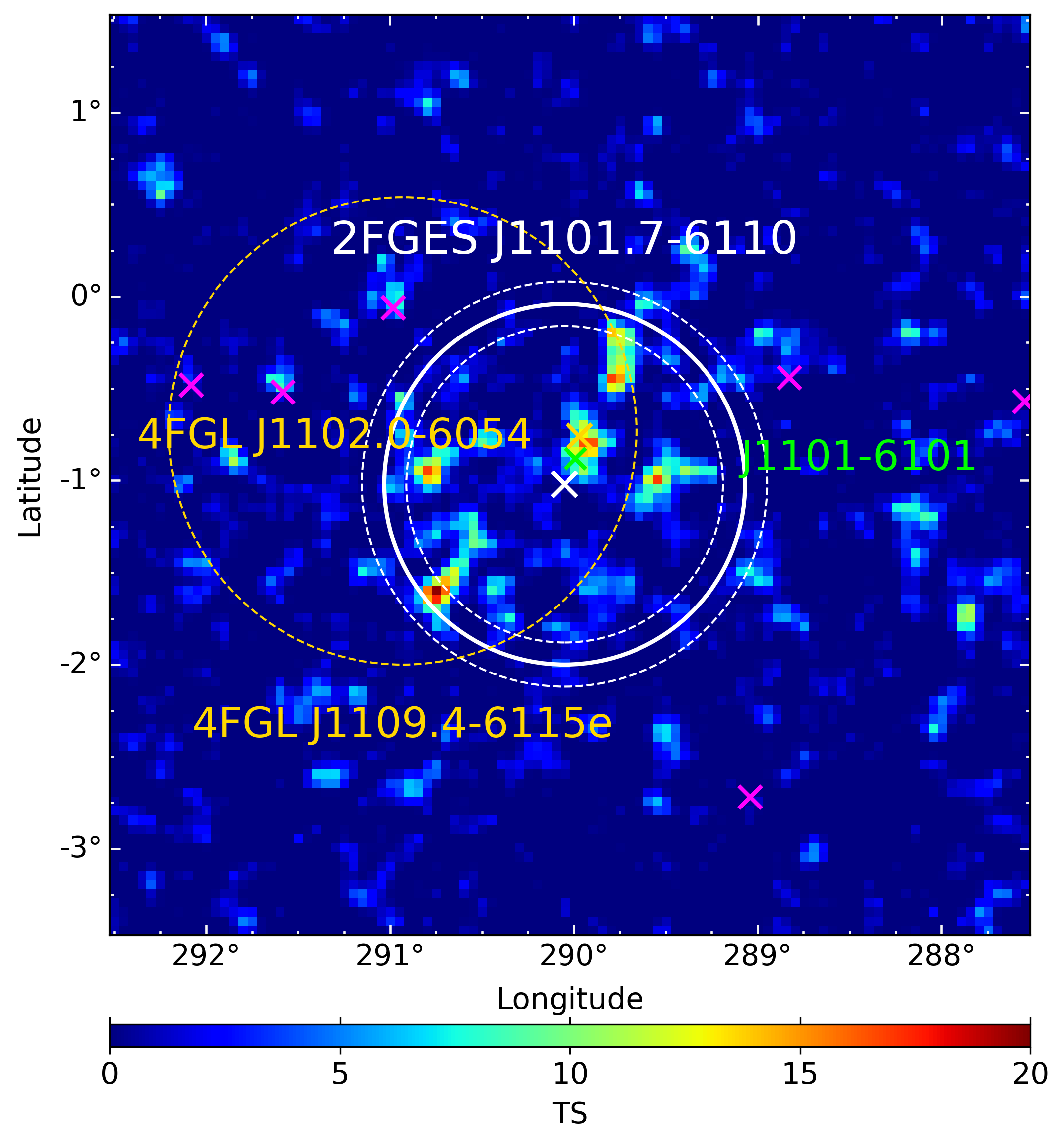}{0.35\textwidth}{}
          \fig{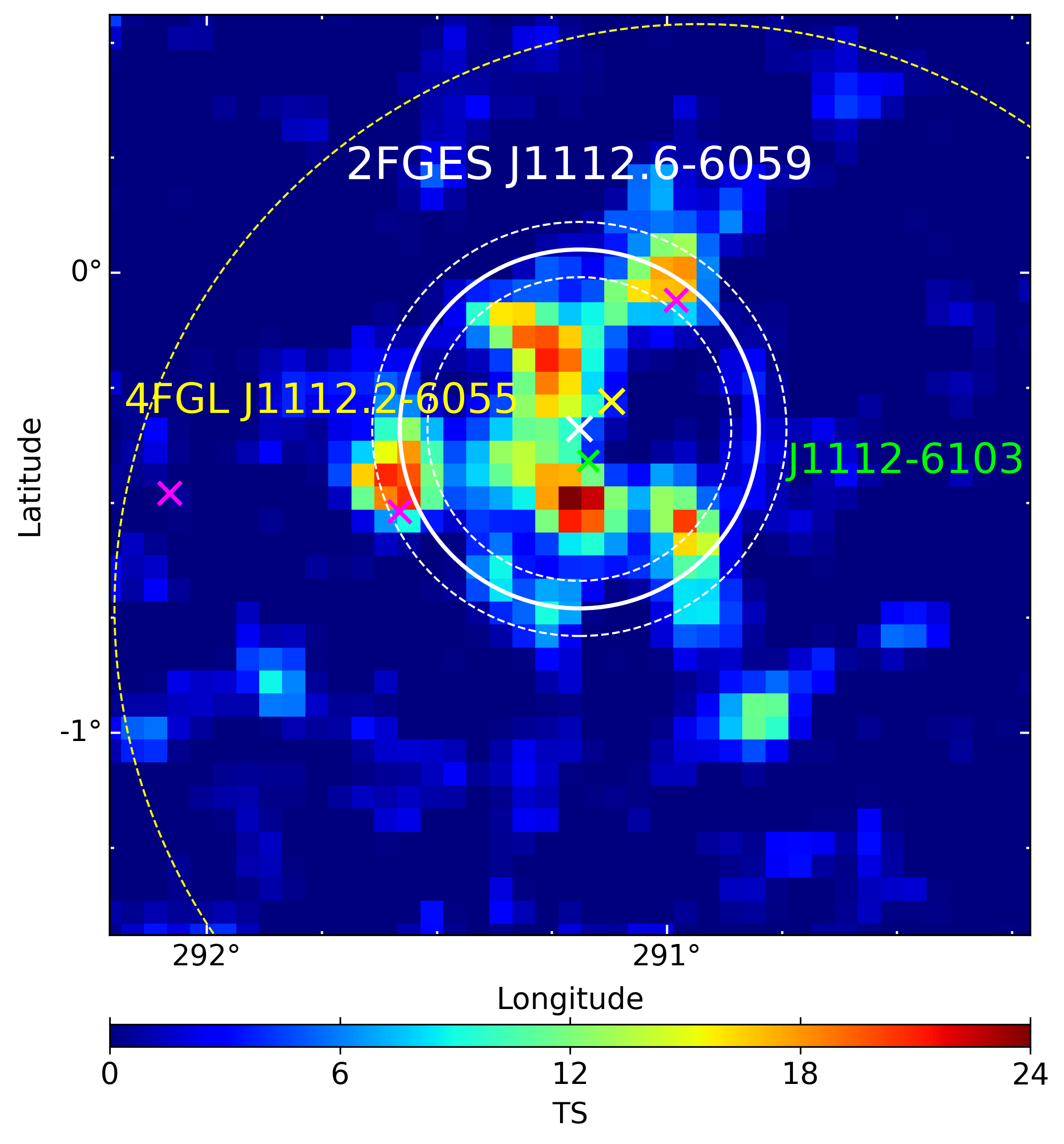}{0.35\textwidth}{}}
\caption{TS maps of 2FGES sources associated with a bright pulsar but not with any TeV source.~Graphical elements are defined in the caption of Fig.~\ref{fig:4fgle}.~The 2FGES sources, shown from left to right, are 2FGES J1101.7$-$6110 and 2FGES J1112.6$-$6059.~\label{fig:psr}}
\end{figure}

\clearpage

\clearpage

\clearpage

\clearpage

\clearpage

\clearpage
\section{Discussion of Candidate Pulsar-Related Sources} \label{sec:Discussion}

Returning to the primary motivation for this work, namely the search for pulsar halos in the GeV range, we now discuss those extended sources that may be the best candidates for such phenomena.~We found six associations between the newly detected extended sources and bright pulsars (defined as having $\dot{E}/d^{2}\,\geq\,10^{34}$ erg s$^{-1}$ kpc$^{-2}$), which are obvious candidates for pulsar-related emission.~Additionally, we included two sources, 2FGES~J1119.3$-$6126 and 2FGES~J1930.3+1838, in the sample of possibly pulsar-related objects because they are positionally well-correlated with pulsars that did not fulfill our bright pulsar selection criterion.~We note that 2FGES J1714.3$-$3829, associated with $\gamma$-ray pulsar 4FGL J1714.4$-$3830 (PSR J1714$-$3830), could have been part of the sample, but there is currently no reliable distance estimate for this pulsar, and the source was recently interpreted as an interacting SNR~\citep{Abdollahi20}.~Ultimately, we compile a list of eight candidates, which is summarized in Table~\ref{tab:psr_powered_properties}.~In the following subsections, we first provide some contextual information for each source and then assess more quantitatively the likelihood that these sources are indeed pulsar-related.

\subsection{Contextual Information on Selected Sources} \label{subsec:psr_related_candidates}

\textit{2FGES J0006.0+7319} is a newly detected emission with a 68\% containment radius of $0.85^{\circ}\pm0.12^{\circ}$ and a hard spectrum with a photon index $\Gamma\mathrel{=}1.85\pm0.16$.~It surrounds the $\gamma$-ray pulsar 4FGL J0007.0+7303 (PSR J0007+7303), which is associated with the radio shell-type SNR G119.5+10.2 (also named CTA 1).~The extension of the 2FGES source is consistent with the size of the radio SNR, however it is offset by 0.57$^{\circ}$ from its centroid.~Recently, the LHAASO collaboration reported the detection of a very-high-energy source, 1LHAASO J0007+7303u, associated with the $\gamma$-ray pulsar and CTA 1~\citep{Cao24}, presenting it as a TeV pulsar halo candidate.

\textit{2FGES J1101.7$-$6110} is a rather large source with a 68\% containment radius of $0.98^{\circ}\pm0.12^{\circ}$ and a very soft spectrum with $\Gamma\mathrel{=}2.79\pm0.31$.~It is associated with the energetic young PSR J1101$-$6101, located at a 0.15$^{\circ}$ offset from the centroid of the 2FGES source, as well as with 4FGL J1102.0$-$6054, classified as a blazar candidate of uncertain type (bcu class in the 4FGL catalog).~It has no TeV counterpart but partially overlaps the unassociated extended source 4FGL J1109.4$-$6115e with an offset of 0.93$^{\circ}$.~The associated pulsar is known as a radio- and gamma-quiet pulsar, which powers the Lighthouse PWN.~Recently,~\citet{Klingler23} reported \textit{NuSTAR} and \textit{Chandra} observations of the PSR J1101$-$6101 field and proposed a serendipitously detected field X-ray source as the counterpart to 4FGL J1102.0$-$6054, supporting its classification as a blazar.~The latter source may therefore be a background object unrelated to the pulsar and its system, but our detection of significantly extended GeV emission invites reconsideration of that scenario.~Moreover, both the PSR and the blazar are consistent with the central TS peak in our analysis (see Figure~\ref{fig:psr}, left panel), suggesting that any contribution from the blazar to the 2FGES emission is likely confined to this region.

\textit{2FGES J1112.6$-$6059} is a relatively compact source with a size $r_{68}\mathrel{=}0.39^{\circ}\pm0.06^{\circ}$ and a hard spectrum with photon index $\Gamma\mathrel{=}1.95\pm0.15$.~It is embedded in the much more extended source 4FGL J1109.4$-$6115e, yet unidentified but positionally coincident with the young massive cluster NGC 3603.~The 2FGES source is associated with the $\gamma$-ray pulsar PSR J1112$-$6103 and an unidentified GeV source 4FGL J1112.2$-$6055.~Its properties are in good agreement with the previously detected extended emission 2FHL J1112.1$-$6101e.~As argued in~\citet{Ackermann16}, the hard GeV extended emission may be connected with the PWN of PSR J1112$-$6103.~However, the large uncertainty on the distance of the pulsar puts this scenario into question.

\textit{2FGES J1119.3$-$6126} is another relatively compact source with a size $r_{68}\mathrel{=}0.32^{\circ}\pm0.05^{\circ}$ and a hard spectrum with photon index $\Gamma\mathrel{=}2.15\pm0.14$.~It is associated with the $\gamma$-ray pulsar 4FGL J1119.1$-$6127 (PSR J1119$-$6127) and HESS J1119$-$614, which has been firmly identified as the composite SNR G292.2$-$00.5, containing both the above-mentioned pulsar and PWN G292.15$-$0.54.~The offset between the centroid of the 2FGES source and the radio SNR is negligible (0.03$^{\circ}$), but the detected GeV emission is approximately twice as large as the radio extension of SNR G292.2$-$00.5.~\citet{Acero16} reported the detection of extended emission with a larger angular separation of 0.10$^{\circ}$ to the radio remnant and marginally classified it as SNR G292.2$-$00.5; however, their spatial extension is consistent with our results.~This finding is also supported by the recent addition of a new extended source, 4FGL J1119.0$-$6127e (classified as SPP), to the 4FGL-DR4 catalog~\citep{Ballet23}, on top of the already known pulsar, where the TeV morphology of HESS J1119$-$614 has been used as the spatial template for this source~\citep{Abdalla18}.

\textit{2FGES J1417.5$-$6057} is a small but significantly extended source with $r_{68}\mathrel{=}0.17^{\circ}\pm0.04^{\circ}$.~It is positionally coincident with the unidentified source 4FGL J1417.7$-$6057, which is associated with the energetic $\gamma$-ray pulsar PSR J1418$-$6058 at an angular offset of 0.15$^{\circ}$.~This pulsar powers the PWN HESS J1418$-$609, also known as the Rabbit nebula, whose position and extent closely match the 2FGES source.~The Rabbit nebula is located within the Kookaburra complex region, which also contains another PWN, HESS J1420$-$607 (sometimes referred to as the K3 nebula), powered by the bright pulsar PSR J1420$-$6048.~\citet{Ackermann17} reported the GeV detection of the PWN HESS J1420$-$607, but did not detect significant extended emission at the position of HESS J1418$-$609 in the Kookaburra region.~A multi-wavelength study of the Rabbit nebula by~\citet{Park23} has shown extended emission above 30\,GeV at the position of 4FGL J1417.7$-$6057, but had difficulties in determining the spatial parameters due to a lack of statistics.~The hard spectrum of the 2FGES source, with a photon index $\Gamma\mathrel{=}1.68\pm0.12$, and the good agreement between the morphologies of the GeV and TeV emission toward HESS J1418$-$609, suggest that the newly detected 2FGES source is most likely the GeV detection of the Rabbit nebula.

\textit{2FGES J1647.6$-$4551} is a flat $\gamma$-ray source with a photon index of $\Gamma\mathrel{=}2.15\pm0.15$ and a size of $r_{68}\mathrel{=}0.54^{\circ}\pm0.07^{\circ}$.~It is associated with the $\gamma$-ray pulsar 4FGL~J1648.4$-$4611 (PSR J1648$-$4611), which is offset by 0.36$^{\circ}$ from the 2FGES source centroid.~This spatial coincidence could suggest pulsar-related emission for this 2FGES source, but the latter also partially overlaps with the complex source HESS J1646$-$458, located in the vicinity of the young massive stellar cluster Westerlund 1.~Recent observations towards HESS J1646$-$458 by~\citet{Aharonian22}, based on significantly increased H.E.S.S. exposure, revealed a shell-like, energy-independent morphology with a diameter of $\sim2.0^{\circ}$ and interpreted the TeV emission as due to CR acceleration at the cluster wind termination shock.~\citet{Harer23} further investigated this scenario and indicated that a leptonic model is consistent with the available data.

\textit{2FGES J1913.5+1024} is a moderately extended source with a size $r_{68}\mathrel{=}0.77^{\circ}\pm0.08^{\circ}$ and a relatively flat spectrum with a photon index $\Gamma\mathrel{=}2.27\pm0.16$.~It is offset from the energetic $\gamma$-ray pulsar 4FGL J1913.3+1019 (PSR J1913+1011) by 0.09$^{\circ}$ and encompasses the extended source HESS J1912+101.~The TeV source, located in a complex region, is a shell-like SNR candidate with no clear counterpart that has been detected by several TeV observatories.~A recent study by~\citet{Reich.Sun19}, however, has found an excessive polarized radio emission at a partial shell indicating the existence of synchrotron emission related to HESS J1912+101.~The detection of extended GeV emission from HESS J1912+101 using the $Fermi$-LAT data has been reported recently in the literature, but the very origin of the $\gamma$-ray emission remains unclear \citep[see, e.g.,][]{Zhang20a, Zeng21, Sun22}.~The presence of PSR J1913+1011 justifies considering a pulsar origin for the source, but it could alternatively well be classified as an SNR candidate.

\textit{2FGES J1930.3+1838} is a hard GeV source with a spectral index of $\Gamma\mathrel{=}1.76$ and a size of $r_{68}\mathrel{=}0.82^{\circ}$.~It is potentially linked with the known TeV source VER J1930+188, which has also been detected by other observatories as HESS J1930+188, 3HWC J1930+188, and 1LHAASO J1929+1846u$^{\ast}$.~The TeV source is identified as composite in the HGPS catalog, due to its association with the Crab-like composite remnant SNR G054.1+00.3 and the nebula surrounding the pulsar PSR J1930+1852, without the possibility to disentangle both contributions to the emission (it is, however, classified as PWN in TeVCat).~The central pulsar is one of the youngest and brightest pulsars, with $\tau_{\rm{c}}\mathrel{=}2.9$\,kyr and $\dot{E}\mathrel{=}1.2\times10^{37}$\,erg s$^{-1}$.~\citet{Cao24} have suggested 1LHAASO J1929+1846u$^{\ast}$ as a TeV halo candidate.

\subsection{Quantitative Assessment of a Pulsar Origin} \label{subsec:discussion_psr_related_assessment}

To assess the pulsar halo nature of these eight extended sources, we first derived the radiation efficiency to estimate the fraction of the pulsar spin-down power that is transferred to 10--1000 GeV luminosity.~The flux to luminosity conversion is done using the distance estimates provided in the ATNF catalog, which are inferred from the dispersion measure~\citep{Yao17}.~Using the same distance estimates, we then compute the typical physical extent of the source, and the physical offset between the 2FGES centroid and the pulsar position.~This information, along with a recap of the main pulsar parameters, are provided in Table~\ref{tab:psr_powered_properties}.

We found that the radiation efficiencies were less than 10\%, with the exception of a higher value of 40\% for 2FGES~J1647.6$-$4551, which coincides with Westerlund 1.~This can be compared to the 5--10\% efficiency range observed for the halo around Geminga~\citep{DiMauro19}.~Given our limited knowledge of halo formation and the small number of such objects that are well-characterized observationally, it is difficult to dismiss candidates based solely on their efficiency.~Variations in radiation efficiency compared to Geminga could be ascribed to changes in particle acceleration efficiency or confinement efficiency (i.e., the level of diffusion suppression).~At the very least, none of the eight sources exceeds the power budget in the context of a pulsar interpretation; however, 2FGES~J1647.6$-$4551 clearly sticks out due to its very high conversion efficiency, which could argue against a pulsar interpretation and instead favour a connection with the star-forming region Westerlund 1~\citep{Aharonian22}.

Regarding the physical sizes of the systems, most of them appear reasonable considering that the halos around J0633+1746 and B0656+14 have been observed in the TeV range to extend up to about 30 pc at least.~The only exceptions are 2FGES~J1101.7$-$6110, with a typical size of 120\,pc, and, to a lesser extent, 2FGES~J1930.3+1838, with a typical size of 90\,pc.~In both cases, it may be challenging to sustain small-scale turbulence at the required level if diffusion is to be strongly suppressed within such large volumes.~Moreover, for 2FGES~J1930.3+1838, the size appears quite large in the context of a pulsar-powered object, considering the small characteristic age of the pulsar.

In GeV halos around middle-aged pulsars, an offset between the centroid of the emission and the position of the pulsar is expected because GeV-emitting particles are long-lived and the bulk of them is released during the initial spin-down timescale of the pulsar, typically over a few hundred to a few thousand years~\citep{Zhang20b}.~For six out of the eight sources discussed here, those with $\gtrsim$\,10 kyr pulsars, the offset distances from pulsars are reasonable as they imply transverse velocities in the range 100--1000\,km\,s$^{-1}$, in agreement with the velocity distribution inferred by~\citet{Verbunt17}.~Yet, in the case of 2FGES~J1119.3$-$6126 and 2FGES~J1930.3+1838, involving young pulsars, the physical offsets are relatively large and imply transverse velocities in excess of 1000\,km\,s$^{-1}$, especially for 2FGES~J1930.3+1838.~This, together with the above argument on source size, would tend to dismiss the latter source as a good pulsar halo candidate.

We emphasize that, although primarily motivated by pulsar halos, the above quantitative arguments can also be used to assess the likelihood of a PWN interpretation of the selected 2FGES sources: radiation efficiency in PWNe can be expected to be comparable to or lower than that of halos because a higher fraction of the particle energy is channeled into synchrotron radiation due to the high magnetic fields within the nebulae; and the typical extent of PWNe is smaller than that of halos, although there are several known objects with radii as large as 20$-$30\,pc \citep{Abdalla18b}.

\begin{deluxetable}{lccccccc}
\tabletypesize{\footnotesize}
\tablecaption{2FGES Sources Considered as Pulsar Halo Candidates~\label{tab:psr_powered_properties}}
\tablenum{10}
\tablehead{[-1.ex]
\colhead{2FGES Name} & \colhead{PSR Name} & \colhead{$d$} & \colhead{$\tau_{\rm{c}}$} & \colhead{$\dot{E}$} & \colhead{PSR Offset} & \colhead{$\eta$} & \colhead{$R_{\rm{GeV}}$} \\ 
\colhead{} & \colhead{} & \colhead{(kpc)} & \colhead{(kyr)} & \colhead{(erg s$^{-1}$)} & \colhead{(pc)} & \colhead{(\%)} & \colhead{(pc)} 
}
\startdata
J0006.0+7319 & J0007+7303 & 1.4 & 13.90 & 4.51e+35 & 6.8$\pm$3.9 & 0.64$\pm$0.14 & 20.8$\pm$2.9 \\
J1101.7$-$6110 & J1101$-$6101 & 7.0 & 116.00 & 1.36e+36 & 18.3$\pm$16.3 & 6.97$\pm$2.03 & 119.7$\pm$14.7 \\
J1112.6$-$6059 & J1112$-$6103 & 4.5 & 32.70 & 4.53e+36 & 5.5$\pm$4.7 & 1.00$\pm$0.21 & 30.6$\pm$4.7 \\
J1119.3$-$6126 & J1119$-$6127 & 6.4 & 1.61 & 2.30e+36 & 3.4$\pm$3.5 & 3.33$\pm$0.65 & 35.7$\pm$5.6 \\
J1417.5$-$6057 & J1418$-$6058 & 1.9 & 10.30 & 4.95e+36 & 5.0$\pm$1.0 & 0.25$\pm$0.04 & 5.6$\pm$1.3 \\
J1647.6$-$4551$^\dag$ & J1648$-$4611 & 4.5 & 110.00 & 2.09e+35 & 26.5$\pm$6.1 & 39.90$\pm$8.23 & 42.1$\pm$5.5 \\
J1913.5+1024 & J1913+1011 & 4.6 & 169.00 & 2.87e+36 & 17.7$\pm$9.5 & 2.71$\pm$0.53 & 62.0$\pm$6.4 \\
J1930.3+1838$^\dag$ & J1930+1852 & 6.2 & 2.89 & 1.20e+37 & 24.8$\pm$13.7 & 1.28$\pm$0.25 & 88.3$\pm$10.8 \\
\enddata
\tablecomments{Columns 2--5 represent the bright pulsars potentially linked to the 2FGES sources, along with their distance ($d$), characteristic age ($\tau_{\rm{c}}$), and spin-down power ($\dot{E}$), respectively.~Column 6 shows the physical offset of each 2FGES source from the associated pulsar.~Column 7 provides the pulsar conversion efficiency ($\eta$).~The physical size of the 2FGES sources in parsec, considering the 68\% containment radius, is given in column 8.~The physical offsets in column 6 and the physical extents of the 2FGES sources in column 8 are computed using the distances provided in column 3.~The quoted uncertainties in columns 6--8 come from the statistical errors on the 2FGES position, the measured energy fluxes, and the 2FGES size.~Sources marked with a dagger are most likely not good pulsar halo candidates for reasons detailed in the text.}
\end{deluxetable}

\section{Summary and Conclusions} \label{sec:Conclusions}

We have conducted a systematic search for extended sources above 10\,GeV using 14 years of \textit{Fermi}-LAT data, focusing on the low-latitude regions of the inner Galactic plane.~Our primary motivation is the exploration of the pulsar halo phenomenon at GeV energies, a range that remains poorly explored.~Yet, the outcome of our work is of much broader interest and can be relevant to the study of SNRs, PWNe, or other source classes, as well as to the identification of GeV counterparts to TeV and PeV sources.

The search resulted in a catalog of 40 significantly extended sources, with a median spatial extent of $r_{68}\mathrel{=}0.6^{\circ}$ and a median photon index of 2.2.~These sources have integrated photon flux above 10\,GeV ranging from 1$\times10^{-10}$ ph cm$^{-2}$ s$^{-1}$ to 6$\times10^{-9}$ ph cm$^{-2}$ s$^{-1}$.~These 40 sources remained significant when analyzed using a set of field-specific alternative interstellar emission models.~This, combined with the relatively hard spectral index of the majority of sources, gives us confidence that most of the detections are robust and not artifacts of the known imperfect modeling of the bright large-scale interstellar emission of the Galaxy.

A visual inspection of the significance maps revealed extended sources with morphologies well captured by the assumed 2D Gaussian intensity distribution, with no significant excesses at comparable spatial scales for 35 out of 40 sources.~In five cases, the modeling of the emission across the entire field was not satisfactory enough, likely due to complex and crowded regions, leading to their classification as dubious and necessitating further dedicated analyses.~Among the 35 reliable sources, 13 correspond to already-known extended objects in the 4FGL-DR3 catalog, which our analysis pipeline has superseded them with slightly different morphological models that better align with the $>10$\,GeV data.~This leaves 22 newly detected extended sources, which were cross-correlated with known sources in the 4FGL-DR3, HGPS, 3HWC, and 1LHAASO catalogs, as well as with bright pulsars having $\dot{E}/d^{2}\,\geq\,10^{34}$ erg s$^{-1}$ kpc$^{-2}$.~This procedure, essentially based on positional coincidence, resulted in the following associations: 13 sources coincide with known TeV sources, including four that also coincide with at least one bright pulsar; another two sources coincide with a bright pulsar but no TeV sources; six sources coincide only with known 4FGL point sources, thereby providing extension measurements of these sources for the first time; and one orphan source lacks counterparts in the scanned catalogs.~Based on these associations and prior identification efforts by other groups, we highlight a selection of the eight particularly interesting sources that may be linked to pulsars, either as classical wind nebulae or pulsar halos.

These results call for more dedicated analyses of the corresponding sources, in particular to refine their morphological description using, for instance, different geometrical models or resolving single sources into multiple smaller neighboring components.

In addition to the main 2FGES catalog, we present a supplementary list of 16 extended sources that exhibit characteristics of potential γ-ray extended emission but fall just below the catalog’s TS criterion (see Appendix~\ref{sec:appendix} for details).

\acknowledgments
$\it Acknowledgements.$~The \textit{Fermi}-LAT Collaboration acknowledges generous ongoing support
from a number of agencies and institutes that have supported both the
development and the operation of the LAT as well as scientific data analysis.~These include the National Aeronautics and Space Administration and the
Department of Energy in the United States, the Commissariat \`a l'Energie Atomique
and the Centre National de la Recherche Scientifique / Institut National de Physique
Nucl\'eaire et de Physique des Particules in France, the Agenzia Spaziale Italiana
and the Istituto Nazionale di Fisica Nucleare in Italy, the Ministry of Education,
Culture, Sports, Science and Technology (MEXT), High Energy Accelerator Research
Organization (KEK) and Japan Aerospace Exploration Agency (JAXA) in Japan, and
the K.~A.~Wallenberg Foundation, the Swedish Research Council and the
Swedish National Space Board in Sweden.~Additional support for science analysis during the operations phase is gratefully
acknowledged from the Istituto Nazionale di Astrofisica in Italy and the Centre
National d'\'Etudes Spatiales in France.~This work performed in part under DOE
Contract DE-AC02-76SF00515.~We acknowledge financial support by CNES for the exploitation of \textit{Fermi}-LAT observations.~We acknowledge financial support from the French Agence Nationale de la Recherche under reference ANR-19-CE31-0014 (GAMALO project).
\facility{\textit{Fermi}-LAT.}
\software{$\mathtt{fermipy}$ python package~\citep{Wood17} version 1.0.1, $\mathtt{Fermitools}$ version 2.0.8, $\mathtt{Astropy}$~\citep{Astropy13}}.

\appendix
\renewcommand{\thetable}{A\arabic{table}}
\setcounter{table}{0}

\section{Supplementary List of Extended Sources with Intermediate TS Values (16--25)} \label{sec:appendix}
We present a supplementary list of 16 extended sources with TS$_{\rm{ext}}$ above 16 and TS values between 16 and 25, using the standard IEM.~Notably, eight of these sources fall below the TS $=$ 16 limit when evaluated with alternative IEMs.~Although these sources do not meet the stricter criteria (TS $\geq 25$) applied in the final 2FGES catalog, they represent potentially interesting $\gamma$-ray emission candidates just below the established threshold for robust detection.~This list is intended to guide further investigation through future observations or refined analytical techniques.~Detailed information on these sources is provided in Table~\ref{tab:A1}.

\startlongtable
\tabletypesize{\scriptsize}
\begin{deluxetable*}{lccccccc}
\tablecaption{Summary of the Properties of the Potential Extended Sources with Intermediate TS Values (16--25) \label{tab:A1}}
\tablecolumns{9}
\tablehead{[-1.ex]
\colhead{2FGES Name} & \colhead{$l$} & \colhead{$b$} & \colhead{$r_{68}$} & \colhead{$\Gamma$ (\%)} & \colhead{$F_{>\,\rm{10\,GeV}}$ (\%)} & \colhead{TS$_{\rm{ext}}$} & \colhead{TS$_{\rm{std.}}$ (TS$_{\rm{alt.}}$)} \\ 
[-1ex]
\colhead{} & \colhead{($^{\circ}$)} & \colhead{($^{\circ}$)} & \colhead{($^{\circ}$)} & \colhead{} & \colhead{($\times\,10^{-10}\,\rm{cm}^{-2}\,\rm{s}^{-1}$)} & \colhead{} & \colhead{}}
\startdata
J0730.0$-$1456 & 230.60$\pm$0.17 & 1.52$\pm$0.12 & 0.55$\pm$0.14 & 1.68$\pm$0.25 (0) & 0.62$\pm$0.21 ($-$3) & 18.60 & 20.97 (18.26) \\
J0852.8$-$4708 & 266.98$\pm$0.11 & $-$1.63$\pm$0.12 & 0.79$\pm$0.16 & 1.63$\pm$0.16 (4) & 2.27$\pm$0.51 ($-$10) & 26.95 & 28.87 (22.12) \\
J0855.8$-$4425$^{\myhollowstar}$ & 265.23$\pm$0.09 & 0.53$\pm$0.19 & 1.04$\pm$0.09 & 3.03$\pm$0.59 (6) & 2.36$\pm$0.56 ($-$19) & 21.43 & 21.46 (14.41) \\
J1552.3$-$5734$^{\myhollowstar}$ & 325.33$\pm$0.20 & $-$2.80$\pm$0.70 & 3.47$\pm$0.81 & 2.78$\pm$0.43 ($-$15) & 6.80$\pm$1.37 ($-$61) & 26.51 & 26.83 (3.98) \\
J1627.1$-$4906 & 334.88$\pm$0.07 & $-$0.10$\pm$0.05 & 0.24$\pm$0.04 & 1.69$\pm$0.21 (2) & 1.45$\pm$0.41 (3) & 22.28 & 24.25 (24.09) \\
J1647.5$-$4643$^{\myhollowstar}$ & 338.94$\pm$0.10 & $-$1.04$\pm$0.11 & 3.10$\pm$0.21 & 2.47$\pm$0.14 (0) & 22.10$\pm$3.27 ($-$59) & 46.78 & 48.12 (7.50) \\
J1707.8$-$4413 & 343.11$\pm$0.13 & $-$2.26$\pm$0.15 & 0.72$\pm$0.14 & 2.18$\pm$0.20 (1) & 3.26$\pm$0.61 ($-$26) & 38.00 & 38.92 (20.42) \\
J1731.8$-$3354 & 354.23$\pm$0.19 & $-$0.17$\pm$0.09 & 3.50$\pm$0.23 & 2.29$\pm$0.09 (0) & 39.50$\pm$3.65 ($-$55) & 124.71 & 125.14 (23.47) \\
J1820.3$-$1159 & 18.73$\pm$0.31 & 1.31$\pm$0.27 & 2.28$\pm$0.32 & 2.49$\pm$0.19 (0) & 14.00$\pm$1.89 ($-$45) & 57.95 & 59.44 (16.85) \\
J1848.6$-$0131$^{\myhollowstar}$ & 31.25$\pm$0.13 & $-$0.07$\pm$0.08 & 0.65$\pm$0.10 & 2.17$\pm$0.17 (0) & 5.68$\pm$1.03 ($-$39) & 35.10 & 35.41 (13.22) \\
J1855.8+0135$^{\myhollowstar}$ & 34.85$\pm$0.07 & $-$0.26$\pm$0.06 & 0.57$\pm$0.09 & 2.95$\pm$0.35 ($-$3) & 5.26$\pm$1.28 ($-$86) & 18.22 & 18.22 (0.36) \\
J1925.7+1637$^{\myhollowstar}$ & 51.58$\pm$0.15 & 0.18$\pm$0.14 & 0.78$\pm$0.12 & 2.14$\pm$0.20 (1) & 3.20$\pm$0.67 ($-$43) & 27.35 & 27.34 (8.26) \\
J2005.8+3413$^{\myhollowstar}$ & 71.32$\pm$0.21 & 1.25$\pm$0.17 & 0.77$\pm$0.13 & 1.62$\pm$0.21 (0) & 1.32$\pm$0.43 ($-$30) & 19.61 & 19.72 (8.92) \\
J2028.4+4118 & 79.69$\pm$0.08 & 1.52$\pm$0.06 & 2.39$\pm$0.06 & 2.19$\pm$0.04 (0) & 53.70$\pm$8.04 ($-$44) & 44.10 & 44.08 (18.66) \\
J2034.2+3840$^{\myhollowstar}$ & 78.22$\pm$0.08 & $-$0.95$\pm$0.11 & 0.56$\pm$0.10 & 2.45$\pm$0.25 (0) & 2.61$\pm$0.46 ($-$42) & 35.95 & 39.17 (12.60) \\
J2107.8+5152 & 92.15$\pm$0.15 & 2.90$\pm$0.15 & 0.83$\pm$0.15 & 2.10$\pm$0.22 (0) & 2.00$\pm$0.45 ($-$18) & 28.28 & 28.42 (18.78) \\
\enddata
\tablecomments{Columns 2--4 provide the Galactic longitude, Galactic latitude, and the spatial extension ($r_{68}$) of the sources, respectively, derived from a 2D Gaussian fit used as the spatial model.~The photon index $\Gamma$ and the 10\,GeV--1\,TeV photon flux, along with their statistical uncertainties, are presented in columns 5 and 6, respectively.~The values in parentheses indicate the relative impact of using the alternative IEM, expressed as a percentage and derived as $\nicefrac{(X_{\rm{alt.}} - X_{\rm{std.}})}{X_{\rm{std.}}}$, where $X_{\rm{std.}}$ and $X_{\rm{alt.}}$ represent the measured quantities using the standard and alternative IEMs, respectively.~Columns 7 and 8 provide the TS$_{\rm{ext}}$ and TS values obtained using the standard IEM in the analysis, with the TS values obtained using the alternative IEM shown in parentheses.~Sources that fall below the TS $=$ 16 limit when evaluated with alternative IEMs are marked with an open star ($\myhollowstar$).}
\end{deluxetable*}


\begin{thebibliography}{}
\bibitem[Abdalla et al.(2020)]{Abdalla20} Abdalla, H., Adam, R., Aharonian, F., et al.\ 2020, A\&A, 635, A167 
\bibitem[Abdalla et al.(2018a)]{Abdalla18} Abdalla, H., Abramowski, A., Aharonian, F., et al.\ 2018, A\&A, 612, A1 
\bibitem[Abdalla et al.(2018b)]{Abdalla18b} Abdalla, H., Abramowski, A., Aharonian, F., et al.\ 2018, A\&A, 612, A2 
\bibitem[Abdo et al.(2009a)]{Abdo09} Abdo, A.~A., Ackermann, M., Ajello, M., et al.\ 2009a, APh, 32, 193 
\bibitem[Abdollahi et al.(2020)]{Abdollahi20} Abdollahi, S., Ballet, J., Fukazawa, Y., Katagiri, H., \& Condon, B.\ 2020, ApJ, 896, 76 
\bibitem[Abdollahi et al.(2022)]{Abdollahi22} Abdollahi, S., Acero, F., Baldini, L., et al.\ 2022, ApJS, 260, 53 
\bibitem[Abeysekara et al.(2017)]{Abeysekara17} Abeysekara, A.~U., Albert, A., Alfaro, R., et al.\ 2017, Science, 358, 911 
\bibitem[Acero et al.(2016)]{Acero16} Acero, F., Ackermann, M., Ajello, M., et al.\ 2016, ApJS, 224, 8 
\bibitem[Ackermann et al.(2012)]{Ackermann12} Ackermann, M., Ajello, M., Albert, A., et al.\ 2012, ApJS, 203, 4 
\bibitem[Ackermann et al.(2016)]{Ackermann16} Ackermann, M., Ajello, M., Atwood, W.~B., et al.\ 2016, ApJS, 222, 5 
\bibitem[Ackermann et al.(2017)]{Ackermann17} Ackermann, M., Ajello, M., Baldini, L., et al.\ 2017, ApJ, 843, 139 
\bibitem[Aharonian et al.(2021)]{Aharonian21} Aharonian, F., An, Q., Axikegu, Bai, L.X., et al.\ 2021, Phys. Rev. Lett., 126, 241103 
\bibitem[Aharonian et al.(2022)]{Aharonian22} Aharonian, F., Ashkar, H., Backes, M., et al.\ 2022, A\&A, 666, A124 
\bibitem[Ajello et al.(2021)]{Ajello21} Ajello, M., Atwood, W.~B., Axelsson, M., et al.\ 2021, ApJS, 256, 12 
\bibitem[Akaike(1974)]{Akaike74} Akaike, H.\ 1974, IEEE Trans. Autom. Control, 19, 716 
\bibitem[Albert et al.(2020)]{Albert20} Albert, A., Alfaro, R., Alvarez, C., et al.\ 2020, ApJ, 905, 76 
\bibitem[Archer(2018)]{Archer18} Archer, A., Benbow, W., Bird, R., et al.\ 2018, ApJ, 862, 41 
\bibitem[$\rm{Astropy}$~Collaboration(2013)]{Astropy13} The Astropy Collaboration, Robitaille, T.~P., Tollerud, E.~J., Greenfield, P., et al.\ 2013, A\&A, 558, A33 
\bibitem[Atwood et al.(2009)]{Atwood09} Atwood, W.~B., Abdo, A.~A., Ackermann, M., et al.\ 2009, ApJ, 697, 1071
\bibitem[Atwood et al.(2013)]{Atwood13} Atwood, W., Albert, A., Baldini, L., et al.\ 2012 \textit{Fermi} Symp. Proceedings, eConf, C121028 (2013); arXiv:1303.3514 
\bibitem[Ballet et al.(2023)]{Ballet23} Ballet, J., Bruel, P., Burnett, T.~H., Lott, B., \& the Fermi-LAT collaboration\ 2023, arXiv:2307.12546 
\bibitem[Bruel et al.(2018)]{Bruel18} Bruel, P., Burnett, T.~H., Digel, S.~W., et al.\ 2018, arXiv:1810.11394 
\bibitem[Cao et al.(2024)]{Cao24} Cao, Z., Aharonian, F., An, Q., et al.\ 2024, ApJS, 271, 25 
\bibitem[Chernoff(1954)]{Chernoff54} Chernoff, H.\ 1954, Ann. Math. Statist., 25, 573  
\bibitem[De La Torre Luque et al.(2022)]{DeLaTorreLuque22} De~La~Torre~Luque, P., Fornieri, O., \& Linden, T.\ 2022, Phys. Rev. D, 106, 123033 
\bibitem[Di Mauro et al.(2019)]{DiMauro19} Di Mauro, M., Manconi, S., \& Donato, F.\ 2019, Phys. Rev. D, 100, 123015 
\bibitem[Di Mauro et al.(2021)]{DiMauro21} Di Mauro, M., Manconi, S., Negro, M., \& Donato, F.\ 2021, Phys. Rev. D, 104, 103002 %
\bibitem[Evoli et al.(2018)]{Evoli18} Evoli, C., Linden, T. \& Morlino, G.\ 2018 Phys. Rev. D, 98, 063017 
\bibitem[Fang et al.(2018)]{Fang18} Fang, K., Bi, X.-J.,Yin, P.-F., \& Yuan, Q.\ 2018, ApJ, 863, 30 
\bibitem[Fang et al.(2019)]{Fang19} Fang, K., Bi, X.-J., \& Yin, P.-F.\ 2019, Mon. Not. R. Astron. Soc., 488, 4074 %
\bibitem[Ge et al.(2022)]{Ge22} Ge, T.-T., Sun, X.-N.,Yang, R.-Z., Liang, Y.-F., \& Liang, E.-W\ 2022, Mon. Not. R. Astron. Soc., 517, 5121 
\bibitem[$\rm{H.E.S.S.}$~Collaboration(2020)]{HESS20} H.E.S.S. Collaboration, Abdalla, H., Adam, R., Aharonian, F., et al.\ 2020, A\&A, 635, A167 
\bibitem[H\"{a}rer et al.(2023)]{Harer23} H\"{a}rer, L.~K., Reville, B., Hinton, J., Mohrmann, L., \& Vieu, T.\ 2023, A\&A, 671, A4 
\bibitem[Hofverberg et al.(2010)]{Hofverberg10} Hofverberg, P., Chaves, R.~C.~G., Fiasson, A., et al., 2010, in 25$^{th}$ Texas Symposium on Relativistic Astrophysics, 196
\bibitem[Klingler et al.(2023)]{Klingler23} Klingler, N., Hare, J., Kargaltsev, O., Pavlov, G.~G., \& Tomsick, J.\ 2023, ApJ, 950, 177 
\bibitem[Linden et al.(2017)]{Linden17} Linden, T., Auchettl, K., Bramante, J., et al.\ 2017, Phys. Rev. D, 96, 103016 
\bibitem[Linden et al.(2018)]{Linden18} Linden, T., \& Buckman, B. J.\ 2018, Phys. Rev. Lett., 120, 121101 
\bibitem[Liu et al.(2019)]{Liu19} Liu, R.-Y., Yan, H., \& Zhang. H.\ 2019, Phys. Rev. Lett., 123, 221103 
\bibitem[L\'{o}pez-Coto \& Giacinti(2018)]{LopezCoto18} L\'{o}pez-Coto, R., \& Giacinti, G.\ 2018, Mon. Not. R. Astron. Soc., 479, 4526 %
\bibitem[Manchester et al.(2005)]{Manchester05} Manchster, R. N., Hobbs, G. B., Teoh, A., \& Hobbs, M.\ 2005, AJ, 129, 1993 
\bibitem[Mart\'{i}-Devesa \& Reimer(2021)]{MartiReimer21} Mart\'{i}-Devesa, G. \& Reimer, O.\ 2021, A\&A, 654, A44 
\bibitem[Martin et al.(2022a)]{Martin22a} Martin, P., Marcowith, A., \& Tibaldo, L.\ 2022, A\&A, 665, A132 
\bibitem[Martin et al.(2022b)]{Martin22b} Martin, P., Tibaldo, L., Marcowith, A. \& Abdollahi, S.\ 2022, A\&A, 666, A7 
\bibitem[Martin et al.(2024)]{Martin24} Martin, P., de Guillebon, L., Collard, E., et al.\ 2024, A\&A, 690, A116 
\bibitem[Mukhopadhyay \& Linden(2022)]{Mukhopadhyay.Linden22} Mukhopadhyay, P. \& Linden, T.\ 2022, Phys. Rev. D, 105, 123008 
\bibitem[Park et al.(2023)]{Park23} Park, J., Kim, C., Woo, J., et al.\ 2023, ApJ, 945, 66 
\bibitem[Profumo et al.(2018)]{Profumo18} Profumo, S., Reynoso-Cordova, J., Kaaz, N., \& Silverman, M.\ 2018, Phys. Rev. D, 97, 123008 
\bibitem[Reich \& Sun(2019)]{Reich.Sun19} Reich, W. \& Sun, X.-H.\ 2019, Res. Astron. Astrophys., 19, 045 
\bibitem[Schinzel et al.(2017)]{Schinzel17} Schinzel, F.~K., Petrov, L., Taylor, G.~B., \& Edwards, P.~G.\ 2017, ApJ, 838, 139 
\bibitem[Schroer et al.(2021)]{Schroer21} Schroer, B., Pezzi, O., Caprioli, D., Haggerty, C., \& Blasi, P.\ 2021, ApJ, 914, L13 %
\bibitem[Schroer et al.(2022)]{Schroer22} Schroer, B., Pezzi, O., Caprioli, D., Haggerty, C.~C., \& Blasi, P.\ 2021, Mon. Not. R. Astron. Soc., 512, 233 %
\bibitem[Smith et al.(2023)]{Smith23} Smith, D.~A., Abdollahi, S., Ajello, M., et al.\ 2023, ApJ, 958, 191 
\bibitem[Steinmassl et al.(2023)]{Steinmassl23} Steinmassl, S., Breuhaus, M., White, R., Reville, B., \& Hinton, J.~A.\ 2023, A\&A, 679, A118 
\bibitem[Sudoh et al.(2019)]{Sudoh19} Sudoh, T., Linden, T., \& Beacom, J. F.\ 2019, Phys. Rev. D, 100, 043016 
\bibitem[Sun et al.(2022)]{Sun22} Sun, X.-N., Yang, R.-Z, \& Liang, E.-W.\ 2022, A\&A, 659, A83 
\bibitem[Tang \& Piran(2019)]{Tang.Piran19} Tang, X. \& Piran, T.\ 2019, Mon. Not. R. Astron. Soc., 484, 3491 
\bibitem[Verbunt et al.(2017)]{Verbunt17} Verbunt, F., Igoshev, A., \& Cator, E.\ 2017, A\&A, 608, A57 
\bibitem[Wakely \& Horan(2008)]{Wakely.Horan08} Wakely, S.~P. \& Horan, D.\ 2008, 30th ICRC, 3, 1341 
\bibitem[Wood et al.(2017)]{Wood17} Wood, M., Caputo, R., Charles, E., et al.\ 2017, arXiv:1707.09551 
\bibitem[Xi et al.(2019)]{Xi19} Xi, S.-Q., Liu, R.-Y., Huang, Z.-Q., Fang, K., \& Wang, X.-Y.\ 2019, ApJ, 878, 104 
\bibitem[Yao et al.(2017)]{Yao17} Yao, J.~M., Manchester, R.~N., \& Wang, N.\ 2017, ApJ, 835, 29 
\bibitem[Zeng et al.(2021)]{Zeng21} Zeng, H., Xin, Y., Zhang, S., \& Liu., S.\ 2021, ApJ, 910, 78 
\bibitem[Zhang et al.(2020a)]{Zhang20a} Zhang, H.-M., Xi, S.-Q, Liu, R.-Y., et al.\ 2020, ApJ, 889, 12 
\bibitem[Zhang et al.(2020b)]{Zhang20b} Zhang, Y., Liu, R.-Y., Chen, S.~Z., \& Wang, X.-Y.\ 2020, ApJ, 922, 130 
\end{thebibliography}
\end{document}